\documentclass[11pt]{article}
\usepackage{array}
\usepackage{dsfont}
\usepackage{wrapfig}
\usepackage[small, it]{caption}
\usepackage{latexsym} 
\usepackage{amsmath, mathtools}
\usepackage{amssymb, amsthm}
\usepackage{graphicx}
\usepackage[english]{babel}
\usepackage{layout}
\usepackage{fancyhdr}
\usepackage{slashed}	
\usepackage{bbm}
\usepackage{tabularx}
\usepackage[usenames]{color}
\usepackage[all]{xy}

\makeatletter
\newif\if@preliminary
\@preliminaryfalse
\def\preliminary{\@preliminaryfalse}

\def\bq{\begin{equation}}
\def\eq{\end{equation}}
\def\ba{\begin{eqnarray}}
\def\ea{\end{eqnarray}}

\def\preprintno#1{\def\@preprintno{#1}}
\def\address#1{\def\@address{#1}}
\def\email#1#2{\thanks{\tt #1@{}#2}}
\def\abstract#1{\def\@abstract{#1}}
\renewcommand\abstractname{ABSTRACT}
\newlength\preprintnoskip
\newlength\abstractwidth
\@titlepagetrue
\renewcommand\maketitle{\begin{titlepage}
  \let\footnotesize\small
  \hfill\parbox{\preprintnoskip}{
  \begin{flushright}\@preprintno\end{flushright}}\hspace*{1cm}
  \vskip 60\p@
  \begin{center}
    {\Large\bf\boldmath \@title \par}\vskip 1cm
    {\sc\@author \par}\vskip 3mm
    {\@address \par}
    \if@preliminary
      \vskip 2cm {\large\sf PRELIMINARY DRAFT \par \@date}
    \fi
  \end{center}\par
  \@thanks
  \vfill
  \begin{center}
    \parbox{\abstractwidth}{\centerline{\abstractname}
    \vskip 3mm
    \@abstract}
  \end{center}
  \end{titlepage}
  \setcounter{footnote}{0}
  \let\thanks\relax\let\maketitle\relax
  \gdef\@thanks{}\gdef\@author{}\gdef\@address{}
  \gdef\@title{}\gdef\@abstract{}\gdef\@preprintno{}
}

\def\@citex[#1]#2{\if@filesw\immediate\write\@auxout{\string\citation{#2}}\fi
  \def\@citea{}\@cite{\@for\@citeb:=#2\do
    {\@citea\def\@citea{,\penalty\@m}\@ifundefined
       {b@\@citeb}{{\bf ?}\@warning
       {Citation `\@citeb' on page \thepage \space undefined}}%
\hbox{\csname b@\@citeb\endcsname}}}{#1}}
\def\citerange{\@ifnextchar [{\@tempswatrue\@citexr}{\@tempswafalse\@citexr[]}}
\def\@citexr[#1]#2{\if@filesw\immediate\write\@auxout{\string\citation{#2}}\fi
  \def\@citea{}\@cite{\@for\@citeb:=#2\do
    {\@citea\def\@citea{--\penalty\@m}\@ifundefined
       {b@\@citeb}{{\bf ?}\@warning
       {Citation `\@citeb' on page \thepage \space undefined}}%
\hbox{\csname b@\@citeb\endcsname}}}{#1}}
\long\def\@makecaption#1#2{
  \vskip\abovecaptionskip
  \sbox\@tempboxa{#1: \emph{#2}}
  \ifdim \wd\@tempboxa >\hsize
    #1: \emph{#2}\par
  \else
    \hbox to\hsize{\hfil\box\@tempboxa\hfil}
  \fi
  \vskip\belowcaptionskip}

\begin{document}



\preprintno{DESY-12-178  \\[0.5\baselineskip] December 21, 2012} 

\title{A Fat Gluino in Disguise}

\author{
  J.~Reuter\email{juergen.reuter}{desy.de}$^{\,a}$ 
 and D.~Wiesler\email{daniel.wiesler}{desy.de}$^{\,a}$
}

\address{\it$^a$DESY Theory Group, Notkestr. 85, D--22603 Hamburg, Germany}

\vspace{-5cm}
\abstract{
In this paper, we investigate how a sizeable width-to-mass ratio for
a gluino, as is for example realized in GMSB scenarios, could affect the
discovery potential of gluinos at the LHC. More importantly, the
influence of the gluino being ``fat'' on the standard mass and spin
determination methods at the LHC are investigated. For this purpose,
we focus on gluino production at the LHC, where we do not factorize 
the first step in the gluino decay cascade, but treat the following
decay cascades step in factorization, including full spin
correlations. The effects of sizeable width-to-mass ratios from a few
up to 15-20 per cent on the endpoint of several mass determination
methods as well as on means for discrimination between BSM spin
paradigms like SUSY and UED are studied.
}

\maketitle
   
%
\section{Introduction and Finite Width Effects}

The Standard Model (SM) of particle physics describes all data in the
field since today with a very good accuracy. Nevertheless, there are 
many reasons for physics beyond the SM (BSM), namely the non-existence
of a dark matter particle within the SM, the insufficent amount of CP
violation to explain the baryon-antibaryon asymmetry in the universe,
and the instability of mass terms of fundamental scalar particles
against radiative corrections. Supersymmetry has been one of the most
favorable candidates to cure these problems. One of its main
predictions to be tested best at a hadron collider is the existence of
strongly interacting supersymmetric partner particles of the quarks and
the gluon, namely squarks and the gluino. Up to now, in the runs at 7
and 8 TeV center-of-mass energy, the Large Hadron Collider (LHC) has
not found any traces of supersymmetric particles, which means that
these particles cannot be too light. Supersymmetry is part of a large
class of BSM models, where new particles around the TeV scale have
only weak interactions, like in models with extra dimensions, or
Little Higgs models. These kind of models seems to be favored by
electroweak precision data, as there are no (big) deviations or
inconsistencies in the fit of the electroweak data to the
SM. Generically, new particles in weakly interacting models show up as
narrow resonances, where in most cases their width is below the
detector resolution of the LHC experiments. This is different from
strongly interacting scenarios like technicolor, composite Higgs
models or conformal sectors where new particles or more like broad
resonances (e.g. similar to the $\rho$ resonance) or even similar to
continuum-like excitations like in QCD or condensed matter physics. 

In the minimal supersymmetric SM (MSSM) and its most simple
extensions, all new particles are consequently rather narrow
resonances with a width-to-mass ratio, $\gamma := \Gamma/M$, of the
order of half a per cent or less. Besides the heavy Higgs bosons in
certain parts of parameter space, the gluino $\tilde{g}$ is the only
particle that can get a sizeable width-to-mass ratio. In the case, the
gluino is heavier than the squarks, there are many decay channels open
for the gluino, such that the gluino can easily access a width-to-mass
ratio of several, and even up to 15-20 per cent. The theoretical upper
limit for this ratio (taking the gluino mass to inifinity or
equivalently assuming quasi-massless squarks) is 32 per cent. More
details under which conditions gluinos can become ``fat'' are
summarized below, when we discuss the model setup for this study.

Since the most severe bottleneck of simulations with multi-particle
final states (at least, but not only at tree level) is the integration
over the high-dimensional phase space, almost all SUSY studies have
been performed with a factorized approach. This is motivated by the
fact that a production process of two SUSY particles is followed by
subsequent two- or three-body decays. The easiest aprroximation relies
on the narrow-width approximation (NWA) which precisely does this
factorization. This can be approved by folding in momentum smearing
according to a Breit-Wigner propagator
$N(q)/(q^2-m^2+im\Gamma)$, where $q$ is the four momentum, $m$
the mass and $\Gamma$ the width of the associated intermediate
particle. In many cases, the most severe effects come from the
numerator factor $N(q)$ which depends upon the quantum numbers of the
intermediate state, and essentially contains the spin density matrix
connecting the production and decay matrix elements. Spin and
polarisation information for the intermediate particle are important
e.g. in top or $W$ decays, or in BSM physics for the discrimination
between different assignments of particle spins, e.g. between SUSY and
UED. 

When the decay matrix element (squared) is integrated out, the small
width $\Gamma$ of the intermediate state particle is neglected, as
well as the (usually) very small dependence of production and decay
matrix elements on the momentum of the intermediate particle $q$. 
In a previous study~\cite{Berdine:2007uv} it was shown if one of the
five following cases is met, the NWA cannot be applied: (1) the
obvious violation, when $\gamma = \Gamma/M$ is not small, e.g. for the
$\rho$ resonance in QCD or a corresponding resonance in composite
models, (2) if the masses of mother and daughter particle are
near-degenerate, $m \sim M$ (decay threshold), e.g. typically in UED,
(3) if the mass of the intermediate is close to the kinematical limit
of the experiment, $M \sim \sqrt{s}$ (particle threshold), (4) if
there are big interferences between different exclusive decay channels
(quasi-combinatorial background), and finally (5) if there is a
propagator non-separable from the matrix element (e.g. in the case of
a non-trivial spectral density of particles or unparticles). In the
SM, there are only a few examples where deviations from the NWA really
matter, as many (fundamental) particles are relatively narrow (the $W$
with $\gamma = $2.5 per cent has the largest width-to-mass ratio) in
the SM. However, even here sometimes the NWA had to be given up in
order to get to predictions that are precise enough to match
experimental analyses, e.g. $e^+e^-\rightarrow W^+W^- \rightarrow
4f$~\cite{Denner:2005es}. In general BSM scenarios, such cases appear 
much more often, cf. e.g.~\cite{Hagiwara:2005wg,Reuter:2005us}. 
Though there have been systematic improvements of the
NWA~\cite{Kauer:2007nt,Uhlemann:2008pm} up to the order $\mathcal
O(\Gamma/M)$, it is always preferrable to use full matrix elements for
the processes under investigation. 

In this paper, we study whether and how much finite-width effects of a
heavy gluino do affect its detectability, and its mass and spin
determination at the LHC. Sec.~\ref{sec:simsetup} defines our
benchmark SUSY models, the studied processes and the setup of our
simulation. The results are shown in Sec.~\ref{sec:mass} for the
effects of finite widths on the mass determination and in
Sec.~\ref{sec:spin} for the spin determination. Finally, we conclude
in Sec.~\ref{sec:conclusions}.


\section{Simulation Setup and Benchmark Model}
\label{sec:simsetup}

For the analyses of off-shell effects on endpoint and shape
measurements of invariant mass distributions, we assume a SUSY
scenario where there are considerable branching ratios of the gluino
both into a first- or second-generation squark accompanied by a jet.
Furthermore, we assume the presence of a so-called \textit{golden
chain} with subsequent decays of the squark into the second-lightest
neutralino, which further decays into slepton and lepton, where the
slepton then ends up in another lepton and the lightest neutralino. 
This demands for the specific mass hierarchy in the SUSY particle
spectrum, $m_{\tilde{q}_L} > m_{\tilde{\chi}_2} > m_{\tilde{l}_R} >
m_{\tilde{\chi}_1}$, which however is not uncommon in phenomenological
SUSY models studied in the literature (early cMSSM scenarios such as 
\textit{SPS1a} \cite{Allanach:2002nj} inspired the presence of this
type of hierarchy). For the study of gluino width effects, we overlay
the two main production processes of gluino pair production with the
much more abundant -- in the case of light(er) squarks -- one of
associated gluino-squark production:
\begin{align}
  pp&\rightarrow \tilde{g_1}\tilde{g_2} + X \label{eq:glu_pair}\\
  pp&\rightarrow \tilde{g_1}\tilde{q}_{L/R} + X \label{eq:glu_squ}
  \quad . 
\end{align}
In the first case, Eq.~\eqref{eq:glu_pair}), we take asymmetric decay 
chains, where one of the two signal gluinos decays into two light
(without loss of generality down) quarks and the lightest neutralino, 
while the other one decays into two bottom quarks and the
second-to-lightest neutralino, which  further decays via an
intermediate (right handed) slepton to two 
corresponding leptons and a lightest neutralino: 
\begin{align}
  \tilde{g}_1&\rightarrow b\tilde{b}_i\rightarrow
  b\bar{b}\tilde{\chi}^0_2 \rightarrow b\bar{b} l^\pm \tilde{l}^\mp_R
  \rightarrow b\bar{b} l^\pm l^\mp
  \tilde{\chi}^0_1 \label{eq:benchmark_cascade}\\ 
  \tilde{g}_2&\rightarrow d\tilde{d}_L\rightarrow
  d\bar{d}\tilde{\chi}^0_1  
\end{align}
Notice the index $i$ at the bottom squark owing to the fact, that we
include both decay modes of the gluino into $\tilde{b}_1$ and
$\tilde{b}_2$. In the second case of squark-gluino associated
production (Eq.~\eqref{eq:glu_squ}), we simulate the prompt
squark decay into a quark and the lightest neutralino  
\begin{align}
  \tilde{q}_{L/R} \rightarrow q \tilde{\chi}^0_1.
\end{align}
With the focus on this particular exclusive final state being the same
for both production processes we omit additional complications from
combinatorial ambiguities. It allows us to study the consequences of
off-shell effects without having to suffer from SUSY or combinatorial
backgrounds (these have been studied recently in the context of purely
hadronic decay modes in~\cite{Pietsch:2012nu,combamb}. Preliminary
results on distortions due to off-shell effects have been shown
in~\cite{wiesler_phd}, while distortions of parton-level distributions
due to exotic non-standard SUSY particles which could mimic off-shell
and combinatorial effects have been studied in~\cite{Reuter:2010nx}. 

The last decay steps in the cascade of equation,
Eq.~\eqref{eq:benchmark_cascade}, are particularly well-known: they
are identical to the \textit{golden chain} with the replacement of a
(first or second generation) squark by a sbottom: $\tilde{q}_L
\rightarrow \tilde{b}_i$. This exclusive final state allows for a
quite generic study of most of the mass determination methods in the
literature~\cite{massdeterm}, while in principle simultaneously
reducing combinatorial mis-assignments due to the possibility of
$b$-tagging. Furthermore, in the analysis of off-shell gluino width
effects we are able to use methods based on the existence of sbottoms
in the cascade. 

In the following, we introduce a benchmark scenario that allows for
the abovementioned signal decay chain while capturing most of
the relevant phenomenological features like a Higgs mass around 125
GeV, heavy colored states beyond the LHC limits~\cite{expsearches} and
a rich phenomenology of decay patterns. To be as generic as possible,
we decided to make use of the phenomenological MSSM with 19 free
parameters (p19MSSM), without any high-scale relations among the
parameters. All parameters determining the model are given at the
electroweak/TeV scale. Their explicit values for the parameter point
chosen here are given in Table~\ref{tab:model_pars}. The mass
hierarchy for the decay chain above is valid for this parameter point,
and the branching ratios of the four successive two body decay steps
are of considerable size:  
\begin{center}
\begin{tabular}{rlr|rlr|rlr|rlr}
  $\tilde{g}$&$\rightarrow b\tilde{b}_{1} $& 10 \% &   $\tilde{b}_1
  $&$\rightarrow b \tilde{\chi}^0_2 $& 16 \% &  $\tilde{\chi}^0_2
  $&$\rightarrow e^\pm \tilde{e}^\mp_R $& 42 \% &   $\tilde{e}_R^\pm
  $&$\rightarrow e^\pm \tilde{\chi}^0_1 $& 100 \% \\ 
  $\tilde{g}$&$\rightarrow b\tilde{b}_{2} $& 07 \% &   $\tilde{b}_2
  $&$\rightarrow b \tilde{\chi}^0_2 $& 34 \% &  $\tilde{\chi}^0_2
  $&$\rightarrow \mu^\pm \tilde{\mu}^\mp_R $& 42 \% &
  $\tilde{\mu}_R^\pm $&$\rightarrow \mu^\pm \tilde{\chi}^0_1 $& 100
  \%\\ 
\end{tabular}
\end{center}\vspace{3mm}
Given these values, the total branching fraction of our exclusive
final state from signal gluinos decaying through the
benchmark cascade above is roughly 7 \%. This number has to be
taken with a grain of salt, as the underlying concept of factorized
cross sections for exclusive final states into cross sections times
branching ratio as within the NWA might no longer be a good
approximation in the presence of off-shell contributions.
However, in this study we investigate the effects arising from
precisely these kind of contributions far away from the resonant pole
of the propagator. A complete treatment should (at least) take into
account fully differential four- (or more) particle final
states. Since we use these figure merely as rough estimates to determine
the actual number of events we expect from our exclusive 
decay cascade final state, we refrain from such a calculation. All
masses of the spectrum were calculated using
\texttt{SOFTSUSY}~\cite{Allanach:2001kg}, while the particle decay
widths were obtained with \texttt{SUSYHIT}~\cite{Djouadi:2006bz}. An
overview of all model parameters is given in
Table~\ref{tab:model_pars}. 
\begin{table}
\centering  
\begin{tabular}{|c|c|c|c|c|c|c|c|c|c|}
\hline
$ M_1 $&$ M_2 $&$ M_3 $&$ A_t $&$ A_b $&$ A_\tau $&$ \mu $&$ M_A $&$
m_{\tilde{l}_L} $&$ m_{\tilde{\tau}_L} $ \\ 
\hline
150 & 250 & 1200 & 4000 & 4000 & 0 & 1500 & 1500 & 1000 & 1000 \\  
\hline
\hline
$ m_{\tilde{l}_R} $&$ m_{\tilde{\tau}_R} $&$ m_{\tilde{q}_L} $&$
m_{\tilde{q}^3_L} $&$ m_{\tilde{q}^u_R} $&$ m_{\tilde{q}^d_R} $&$
m_{\tilde{t}_R} $&$ m_{\tilde{b}_R} $&$ \tan \beta $ & \\ 
\hline
200 & 1000 & 1000 & 1000 & 1000 & 1000 & 4000 & 1000 & 10 & \\
\hline
\end{tabular}
\caption{Model parameters of the pMSSM parameter point studied in this
  paper. All numbers (except $\tan \beta$) are in units of GeV.}
\label{tab:model_pars}    
\end{table}
Feeding the obtained SLHA file~\cite{Skands:2003cj,Allanach:2008qq}
for the given 
mass spectrum into \texttt{PROSPINO}~\cite{Beenakker:1996ch}, we    
calculated the relevant cross sections for LHC at 14 TeV: for
squark-gluino associated production 376.0 fb, for gluino pair
production 47.6 fb, and neutralino-gluino associated production 3.7
fb, respectively. The   sum we multiply by the total branching
fraction of our exclusive final state (7 \%) to arrive at an event
number of roughly 9,000 for an overall integrated luminosity of 300
fb$^{-1}$. Several diluting effects such as detector acceptance,
b-tagging efficiencies and event selection criteria may further reduce
this figure. For that reason, we took a conservative estimate and
chose to analyse an event number of 5,000. The following simulations 
are based on the most recent version of the event generator
\texttt{WHIZARD}~\cite{whizard}. All generated events were exported to
\texttt{HepMC}~\cite{Dobbs:2001ck} format and afterwards passed
through a C++ analysis framework based on ROOT~\cite{Antcheva:2011zz},
which was specifically developed and tailored to the investigation of
deviations from off-shell contributions in spin and mass determination
methods.  

As we want to investigate effects that come directly from off-shell
propagators and interference effects, we compare full matrix-element
calculations with factorized approaches in the NWA at parton level, in
order to be able to disentangle their effects from pollution that
stems from QCD radiation, hadronization and detector effects. To study
a ``fat'' gluino, we use full matrix elements for the production
including the first decay of the gluino. More precisely, the first
part of Eq.~\eqref{eq:benchmark_cascade} is completely calculated 
in one step,  
\begin{align}
  p p \rightarrow (b \bar b \tilde{\chi}^0_2) +
  (\tilde{g}/\tilde{q}) \label{eq:factorize_prod} \qquad ,
\end{align}
including all interferences. On the other hand, the successive decays
of the second-to-lightest neutralino and the ``spectator'' gluino
or squark are factorised with full spin correlations using the
NWA. Furthermore, the phenomenological width-to-mass ratio $\gamma =
\Gamma / M$ is scanned over the following values  
\begin{align}
\gamma \in \{0.5 \%, 2.5 \%, 5.0 \%, 10.0 \%, 15.0 \%, 20 \%, 25
\%\}. 
\end{align} 
This set is chosen so as to resemble a broad range of possible widths
that might be realised in nature. While the lowest value is a typical
width for standard parameter points, values above 10 \% appear when
the gluino is clearly heavier than the squarks as in GMSB
setups. There is also a particular phenomenological reasoning behind
the range of values, which will be discussed at the beginning of the
next section. Using this relative width ratio $\gamma$ we study the
impact of off-shell contributions on a selected choice of mass and
spin determination methods listed in the next two sections.


\section{Effects on Mass Measurements}
\label{sec:mass}

\begin{figure}
  \includegraphics[width=0.47\textwidth]{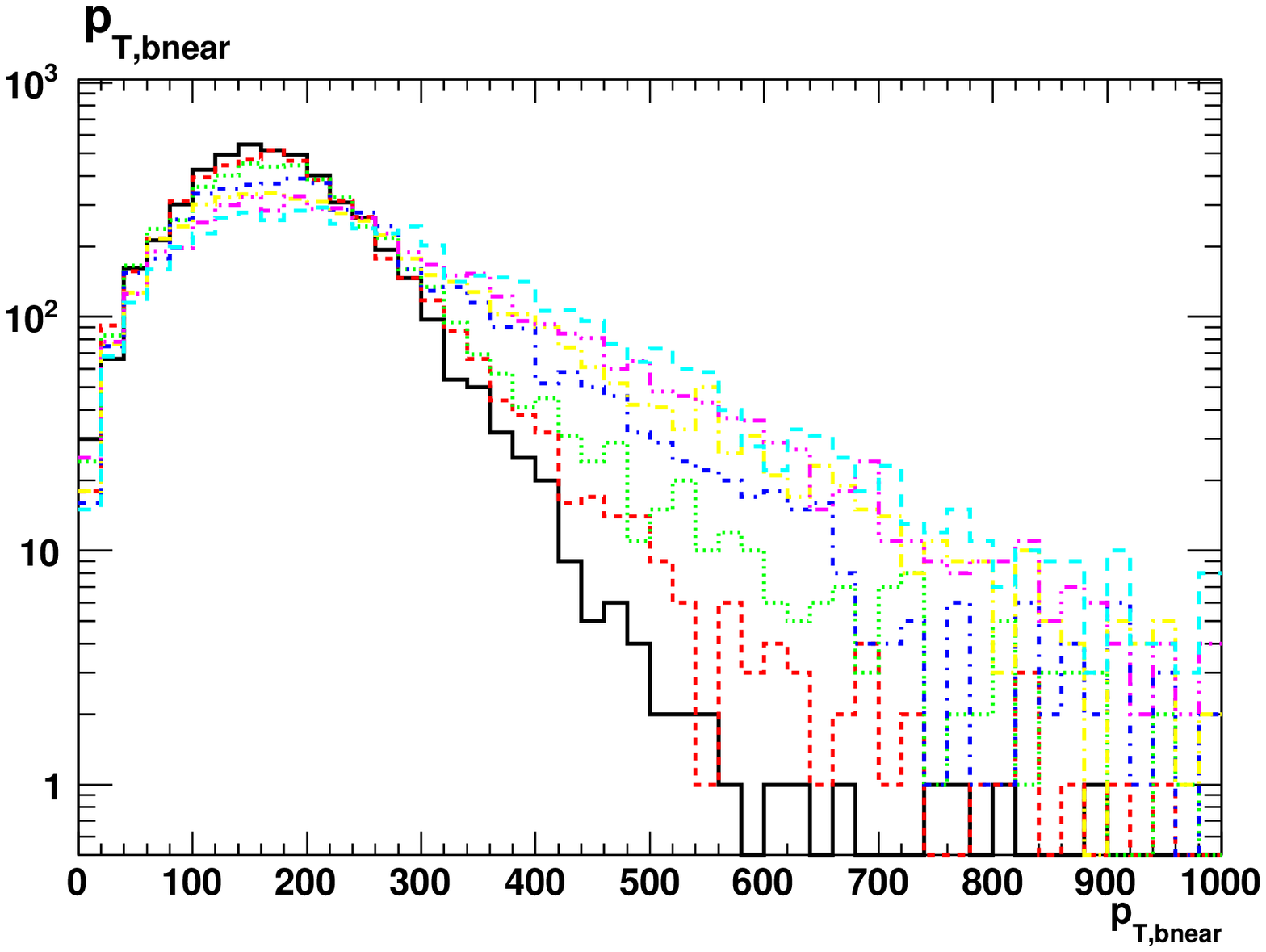}
  \includegraphics[width=0.47\textwidth]{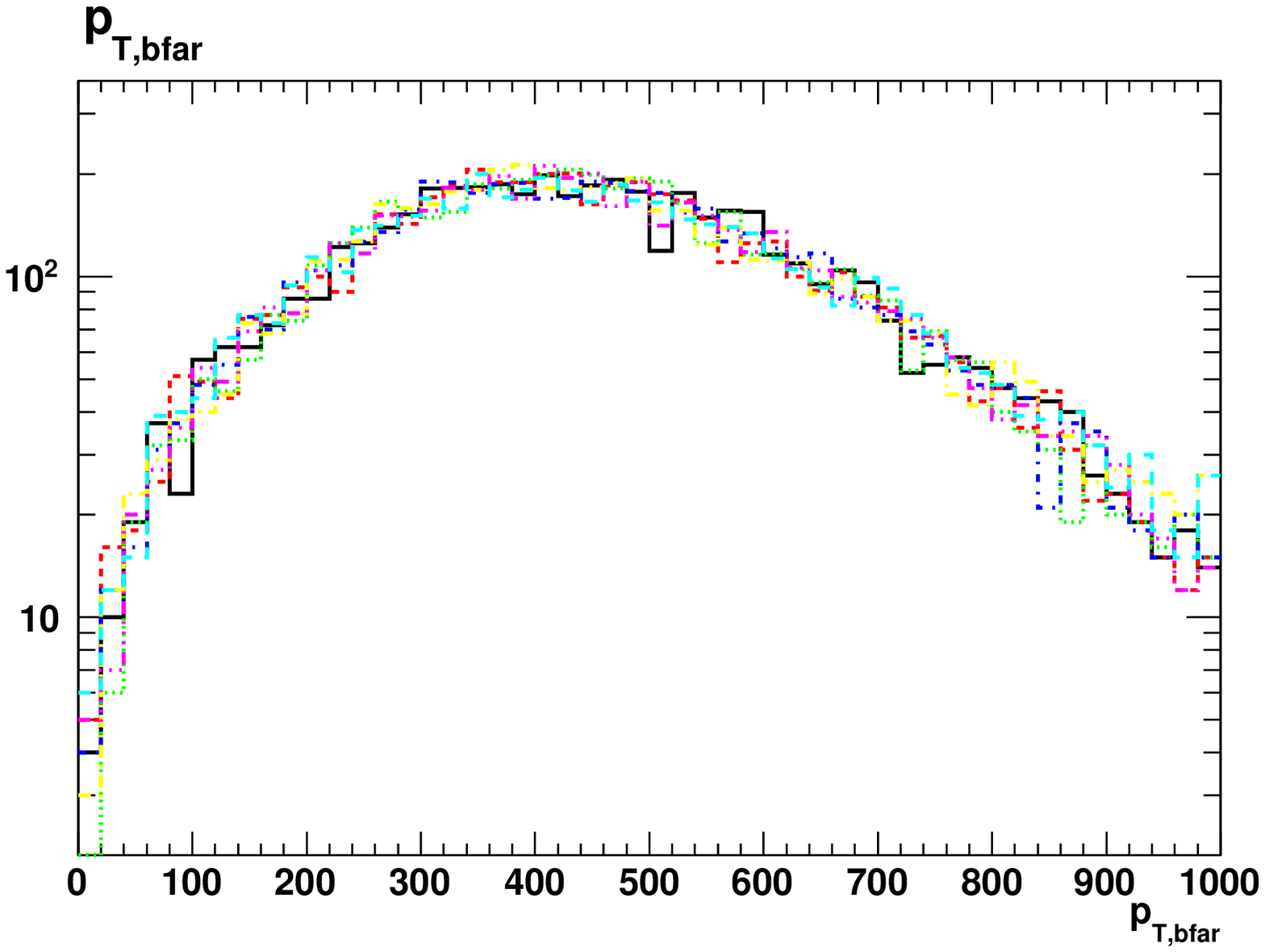}
  \caption{Transverse momentum distributions of the two bottom quarks
    in the signal cascasde (near from the gluino and far from the
    sbottom decay). The black (solid), red (short-dashed), green
    (dotted), blue (short-dashed-dotted), yellow (long-dashed-dotted),
    magenta (long-dashed-double-dotted) and cyan (long-dashed) line
    correspond to $\gamma =$ 0.5 \%, 2.5 \%, 5.0 \%, 10.0 \%, 15.0 \%,
    20.0 \% and 25.0 \%, respectively.} 
  \label{fig:pt_bottom_near_far}
\end{figure}

We begin our investigation with the study of several mass measurement
variables. A large width $\Gamma$ in the gluino propagator affects the
momenta of both the intermediate sbottom and the near (bottom) quark
$b_{n}$, as both are directly resulting out of the gluino
propagator. The far bottom quark $b_{f}$, coming from the sbottom
decay, on the other hand is expected to receive only a minor
modification and thus should not be part of distorted
invariant mass distributions. Fig.~\ref{fig:pt_bottom_near_far}
depicts the transverse momentum distributions of both the near and the
far bottom quark and their distortion with respect due to different
values of $\gamma$. The black (solid), red (short-dashed), green
(dotted), blue (short-dashed-dotted), yellow (long-dashed-dotted),
magenta (long-dashed-double-dotted) and cyan (long-dashed) line
correspond to $\gamma =$ 0.5 \%, 2.5 \%, 5.0 \%, 10.0 \%, 15.0 \%,
20.0 \% and 25.0 \%, respectively. As expected, the far bottom quark,
is almost unaffected by a fat gluino as it originates from the
subsequent decay of the sbottom. On the other hand, we observe that
the near bottom quark exhibits a considerable distortion in its $p_T$
distributions, already visible by eye. While there is an obvious
tendency for a severely increasing distortion within the first five
values of $\gamma$ (up to 15 \%), the two largest effective widths
(20 and 25 \%) exhibit only a slight further increase. These values
are anyhow to some extent academic, since in realistic scenarios they
are quite hard to realize: either these scenarios are already excluded
due to the Higgs boson discovery at 125 GeV, or the gluinos are so
heavy that they are out of reach of a 14 TeV LHC. In the first plot
for the transverse momentum distributions, we included them to
illustrate the effect of a moderate saturation. This observation 
motivates us to leave out the two highest values $\gamma =$ 20.0\%,
25.0 \% and from now on investigate the reduced range of
values up to 15 \%, also in order not to make the plots too
crowded. 

As the far quark/jet is (almost) not affected by the
off-shell effects, we restrict our investigation to that subset of
mass determination observables introduced in~\cite{massdeterm} that
contain the near quark $b_{n}$: $\{m_{bb}, m_{b_n\ell,low},
m_{b_n\ell,high}, m_{bb\ell,low}, m_{bb\ell,high}, m_{b_n\ell\ell}, m_{bb\ell\ell}\}$. 
These are the invariant mass distributions of either two $b$ jets, one
$b$ jet and a lepton (with the softer or the harder jet,
respectively), of two $b$s with a lepton, or finally two jets and two
leptons. All these variables are tailor-made for exclusive decay
cascades, which are shown in the next subsection, while the subsection
after that shows the effect of broad gluinos on inclusive variables
(like e.g. $M_{T_2}$).


\subsection{Exclusive Cascades}

\paragraph{$\boxed{m_{bb}}$}
The first mass edge we study is the classical dijet endpoint of the
first gluino decay step. Due to the scalar propagator of the
intermediate bottom squark, we expect to see no effects of spin
correlation, such that the shape of the distribution should resemble
the well-known triangular nature of the di-lepton edge with a linear
rise from 0 to $m_{bb}^{\text{max}}$, where a sharp cutoff marks the
endpoint.  
\begin{figure}[!ht]
  \centering
  \includegraphics[width=0.47\textwidth]{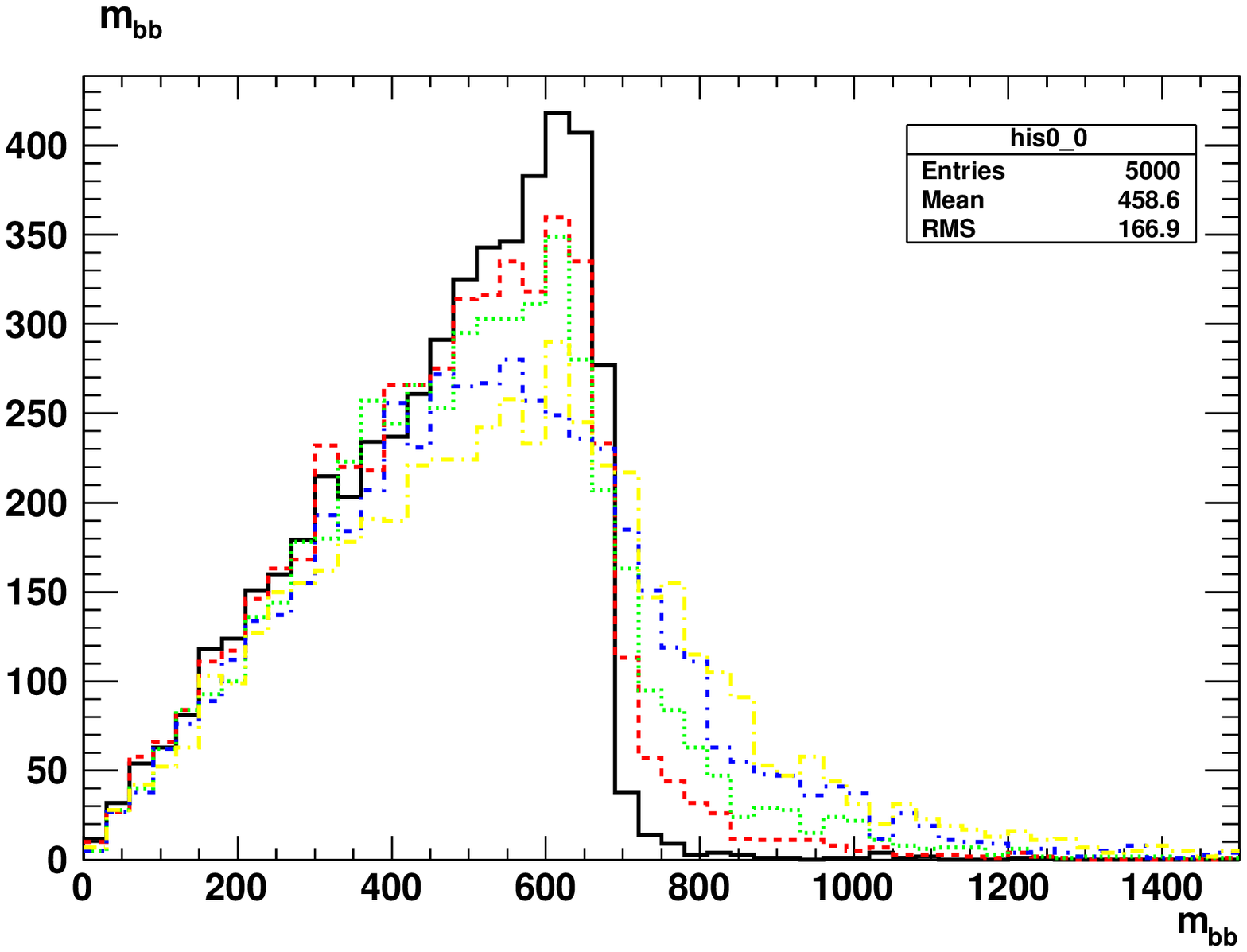}
  \includegraphics[width=0.47\textwidth]{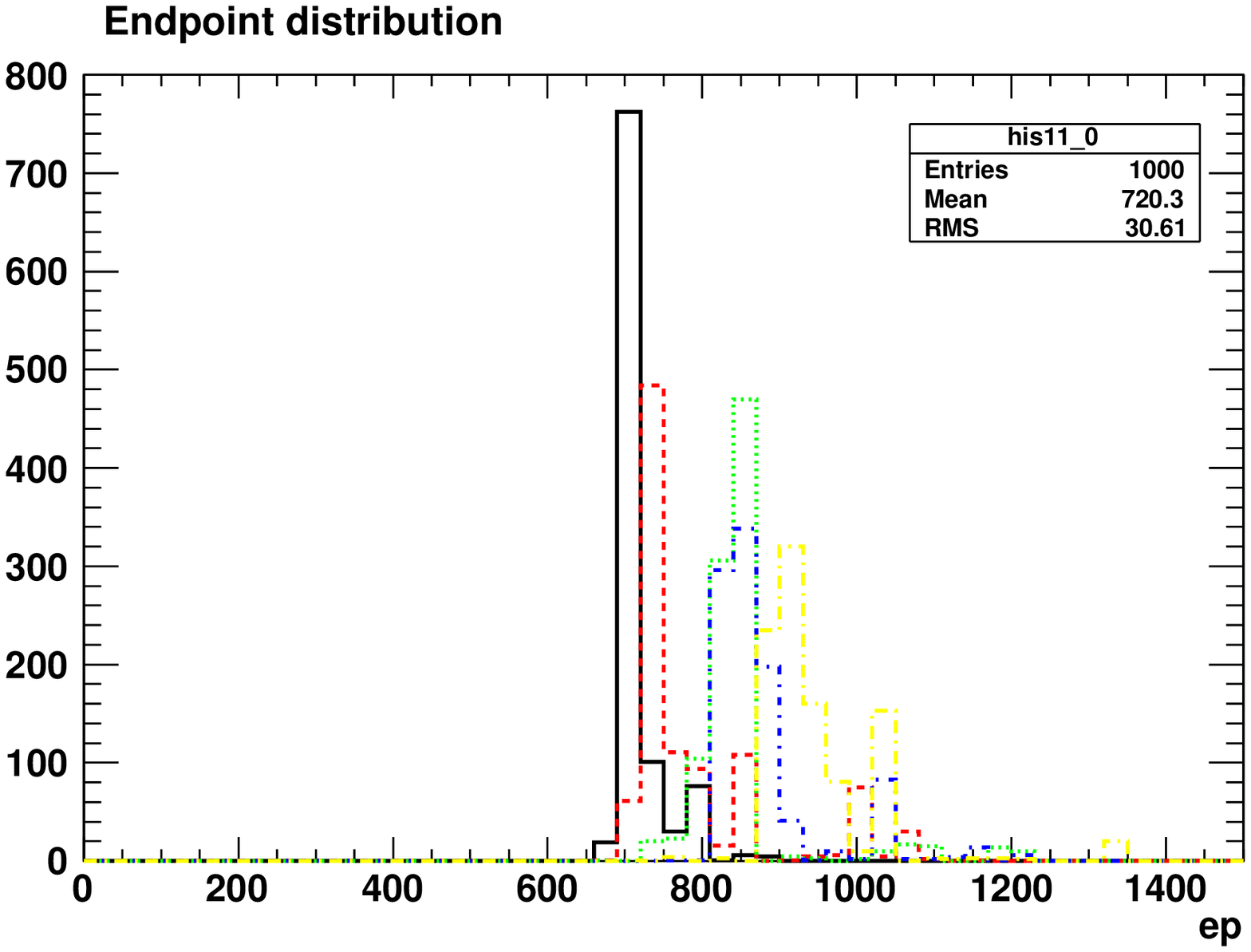}
  \caption{Left hand side: Invariant dijet mass of two bottom quarks,
    $m_{bb}$, from the gluino decay for different values of $\gamma$
    (line styles identical to Fig.~\ref{fig:pt_bottom_near_far}). On
    the right hand side, the distribution of endpoints obtained as fit
    parameters of the edge-to-bump method.}  
  \label{fig:m_bb}
\end{figure}
Figure \ref{fig:m_bb} shows on the left hand side the simulated
distributions for five different values of $\gamma$, where the line
colors are identical to Fig.~\ref{fig:pt_bottom_near_far}. One notices
that the solid black line deviates only very little from the ideal
triangular shape dictated by phase space (one basically has a small
smearing at the edge of the distribution), which is a direct
consequence of the relatively small width: $\Gamma = 0.005 \,M$. The 
situation drastically changes when the width is incrementally
increased: the distortion steadily grows and 
starts to wash out the clear-cut edge structure with increasing
$\gamma$. At high values of $\gamma$ ($\ge$ 5 \%) the distribution
acquires an irreducible tail, which mimics other distorting effects
such as combinatorial mis-assignments or detector smearing already at
this early parton level stage. However, we want to stress the
fact, that this kind of deviation from off-shell contributions is
irreducible in the sense, that it cannot be tuned away e.g. by other
methods designed to minimize combinatorial problems.

In the next step, it is our goal to quantify these intrinsic
contributions by estimating the distortion of the shape in the
vicinity of the endpoint as a function of the effective width-to-mass
ratio $\gamma$. For this study, we use the
edge-to-bump method~\cite{Curtin:2011ng}. It translates the human bias
in fitting different lines to kinematical distribution and hence a
systematic uncertainty into a statistical uncertainty over a sample of
line/edge fit, thereby scanning of a variety of different fit
ranges. It allows us to extract the edge information and to 
discriminate different endpoint behaviours in an unbiased way. In a
nutshell, the approach fits a naive linear kink function $\mathcal O$
(1000) times and returns bumps at the most likely positions of inks
(supposedly physical edges) in the original distribution.  

As a first estimate of the impact of width effects on the measurement,
we consider the shift of the actual endpoint position with respect to
the value of $\gamma$. Table~\ref{tab:m_bb_fitvals} discloses in the
second column these values obtained with our own implementation of the
method described in detail in~\cite{wiesler_phd}. The according
distributions are in Fig.~\ref{fig:m_bb} on the right hand side. While
for small width-to-mass ratios of 0.5\% the obtained value (708 GeV)
is close to the template one (679.6 GeV), the endpoint positions for
large of $\gamma$ are off by more than 200 GeV (920 GeV), using the
same method and settings.  

An alternative, but related measure for the endpoint smearing is given
by the size of the corresponding error estimates. Their increase with
respect to $\gamma$ reflects the observation that the spread of
endpoint values in the right plot of Fig.~\ref{fig:m_bb} is
considerably enhanced for an increased effective width. While for
$\gamma = 0.5\%$ the purely statistical error is small, the sheer
growth of the standard deviation going up to $\gamma = 15.0 \%$ by
nearly two orders of magnitude serves as another good indication for a
huge endpoint smearing. This raises an important point: the overly
high confidence expressed through the small errors of endpoints for
low values of $\gamma$ is a mere binning effect and does not represent
a realistic error estimate for sophisticated endpoint
measurements. Moreover, these error estimates of the edge-to-bump
method are purely statistical and reflect the transformation of a
statistical uncertainty on an endpoint position onto the particular
position of a mean value in a distribution of fit results. Since the
binning of a histogram is bounded by experimental resolutions, a
relatively large minimum bin size results in a systematic
underestimation of the given errors. 

The usage of the endpoint position as a measure for distortion is
based on just one fit parameter (the so-called $p_4$, for details
cf.~\cite{Curtin:2011ng,wiesler_phd}) of the edge-to-bump 
method. A detailed observation of the edge fitting function suggests
to make use of the sampling point parameters $p_2$ and $p_3$, which
are the two linear slopes left and right of a possible edge or
kink. Notwithstanding the fact, that the off-shell contributions tend
to wash out the sharp edge and lead to smoother and longer tails, we
propose to use the absolute difference of the two slopes as well as
the ratio of the two slope parameters as measures to quantify the
amount of edge distortion:  
\begin{align}
  s_d &:= |p_3 - p_2|\\
  s_r &:= |p_2 / p_3|
\end{align}
For each edge fit, both values are calculated from the parameters
returned by the edge-to-bump method. Ideally, for a pure phase space
distribution of triangular shape and no smearing beyond the sharp
cutoff, the first slope is infinite and the second slope zero, hence
the slope difference maximal (namely infinite) and the slope ratio
minimal (namely zero). In that sense, the difference in slopes
measures the sharpness of a kink in the distribution whereas the ratio
returns information about the size of the second slope relative to the
first one (pronouncedness of the kink). A ratio 
close to zero may thus be attributed to a tail-less distribution such
as the triangular shaped one. Keep in mind, that due to the sanity
checks of our method, $|p_2| < |p_3|$ and hence $s_r \in (0,1)$. The
full treatment of large width effects on the other hand introduces a
tail and smears the endpoint behavior, which gives rise to
considerably smaller slope differences and higher slope ratios. If the
returned fit value of such a slope difference is compatible with zero,
the underlying distribution apparently lacks robust kinky
features. Extraction of such shallow endpoints is therefore a very
delicate task. 

\begin{figure}[!ht]
  \includegraphics[width=0.47\textwidth]{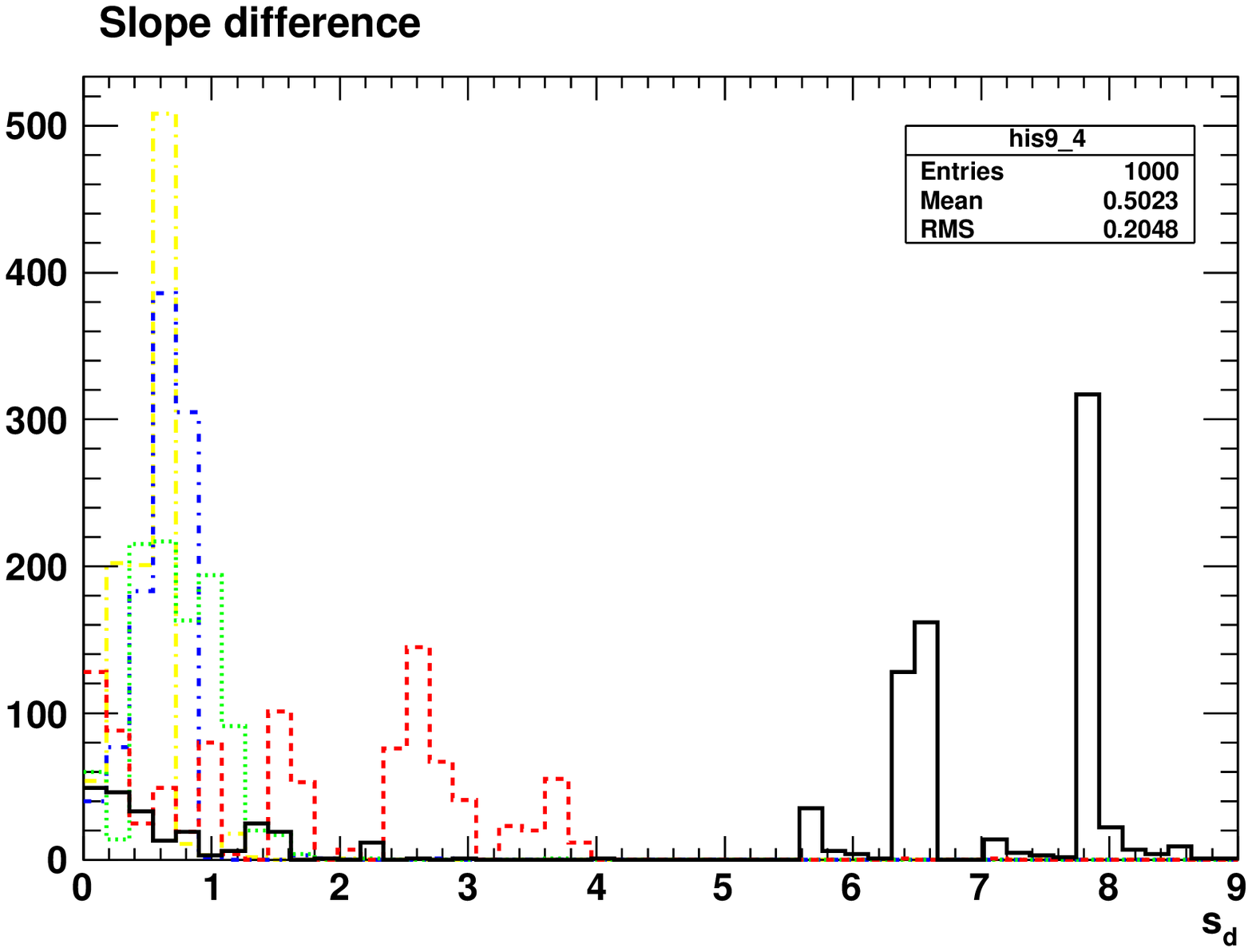}
  \includegraphics[width=0.47\textwidth]{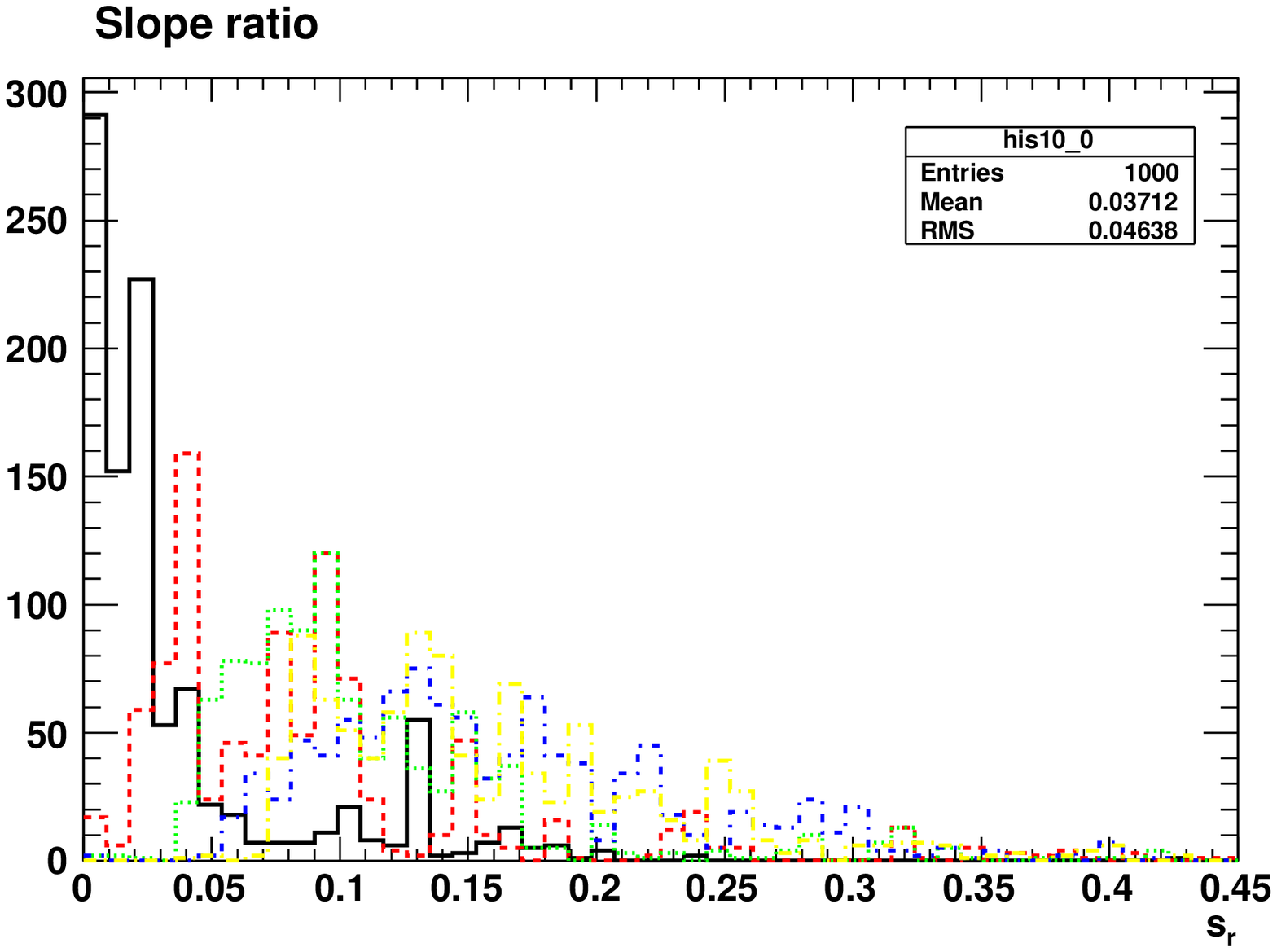}
  \caption{Statistically distributed slope differences (left) and
    ratios (right) for the dijet invariant mass $m_{bb}$ obtained as
    fit parameters with the edge-to-bump method.} 
  \label{fig:m_bb_slopes}
\end{figure}

In the last two columns of Tab.~\ref{tab:m_bb_fitvals} we collect
results for the slope differences and slope ratios of the invariant
di-bottom mass for all five scan values of $\gamma$. In
Fig.~\ref{fig:m_bb_slopes} we show the  
impact of the different values of $\gamma$: while we observe slope
differences well above one and small slope ratios compatible with zero
for small $\gamma$, at an effective width of already 5 \% the mean
$\bar s_d$ is reduced to a value smaller than one. The slope ratio
exhibits an equivalent behavior: the mean value $\bar s_d$ is
increased by a factor of 10. 
\begin{table}[!ht]
\begin{center} 
\begin{tabular}{r|ccc}
  $\gamma$ [\%] & $\bar{m}_{bb}^{\text{max}}$  & $\bar{s}_d$ & $\bar{s}_r$  \\
  0.5     & 708.5 $\pm$ 0.9 & 5.70 $\pm$ 2.90 & 0.014 $\pm$ 0.008 \\ 
  2.5     & 740.9 $\pm$ 2.5 & 1.71 $\pm$ 1.17 & 0.064 $\pm$ 0.028 \\ 
  5.0     & 835.7 $\pm$ 19.2 & 0.78 $\pm$ 0.24 & 0.084 $\pm$ 0.023 \\ 
 10.0     & 886.5 $\pm$ 13.0 & 0.67 $\pm$ 0.08 & 0.141 $\pm$ 0.045 \\ 
 15.0     & 921.3 $\pm$ 25.9 & 0.62 $\pm$ 0.04 & 0.131 $\pm$ 0.035 \\
  \end{tabular}
\end{center}
\caption{Adapted mean values of endpoint positions (in GeV), slope
  differences (in 1/GeV) and slope ratios for invariant di-bottom mass
  $m_{bb}$.} 
\label{tab:m_bb_fitvals}
\end{table}

\paragraph{$\boxed{m_{b_n\ell}}$}
Next, we investigate the mass edges obtained by the minimization and
maximization over two possible lepton combinations with the near
bottom quark. Their definitions are: 
\begin{align}
  m_{b_n\ell,low} = &\min  \left[m_{b_n\ell^+},m_{b_n\ell^-}\right] \\
  m_{b_n\ell,high} =& \max \left[m_{b_n\ell^+},m_{b_n\ell^-}\right] \qquad ,
  \label{eq:b_nlhigh}
\end{align}
while their endpoints given are given by 
\begin{align}
  (m^\text{max}_{b_n\ell_{low}})^2 &=\; \min \left[ 
      (m^\text{max}_{b_n\ell_n})^2,
        (m^\text{max}_{b_n\ell_{eq}})^2 \right]       
      \\        
  (m^\text{max}_{b_n\ell_{high}})^2 &=\; \max \left[ 
      (m^\text{max}_{b_n\ell_{eq}})^2,
        (m^\text{max}_{b_n\ell_f})^2 \right]
\end{align}
with 
\begin{equation*}
  (m^\text{max}_{b_n\ell_{eq}})^2 = \frac{(m_{\tilde{g}}^2 -
        m_{\tilde{b}}^2)(m_{\tilde{\ell}}^2
        -m_{\tilde{\chi}_1^0}^2)}{(2m_{\tilde{\ell}}^2 -
        m_{\tilde{\chi}_1^2}^2)} \quad .
\end{equation*}
\begin{figure}[!ht]
  \centering
  \includegraphics[width=0.47\textwidth]{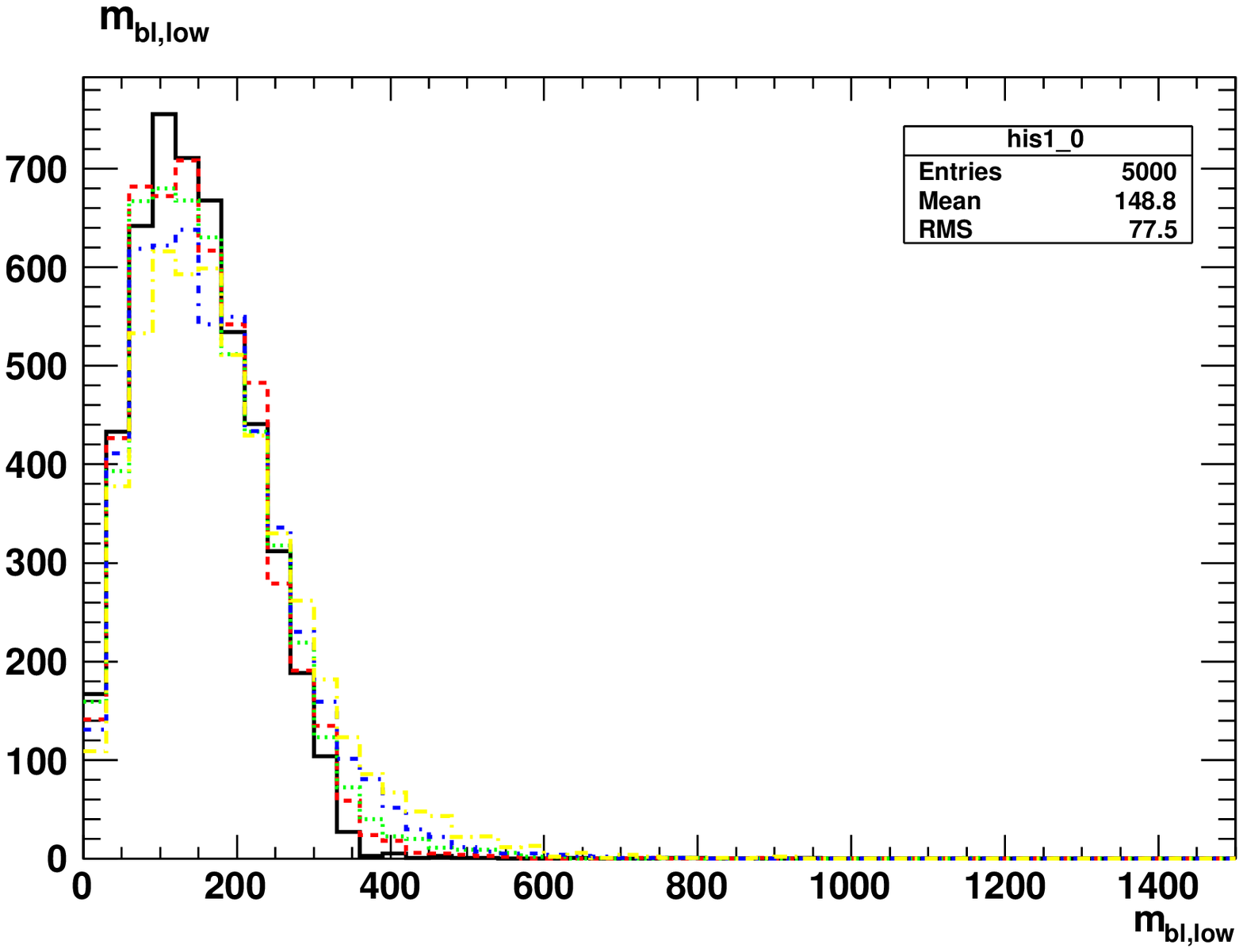}
  \includegraphics[width=0.47\textwidth]{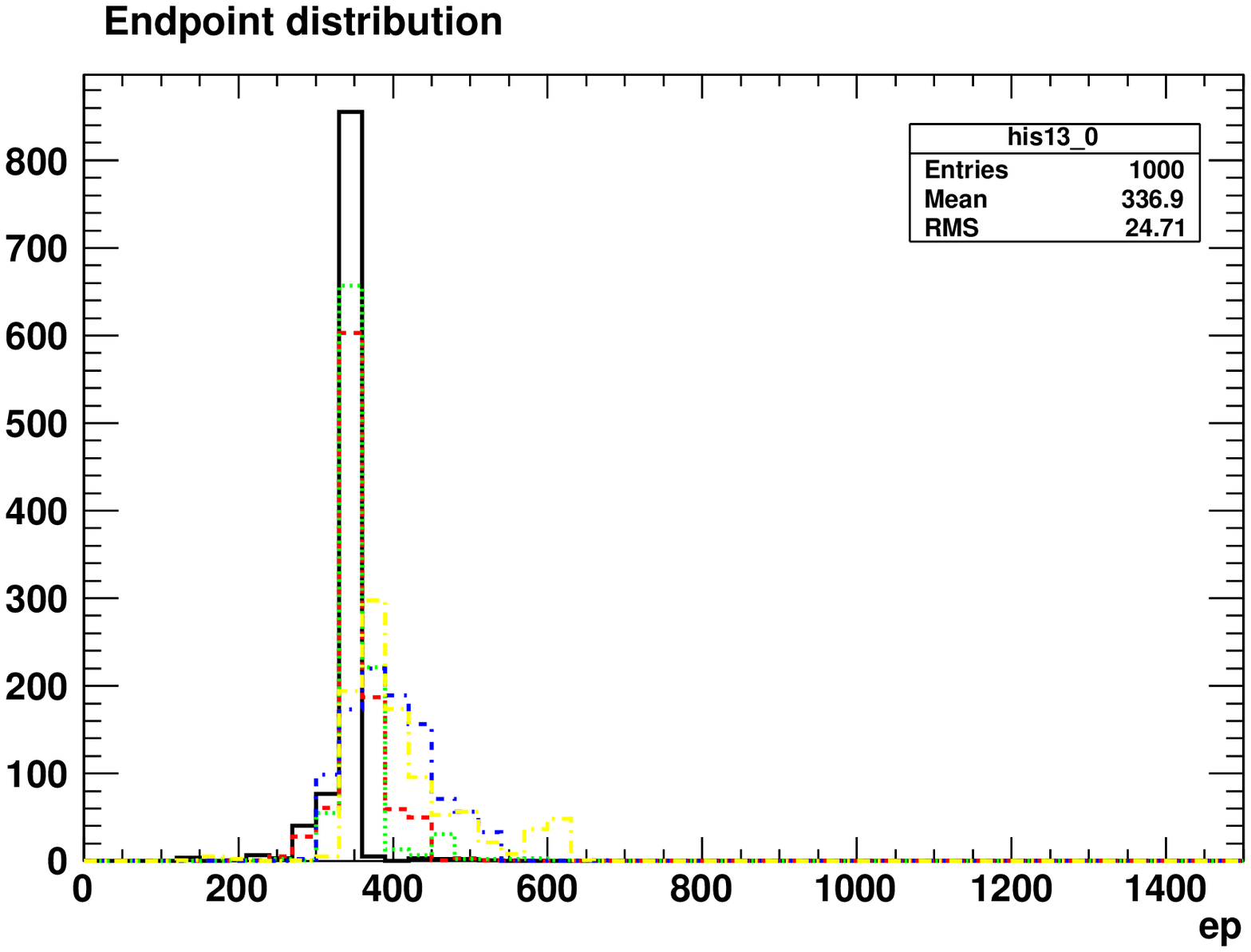}\\

  \includegraphics[width=0.47\textwidth]{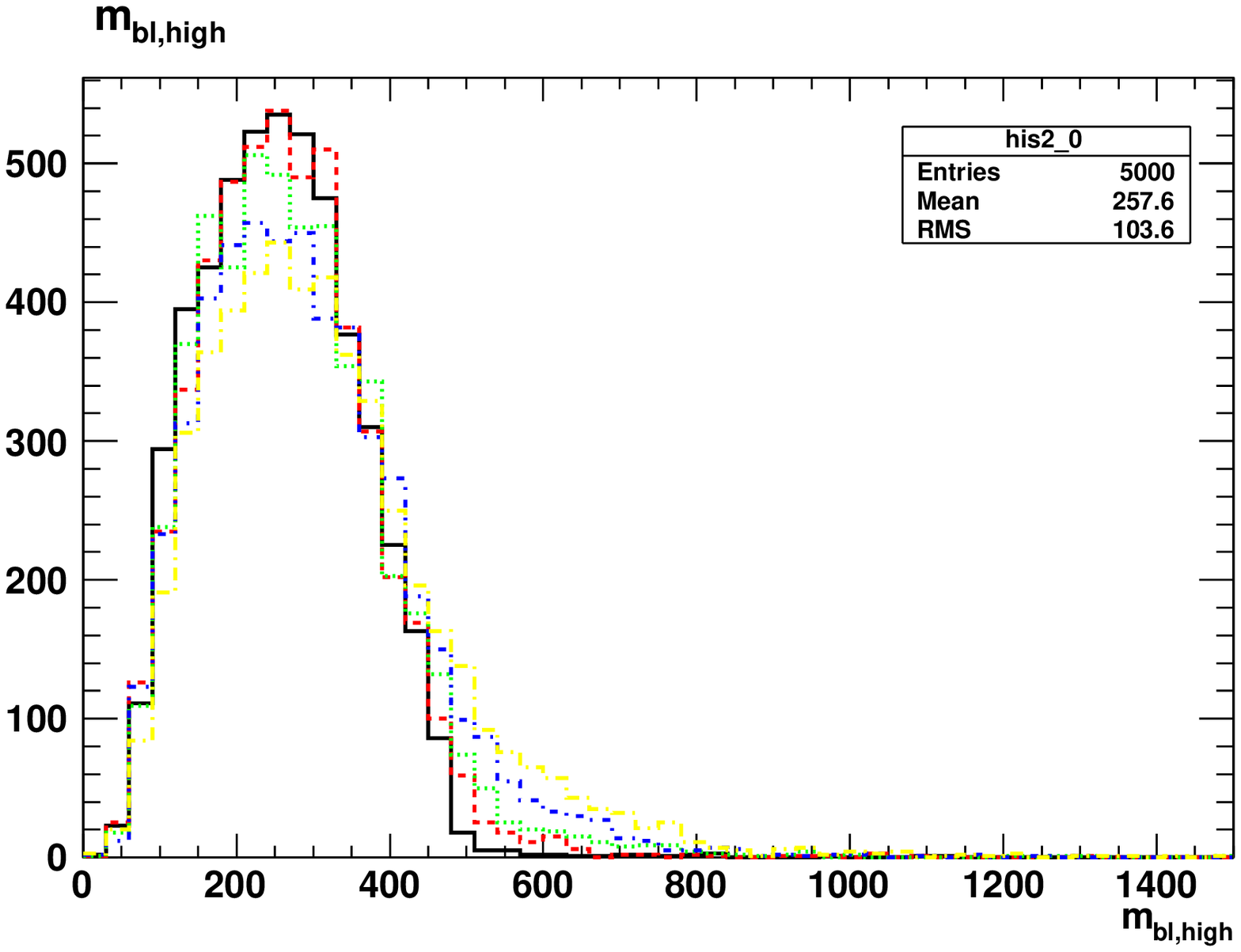}
  \includegraphics[width=0.47\textwidth]{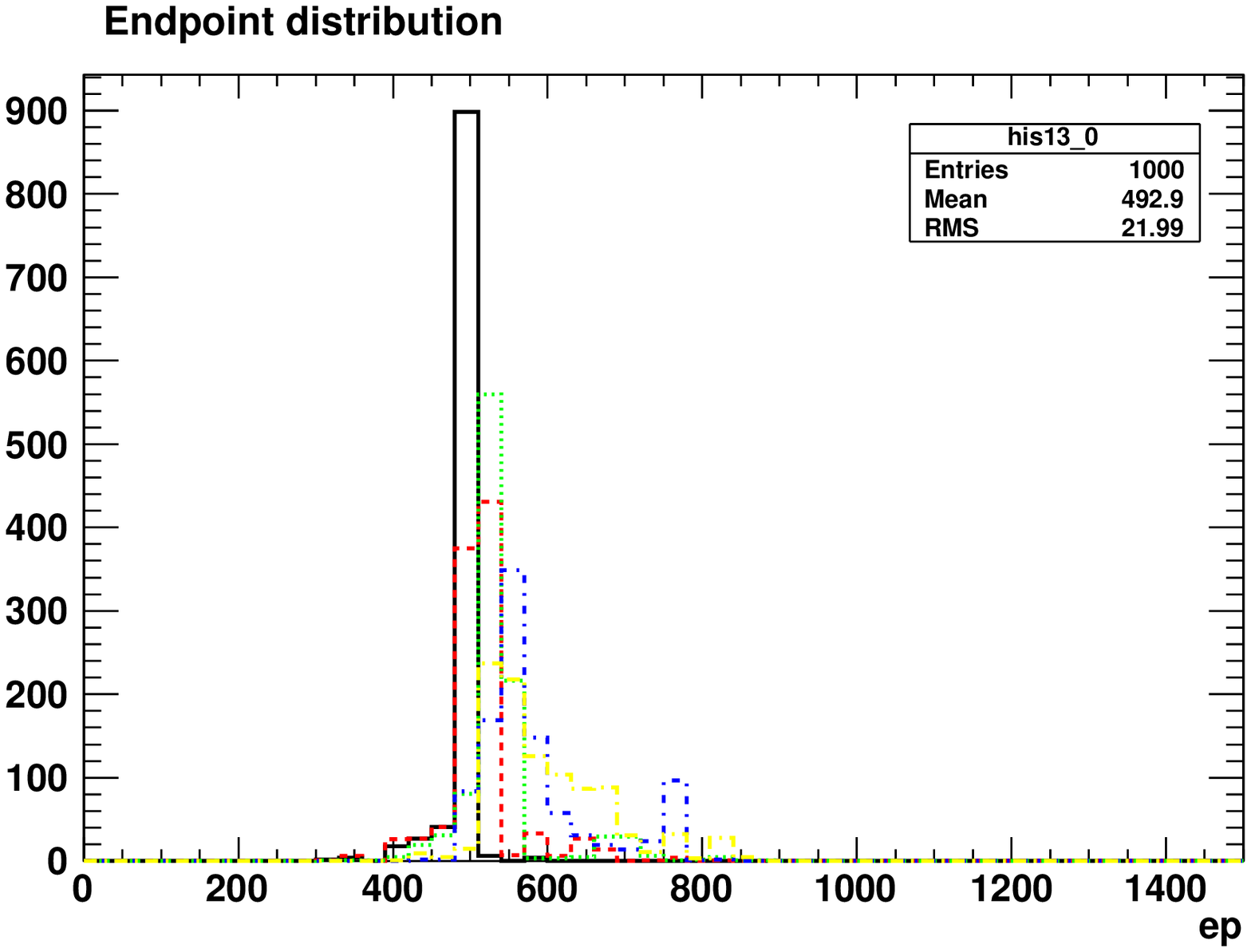}\\
  \caption{Left: Invariant mass of near bottom quark and lepton
    as a function of $m_{b\ell,low}$ (upper plot) and $m_{b\ell,high}$
    (lower plot) for different values of $\gamma$. The black (solid),
    red (short-dashed), green (dotted), blue (short-dashed-dotted),
    and yellow (long-dashed-dotted) correspond to $\gamma =$ 0.5 \%,
    2.5 \%, 5.0 \%, 10.0 \%, and 15.0 \%, respectively. Right: 
    Distribution of the corresponding endpoints obtained as fit
    parameters with the edge-to-bump method for $m_{b\ell,low}$ (upper
    plot) and $m_{b\ell,high}$ (lower plot).} 
  \label{fig:m_bl}
\end{figure}
By the very nature of the intermediate particles, there is a small
correlation of the leptons and the bottom quark, that manifests itself
in the shape of the distributions. Hence, with no pure phase space
shape, we expect the distortion to be less pronounced in comparison to
the invariant di-bottom mass. Especially $m_{b_n\ell,low}$ should not
contain too much excess events in the upper parts, since these
contributions will mostly be omitted due to the minimization
procedure. The variable $m_{b_n\ell,high}$, on the other hand, will
severely be affected for the very same reason. The distributions in
Fig.~\ref{fig:m_bl} on the left hand side (upper and lower plot)
confirm these conjectures. The extracted endpoints depicted on the
corresponding right hand sides of the upper and lower line in
Fig.~\ref{fig:m_bl} and their numerical values in
Table~\ref{tab:m_bl_fitvals} also support this statement: while for
$m_{b_n\ell,low}$ the overall endpoint variation with respect to $\gamma$
is about $40$ GeV, $m_{b_n\ell,high}$ suffers from more than twice the
endpoint shift with a value of about $80$ GeV. Comparing this to the
expected ideal edge positions 
\begin{align}
  m_{b_nl,low}^{\text{max}} &= 364.4 \text{ GeV} \\
  m_{b_nl,high}^{\text{max}} &= 493.1 \text{ GeV}
\end{align}
 calculated from Eq.~\eqref{eq:b_nlhigh}, we
 find that the discrepancy for $m_{b_n\ell,low}$ is indeed small, and
 the determined value is mostly in agreement with the expectation. For
 $m_{b_n\ell,high}$, however, the endpoint shift is as large as 15\%. 
\begin{figure}[!ht]
  \includegraphics[width=0.47\textwidth]{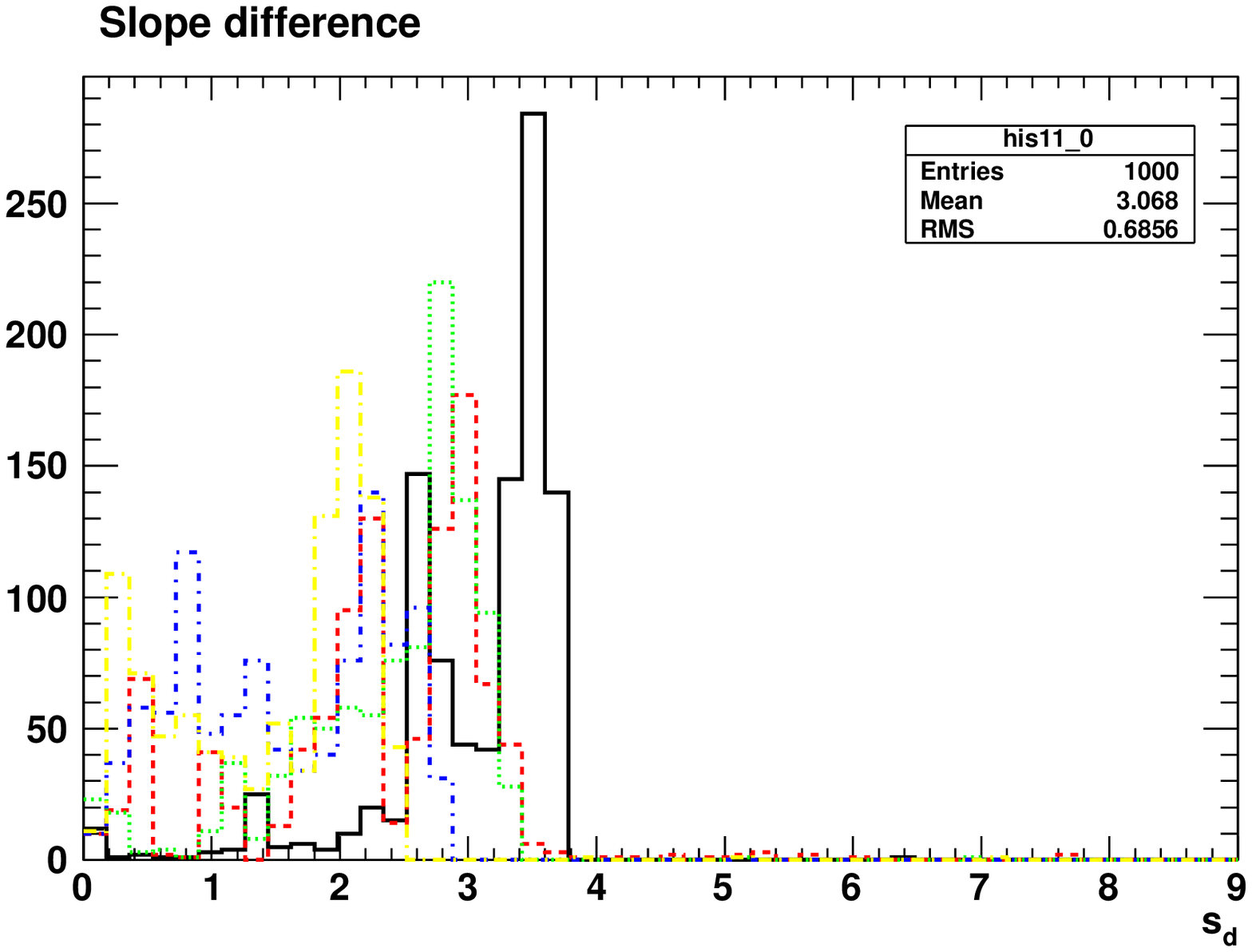}
  \includegraphics[width=0.47\textwidth]{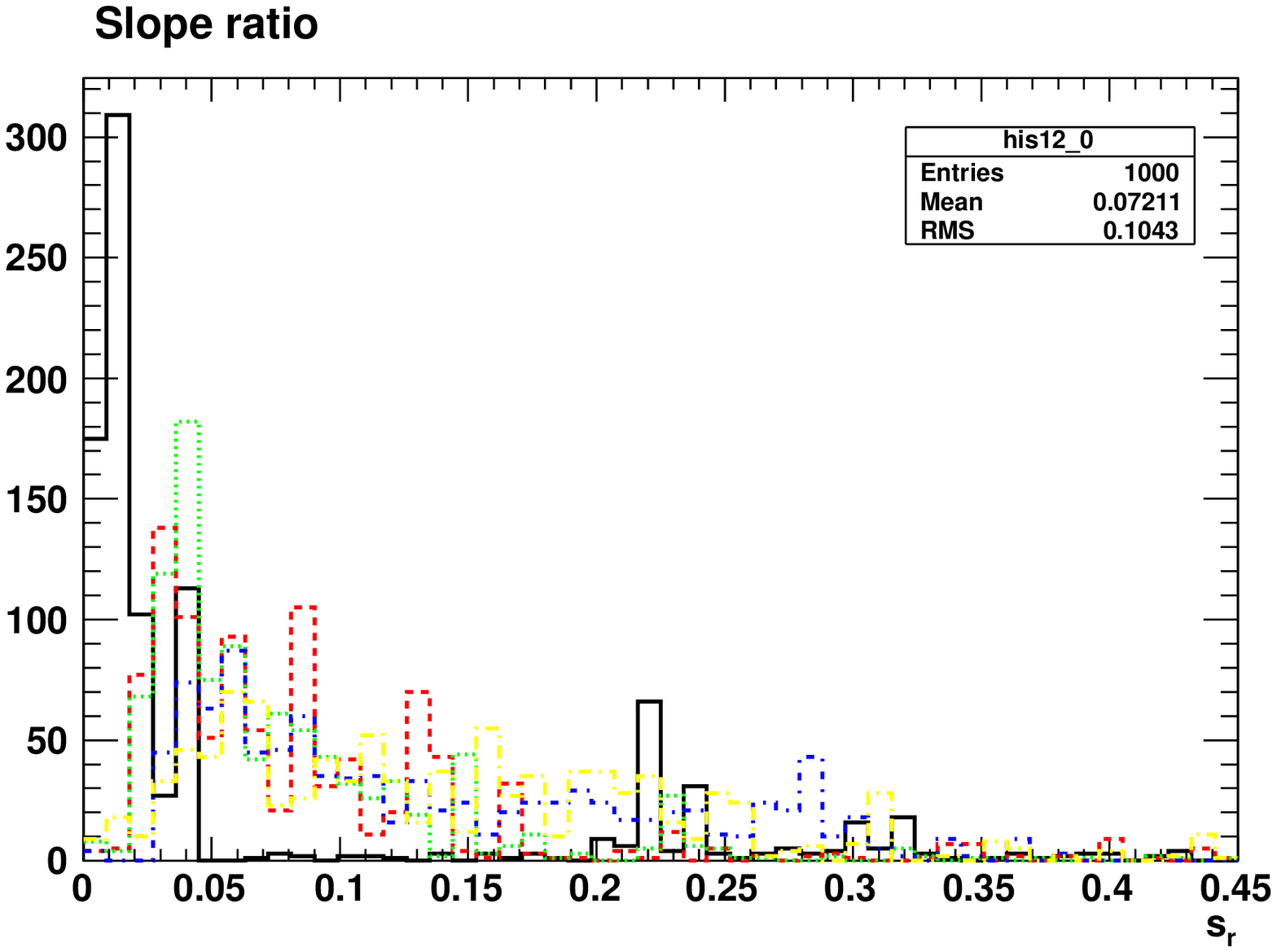}\\
  \includegraphics[width=0.47\textwidth]{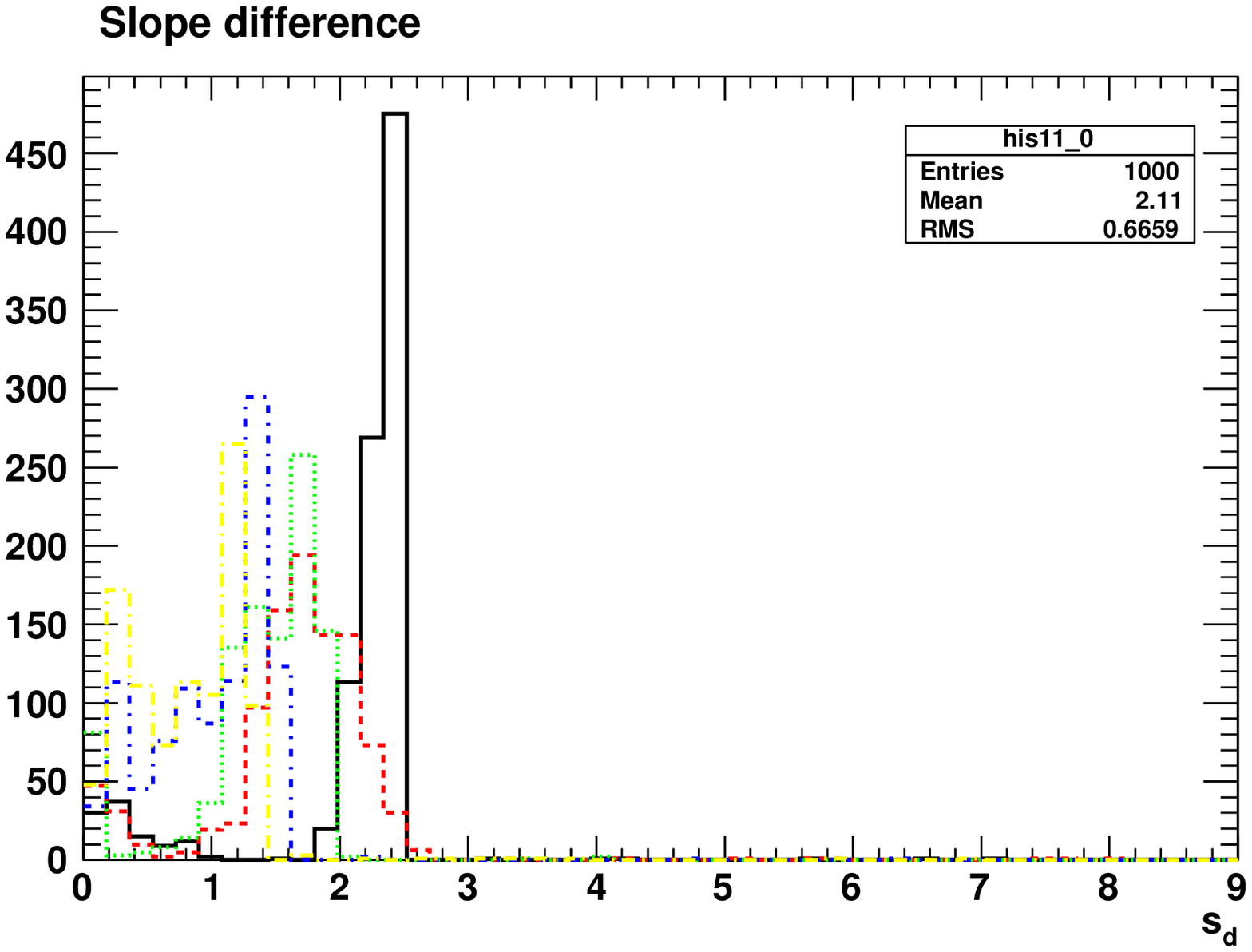}
  \includegraphics[width=0.47\textwidth]{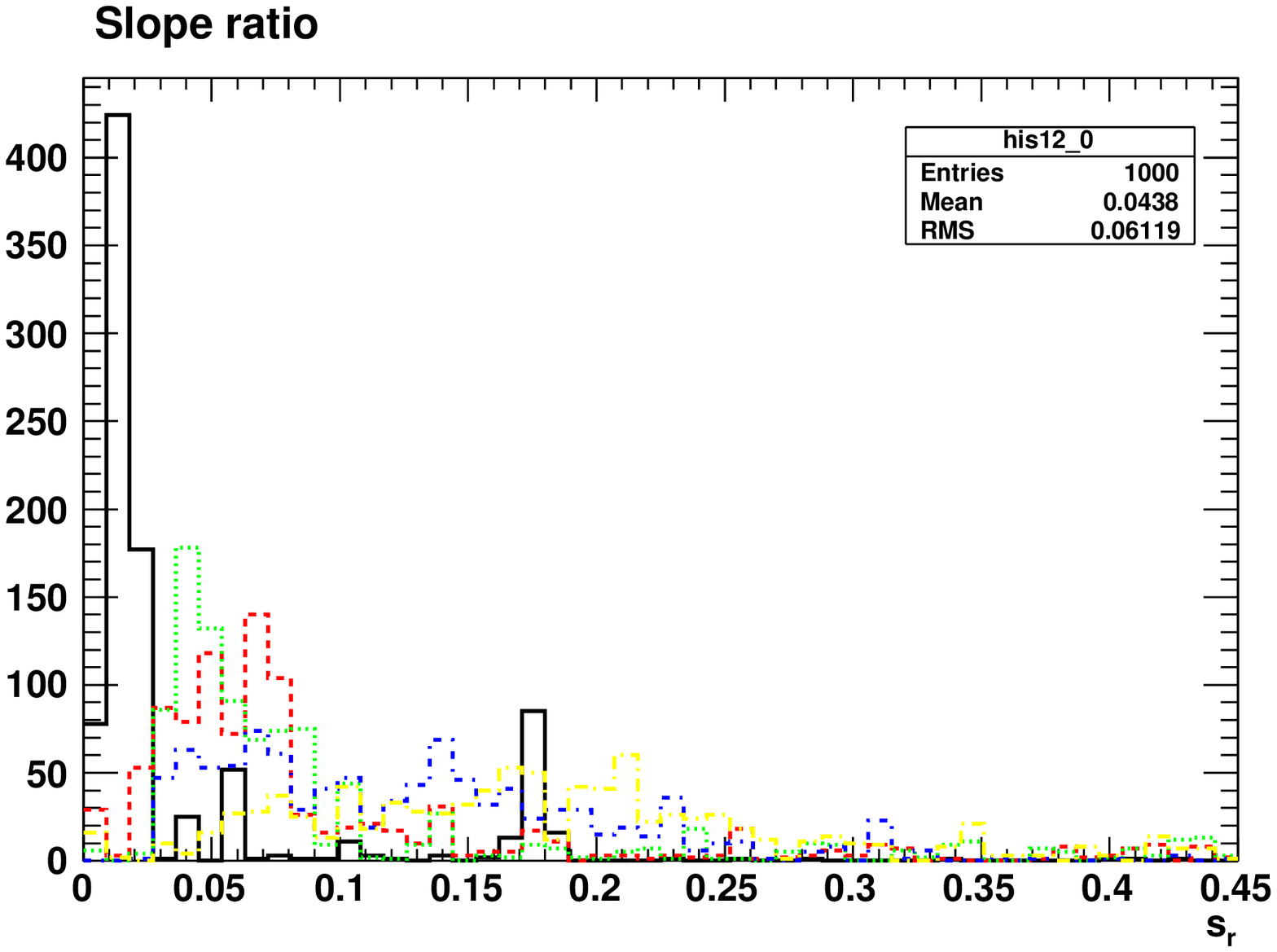}\\
  \caption{Statistically distributed slope differences (left) and
    slope ratios (right) for $m_{b\ell,low}$ (upper plot) and
    $m_{b\ell,high}$ (lower plot), obtained as fit parameters by means
    of the edge-to-bump method.} 
  \label{fig:m_bl_slopes}
\end{figure}
The adapted slope differences $\bar s_d$ show a similar picture: while
the numbers of the minimized distributions exhibit only a mild overall
decline, the maximized version illustrates an apparent tendency of
$\bar s_d$ to decrease with rising $\gamma$. As for the slope ratios
$\bar s_r$, the same holds, but to an even greater extent, in that for
$m_{b\ell,low}$ the three intermediate ratios are all in the same
ballpark, whereas for $m_{b\ell,high}$, $\bar s_r$ increases by nearly one
order of magnitude. 
\begin{table}[!ht]
\begin{center}
\begin{tabular}{r|ccc|ccc}
  $\gamma$ [\%] & $\bar{m}_{b\ell,low}^{\text{max}}$  & $\bar{s}_d$ &
  $\bar{s}_r$ & $\bar{m}_{b\ell,high}^{\text{max}}$  & $\bar{s}_d$ &
  $\bar{s}_r$  \\ 
\hline
  0.5     & 341.4 $\pm$ 6.7 & 3.25 $\pm$ 0.38 & 0.014 $\pm$ 0.007  &
  499.8 $\pm$ 0.3 & 2.35 $\pm$ 0.03 & 0.017 $\pm$ 0.005 \\  
  2.5     & 356.6 $\pm$ 2.7 & 2.58 $\pm$ 0.47 & 0.055 $\pm$ 0.025  &
  511.2 $\pm$ 3.6 & 1.80 $\pm$ 0.26 & 0.055 $\pm$ 0.018 \\  
  5.0     & 355.5 $\pm$ 1.8 & 2.82 $\pm$ 0.25 & 0.041 $\pm$ 0.011  &
  527.0 $\pm$ 11.9 & 1.55 $\pm$ 0.25 & 0.054 $\pm$ 0.017 \\  
  10.0    & 383.0 $\pm$ 39.6 & 1.59 $\pm$ 0.77 & 0.066 $\pm$ 0.024 &
  556.3 $\pm$ 13.4 & 1.00 $\pm$ 0.44 & 0.103 $\pm$ 0.048 \\  
  15.0    & 377.2 $\pm$ 22.3 & 1.47 $\pm$ 0.74 & 0.124 $\pm$ 0.069 &
  579.0 $\pm$ 47.9 & 0.79 $\pm$ 0.40 & 0.158 $\pm$ 0.059 \\  
  \end{tabular}
\end{center}
\caption{Adapted mean values of endpoint positions (in GeV), slope
  differences (in 1/GeV) and slope ratios for $m_{b\ell,low}$ and
  $m_{b\ell,high}$, respectively.} 
\label{tab:m_bl_fitvals}
\end{table}


\paragraph{$\boxed{m_{bb\ell}}$}
Extending the \textit{low}- and \textit{high}-type invariant masses
from above by the additional (far) bottom quark, we arrive at a
three-particle invariant mass, that has a similar feature as the ones
just discussed: an intermediate neutralino propagator communicating
spin correlations and allowing for a similarly altered shape compared
to the triangular phase space.  
\begin{figure}[!ht]
  \centering
  \includegraphics[width=0.47\textwidth]{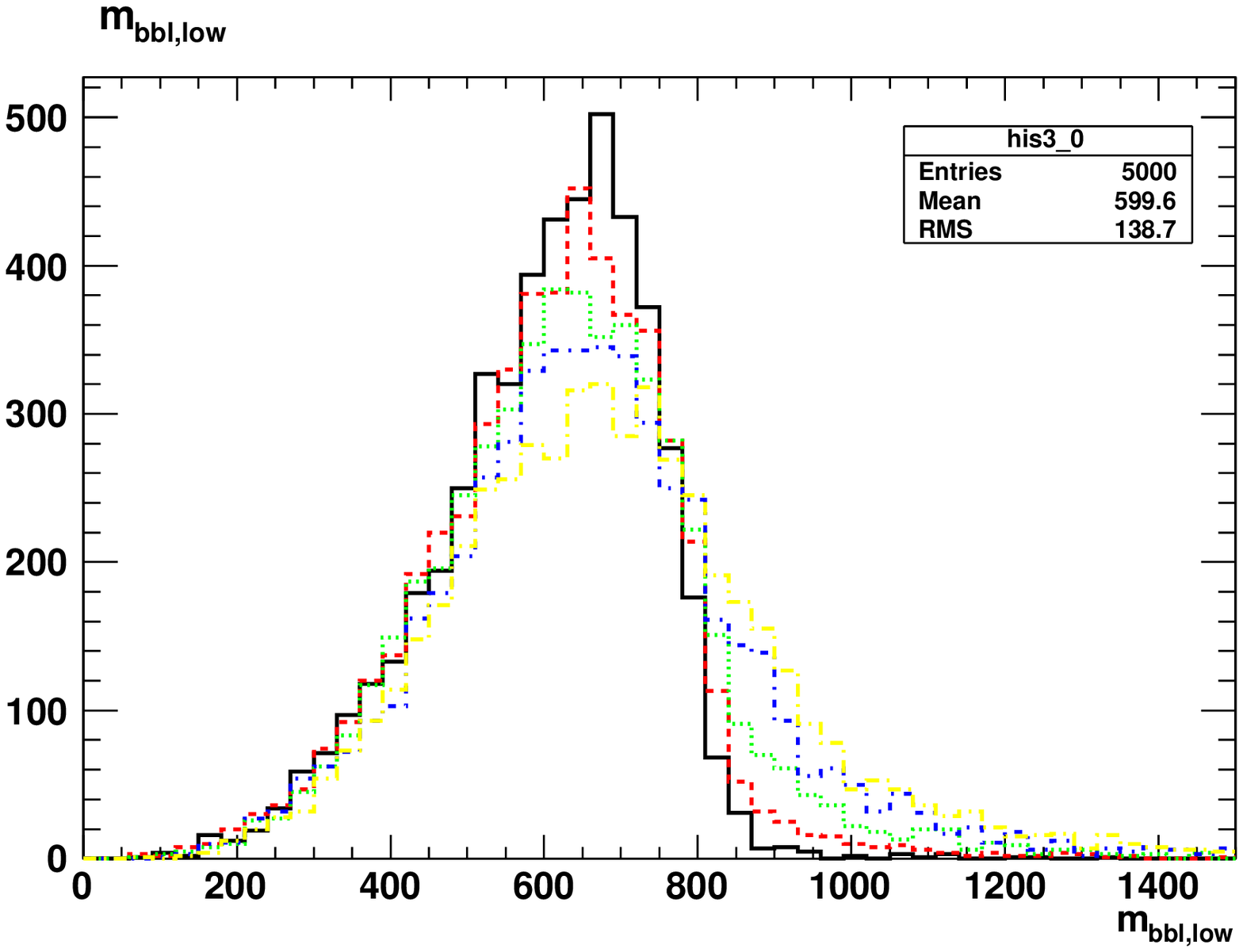}
  \includegraphics[width=0.47\textwidth]{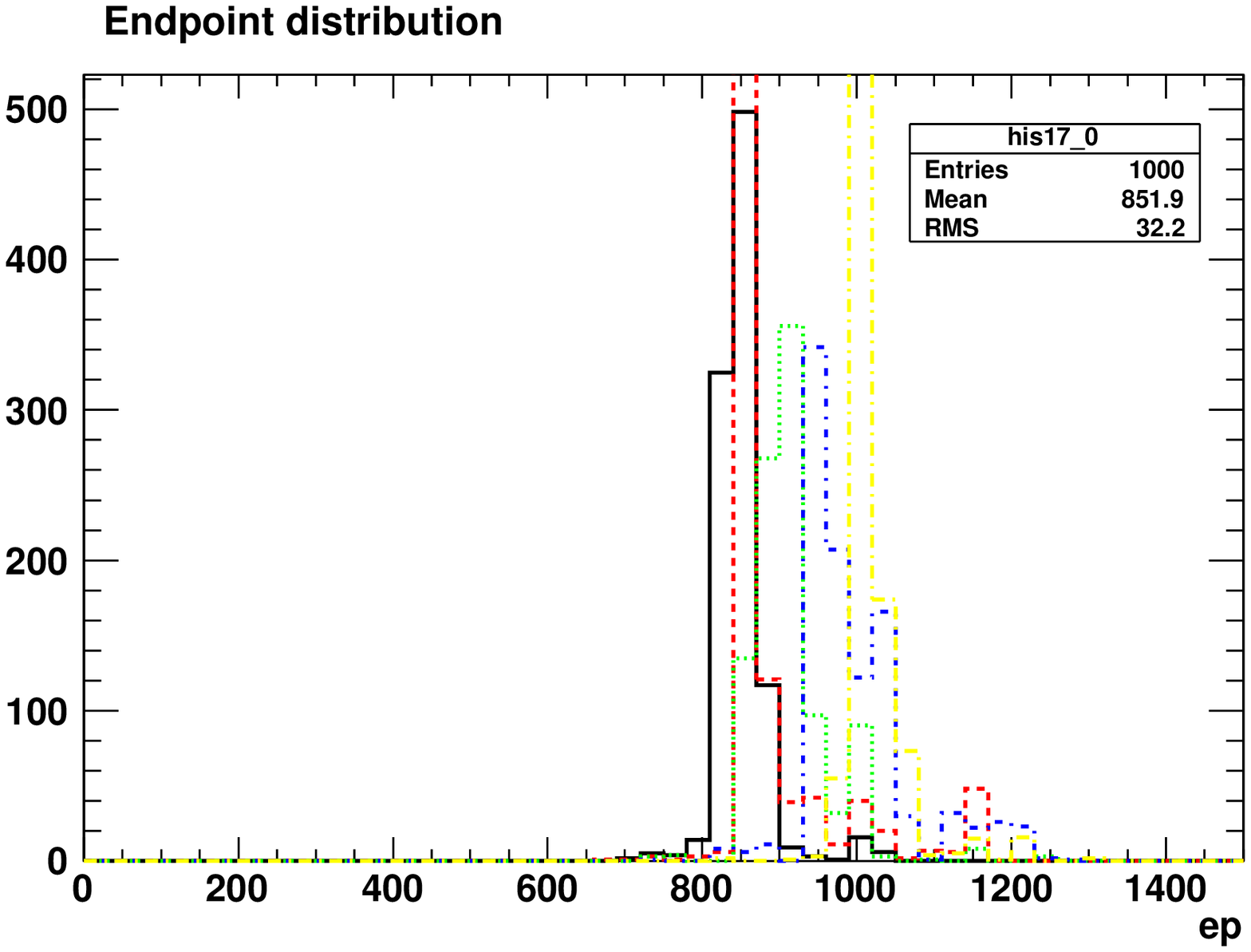}\\
  \includegraphics[width=0.47\textwidth]{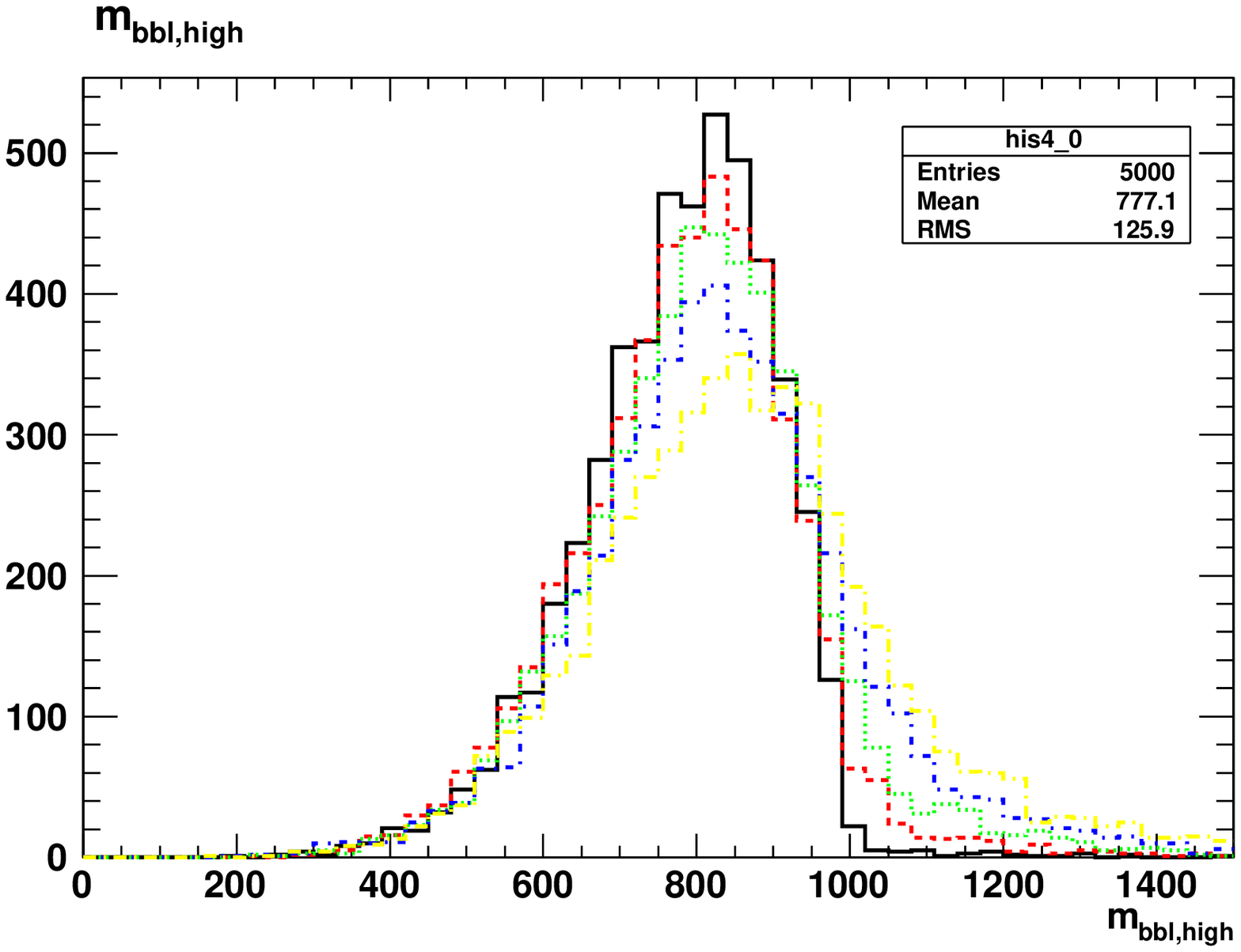}
  \includegraphics[width=0.47\textwidth]{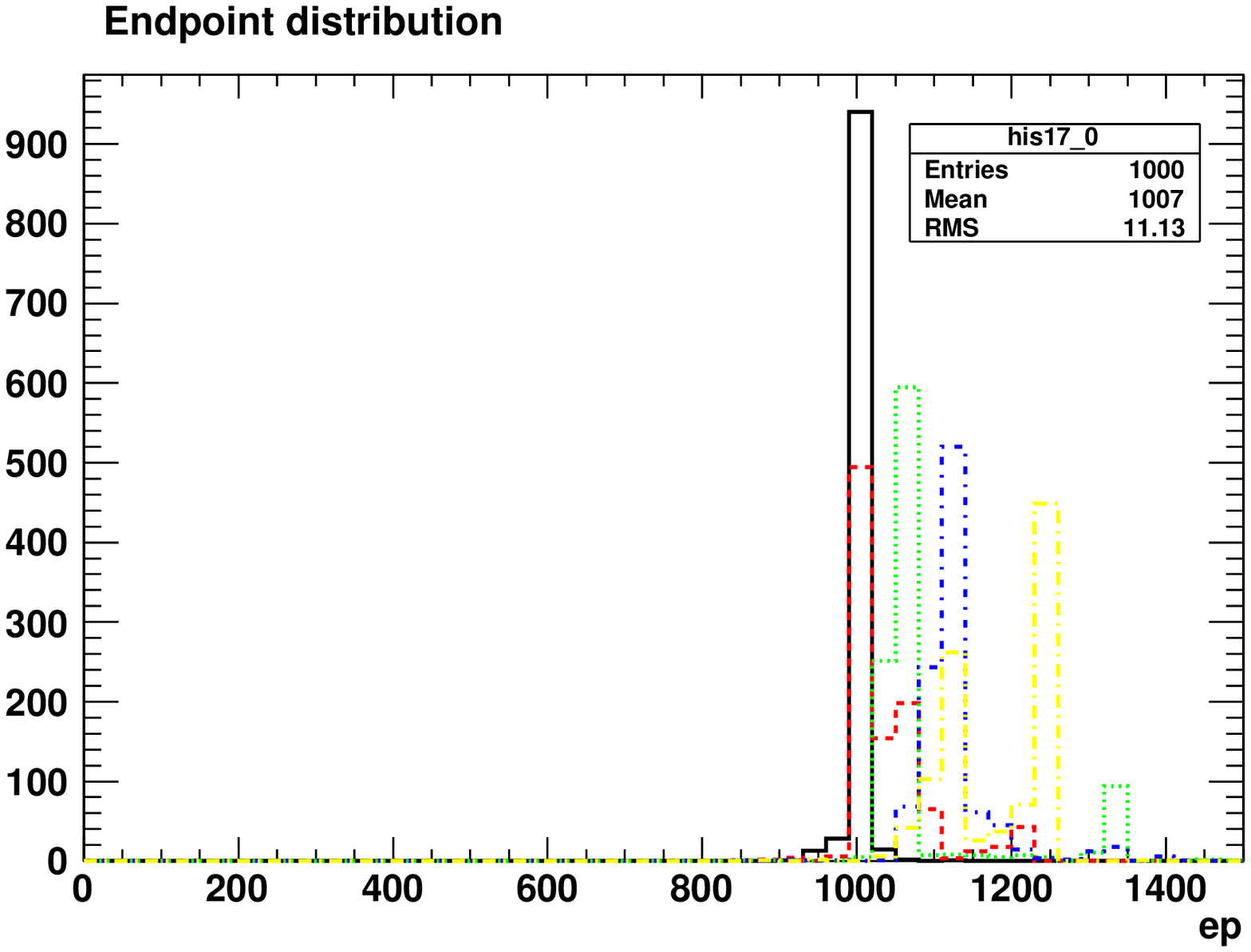}\\
  \caption{Left: Invariant mass of near bottom quark and leptons,
    varied as defined by $m_{bb\ell,low}$ (top) and $m_{bb\ell,high}$
    (bottom) for different values of $\gamma$ (line style identical to
    Fig.~\ref{fig:m_bl}). Right: Distribution of the corresponding
    endpoints obtained as fit parameters with the edge-to-bump method
    for $m_{bb\ell,low}$ (top) and $m_{bb\ell,high}$ (bottom).} 
  \label{fig:m_bbl}
\end{figure}
If we now start to gradually increase the effective width parameter
$\gamma$, the off-shell contributions start to enter the game in a
more severe way as for the two-particle invariant masses,
$m_{b\ell,low}$ and $m_{b\ell,high}$. This can be understood in terms
of the inclusion of nearest neighbors: since the invariant di-bottom
mass is heavily distorted by width effects, and it is always included
in both $m_{bb\ell,low}$ and $m_{bb\ell,high}$, we in fact expect 
to observe large deviations. In that sense -- and in contrast to
$m_{b\ell,low}$ -- the minimization procedure over two possible lepton
combinations is not able to suppress the appearance of these intrinsic
contributions, as can be seen in Fig.~\ref{fig:m_bbl}.  
\begin{figure}[!ht]
  \includegraphics[width=0.47\textwidth]{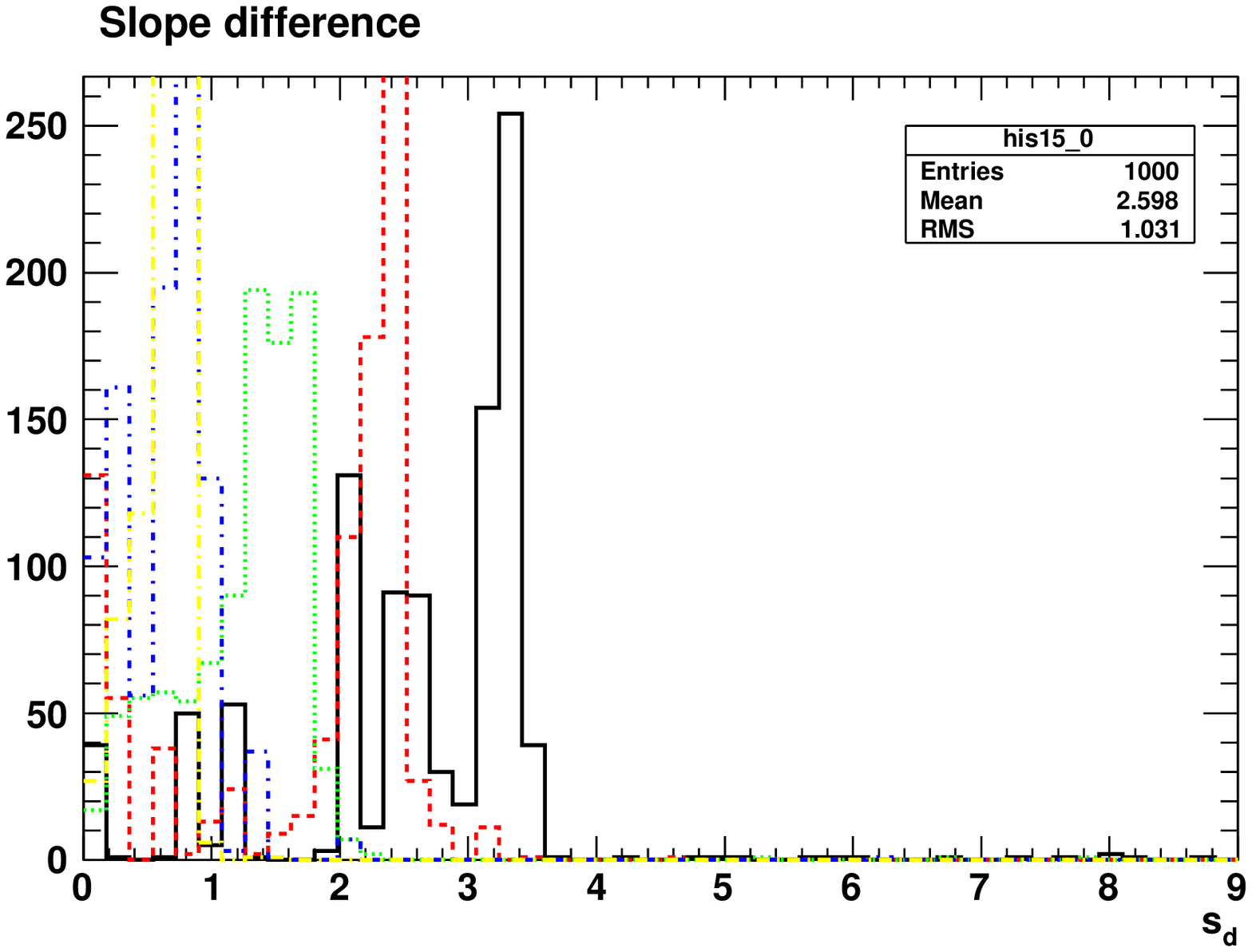}
  \includegraphics[width=0.47\textwidth]{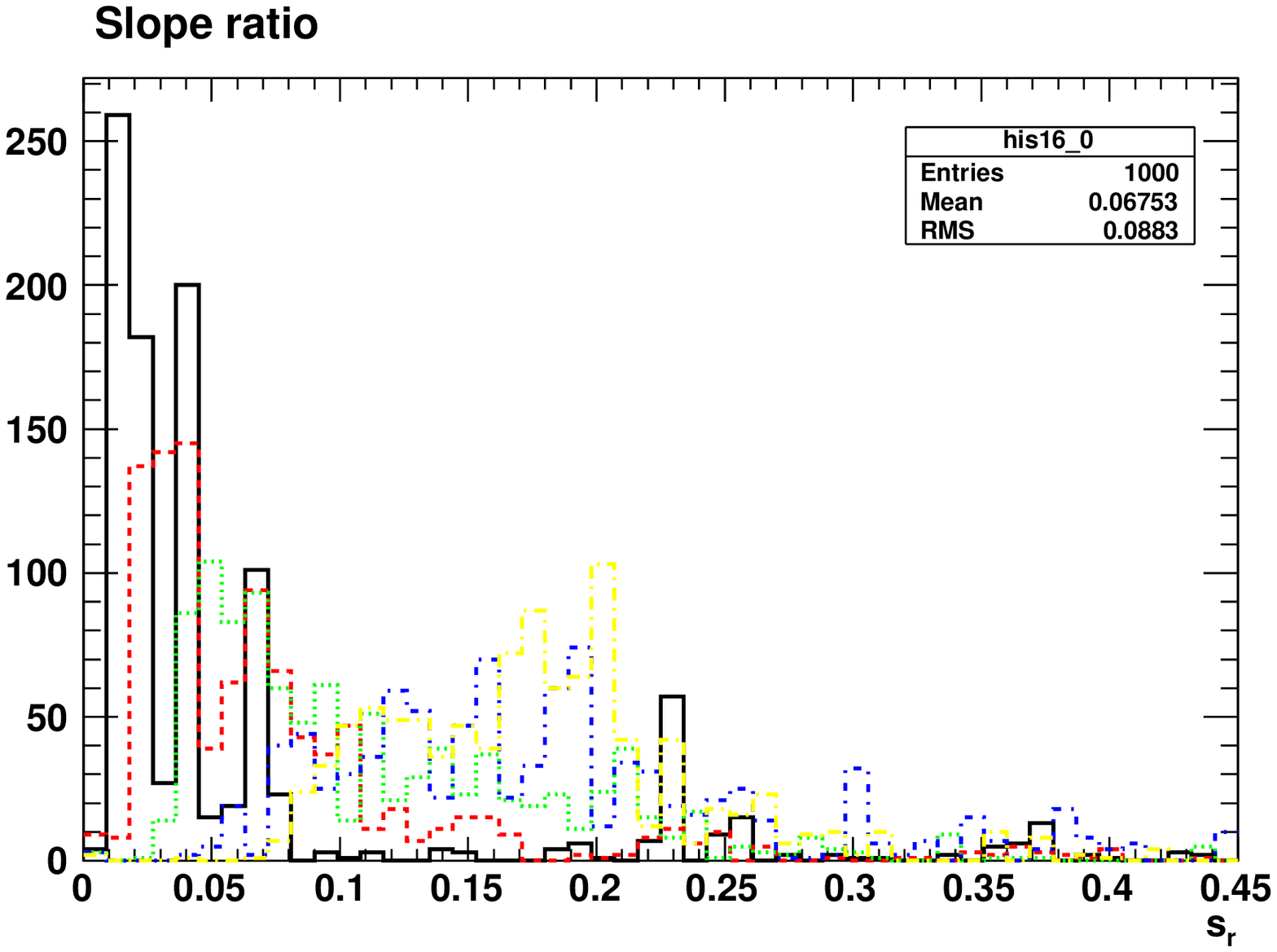}\\
  \\
  \includegraphics[width=0.47\textwidth]{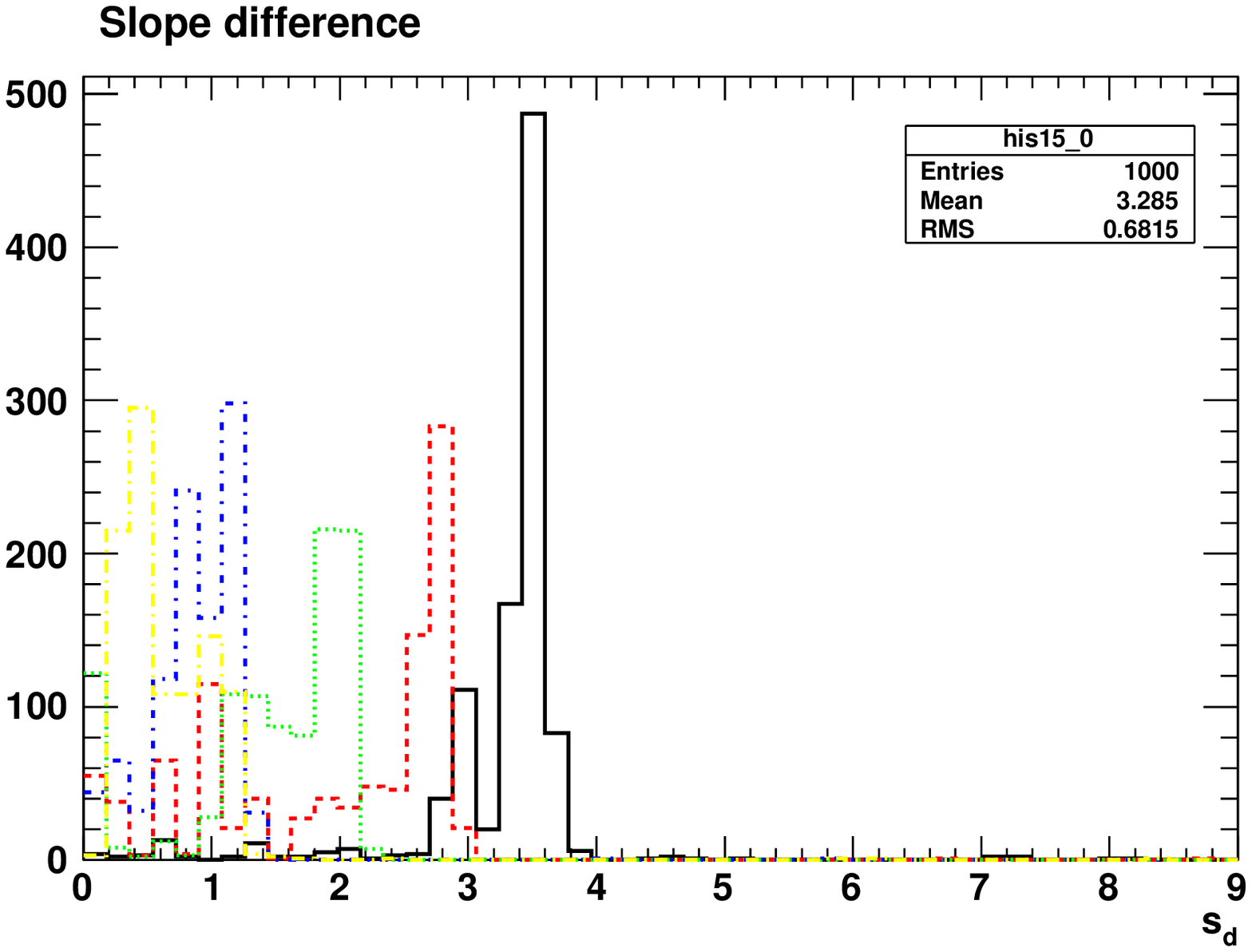}
  \includegraphics[width=0.47\textwidth]{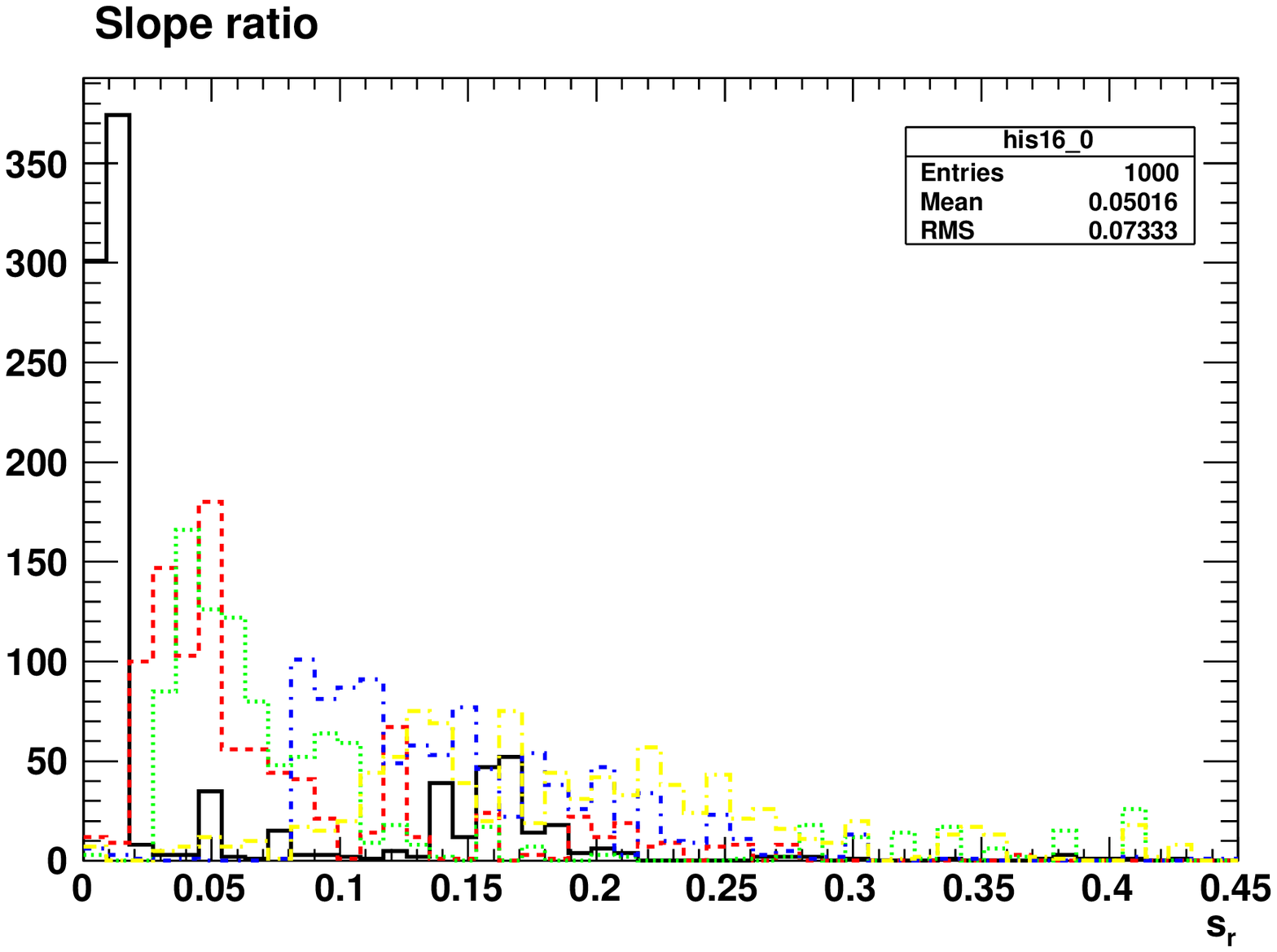}\\
  \caption{Statistically distributed slope differences (left) and
    slope ratios (right) for $m_{bb\ell,low}$ (top) and
    $m_{bb\ell,high}$ (bottom), obtained as fit parameters of the
    edge-to-bump method.}  
  \label{fig:m_bbl_slopes}
\end{figure}
Regarding the shift of edge positions, the situation is comparable to
the case of $m_{bb}$: both $m_{bb\ell,low}$ and $m_{bb\ell,high}$ exhibit
displacements of up to 150 and 180 GeV, respectively. For the highest
value of $\gamma = 0.15$ in Table~\ref{tab:m_bbl_fitvals}, we notice
an exceptionally small error estimate for the endpoint position of
$m_{bb\ell,low}$ in contrast to an unusually large one for
$m_{bb\ell,high}$. The reasoning behind this is a trivial matter of
statistics as is evident from Figure \ref{fig:m_bbl}: while the lower
of the two invariant masses has a sharp drop at the bin corresponding
to the endpoint position, the higher distribution has two such
\textit{fake} kinks at around 1100 and 1250 GeV, respectively. These
are purely statistical issues happening by chance and attributed to
the low overall number of events of 5,000. Hence the gross under- and
overestimation of the error estimates for both edges. The mean value
however still captures the important feature of endpoint translation:
a shift of the returned mean value of the edge-to-bump method of up to
180 GeV. 
\begin{table}[!ht]
\begin{center}
\begin{tabular}{r|ccc|ccc}
  $\gamma$ [\%] & $\bar{m}_{bbl,low}^{\text{max}}$  & $\bar{s}_d$ & $\bar{s}_r$ & $\bar{m}_{bbl,high}^{\text{max}}$  & $\bar{s}_d$ & $\bar{s}_r$  \\
\hline
  0.5   &845.2 $\pm$ 11.6 & 2.87 $\pm$ 0.50 & 0.025 $\pm$ 0.011&1009.7
  $\pm$ 00.4 & 3.47 $\pm$ 0.10 & 0.009 $\pm$ 0.005 \\  
  2.5   &864.4 $\pm$ 02.2 & 1.79 $\pm$ 0.91 & 0.046 $\pm$ 0.020
  &1017.0 $\pm$ 01.7 & 2.11 $\pm$ 0.77 & 0.041 $\pm$ 0.012 \\  
  5.0   &900.8 $\pm$ 23.5 & 1.45 $\pm$ 0.22 & 0.067 $\pm$ 0.022&1060.0
  $\pm$ 10.6 & 1.72 $\pm$ 0.32 & 0.050 $\pm$ 0.012 \\  
  10.0   &979.4 $\pm$ 34.8 & 0.62 $\pm$ 0.29 & 0.156 $\pm$ 0.051&
  1120.8 $\pm$ 11.3 & 0.94 $\pm$ 0.22 & 0.126 $\pm$ 0.031 \\  
  15.0   &1002.3 $\pm$ 4.9 & 0.71 $\pm$ 0.07 & 0.163 $\pm$
  0.041&1184.3 $\pm$ 62.9 & 0.63 $\pm$ 0.30 & 0.175 $\pm$ 0.052 \\  
  \end{tabular}
  \end{center}
\caption{Adapted mean values of endpoint positions (in GeV), slope
  differences (in 1/GeV) and slope ratios for $m_{bb\ell,low}$ and
  $m_{bb\ell,high}$.} 
\label{tab:m_bbl_fitvals}
\end{table}
Comparing these with the theoretically expected values
\begin{align}
  m_{bbl,low}^{\text{max}} = 868.6 \\
  m_{bbl,high}^{\text{max}} = 996.6
\end{align}
we again find agreement for small widths, but large deviations for
high values of $\gamma$. In Fig.~\ref{fig:m_bbl_slopes} and
Table~\ref{tab:m_bbl_fitvals}, the slope differences $\bar s_d$
(ratios $\bar s_r$) of the two variables both display a clear
decreasing (increasing) trend for growing (falling) effective width
factors $\gamma$ and thus confirm our choice as a parameter
quantifying the size of the off-shell effects on the quality of the
mass determination.

\begin{figure}[!ht]
  \centering
  \includegraphics[width=0.47\textwidth]{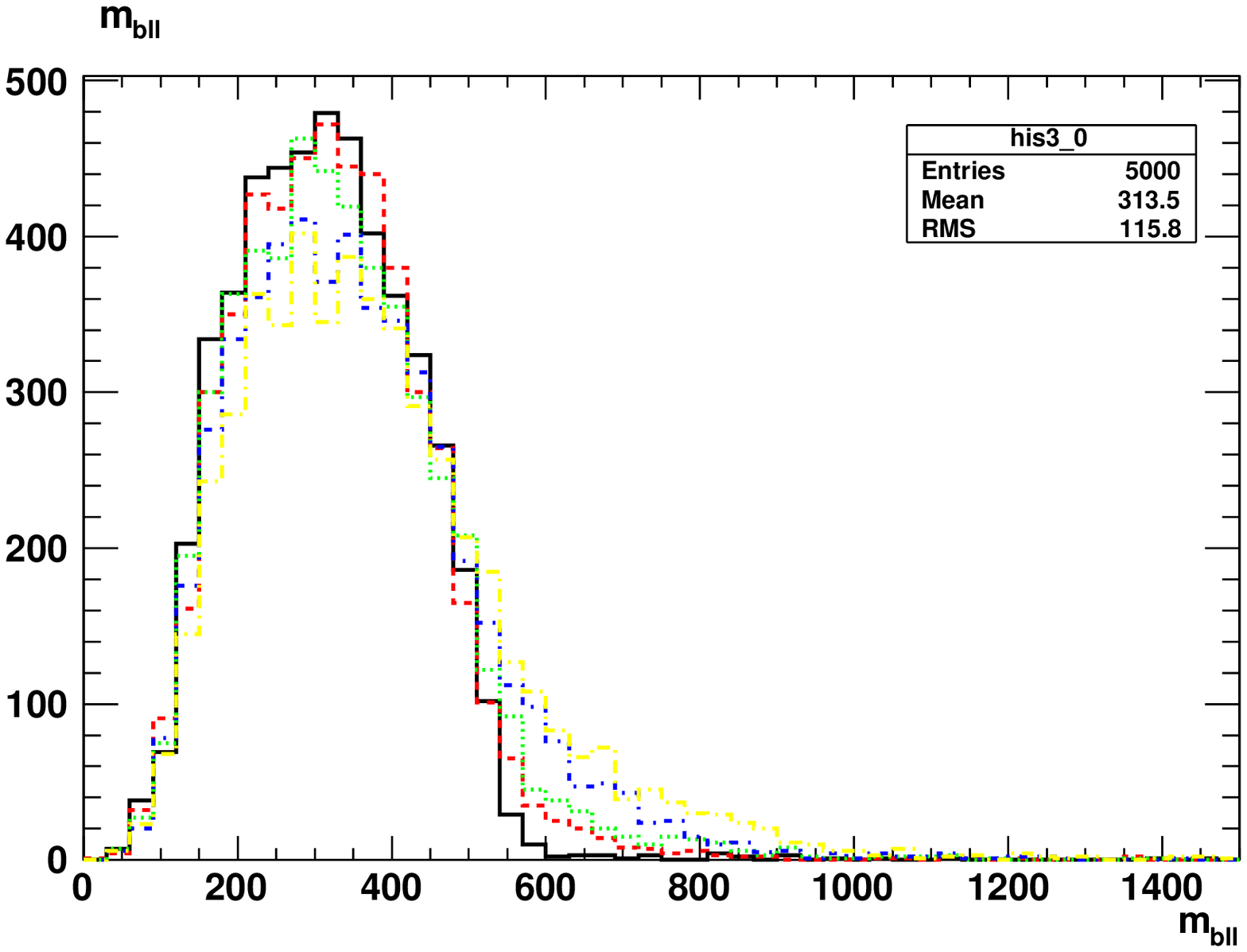}
  \includegraphics[width=0.47\textwidth]{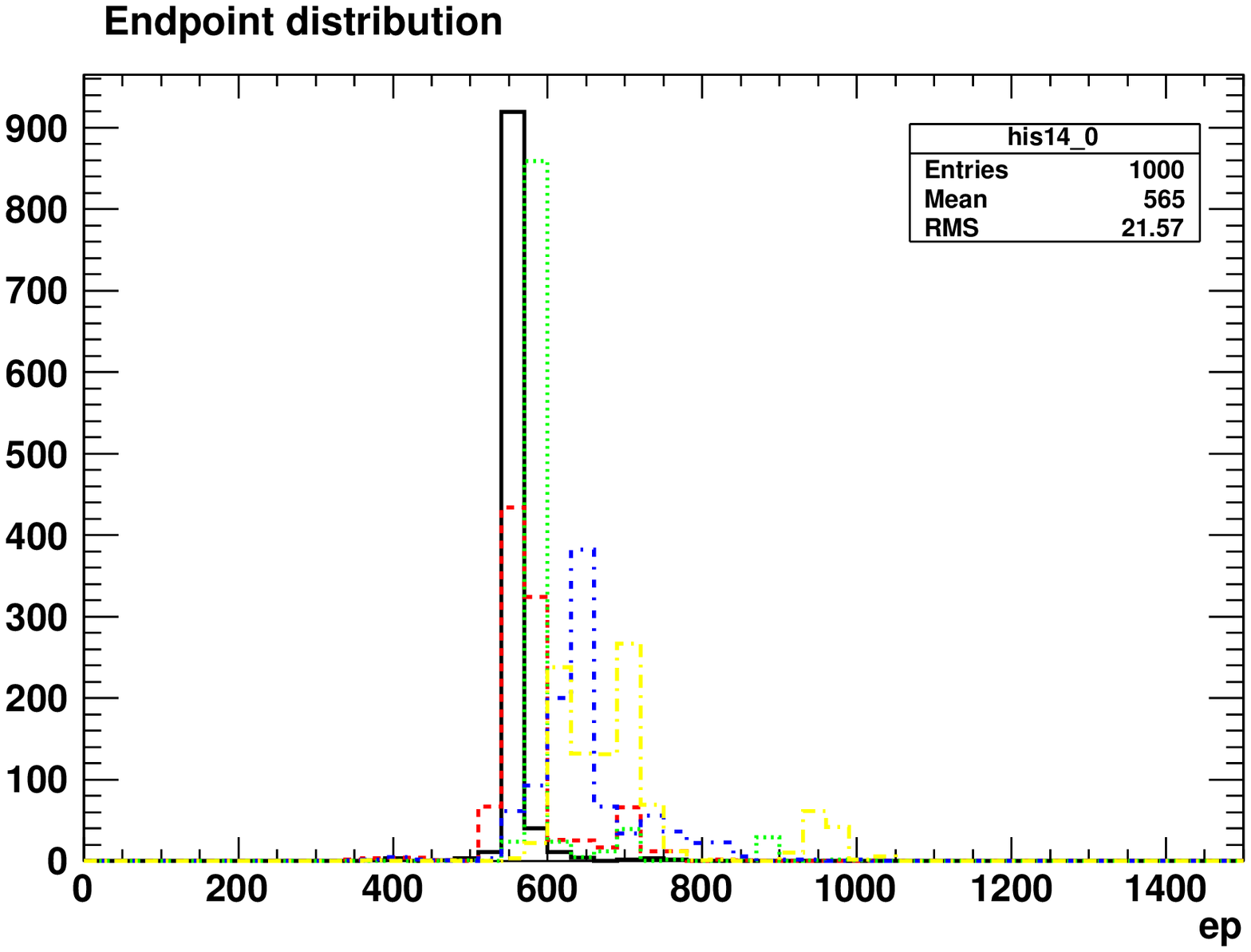}\\
  \includegraphics[width=0.47\textwidth]{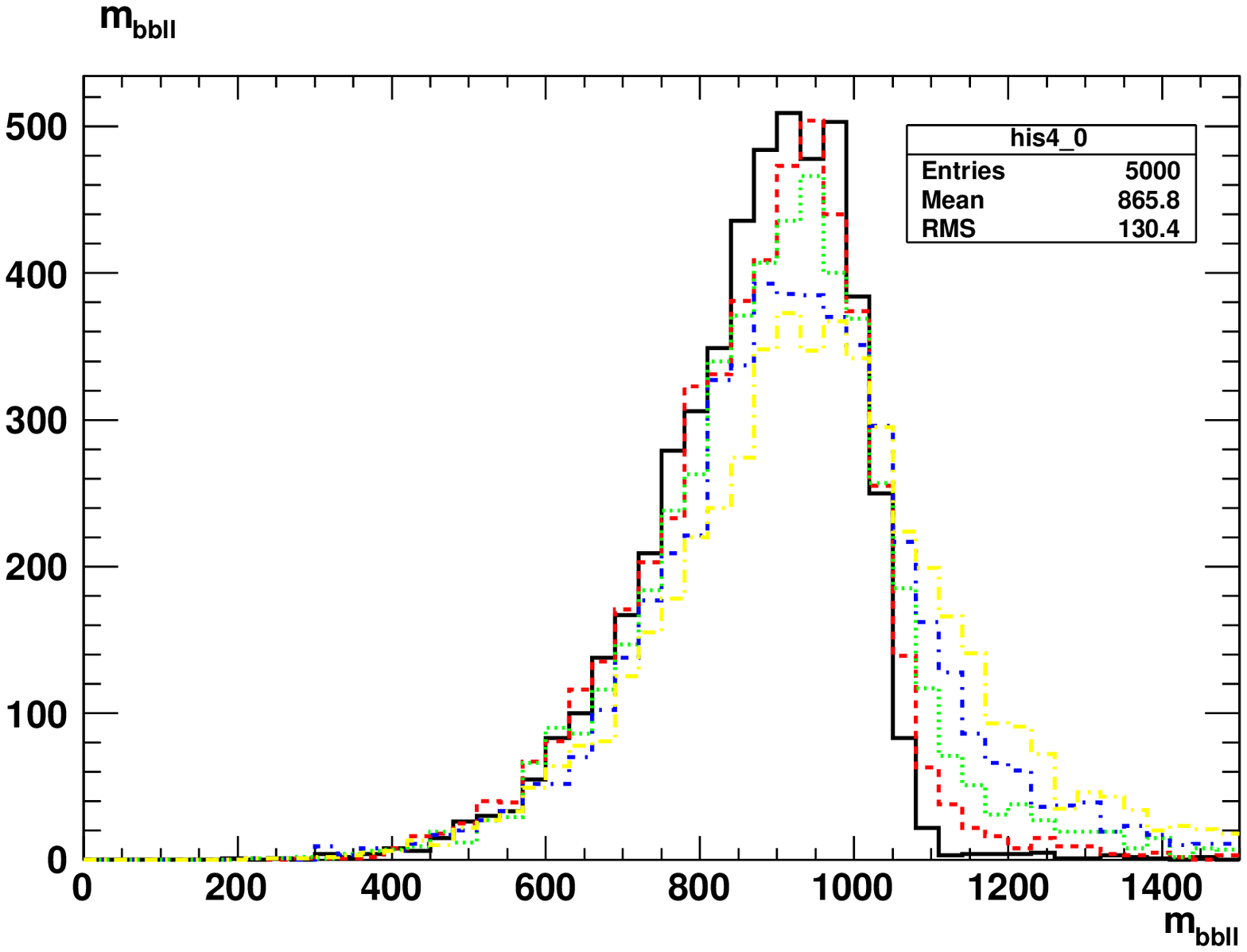}
  \includegraphics[width=0.47\textwidth]{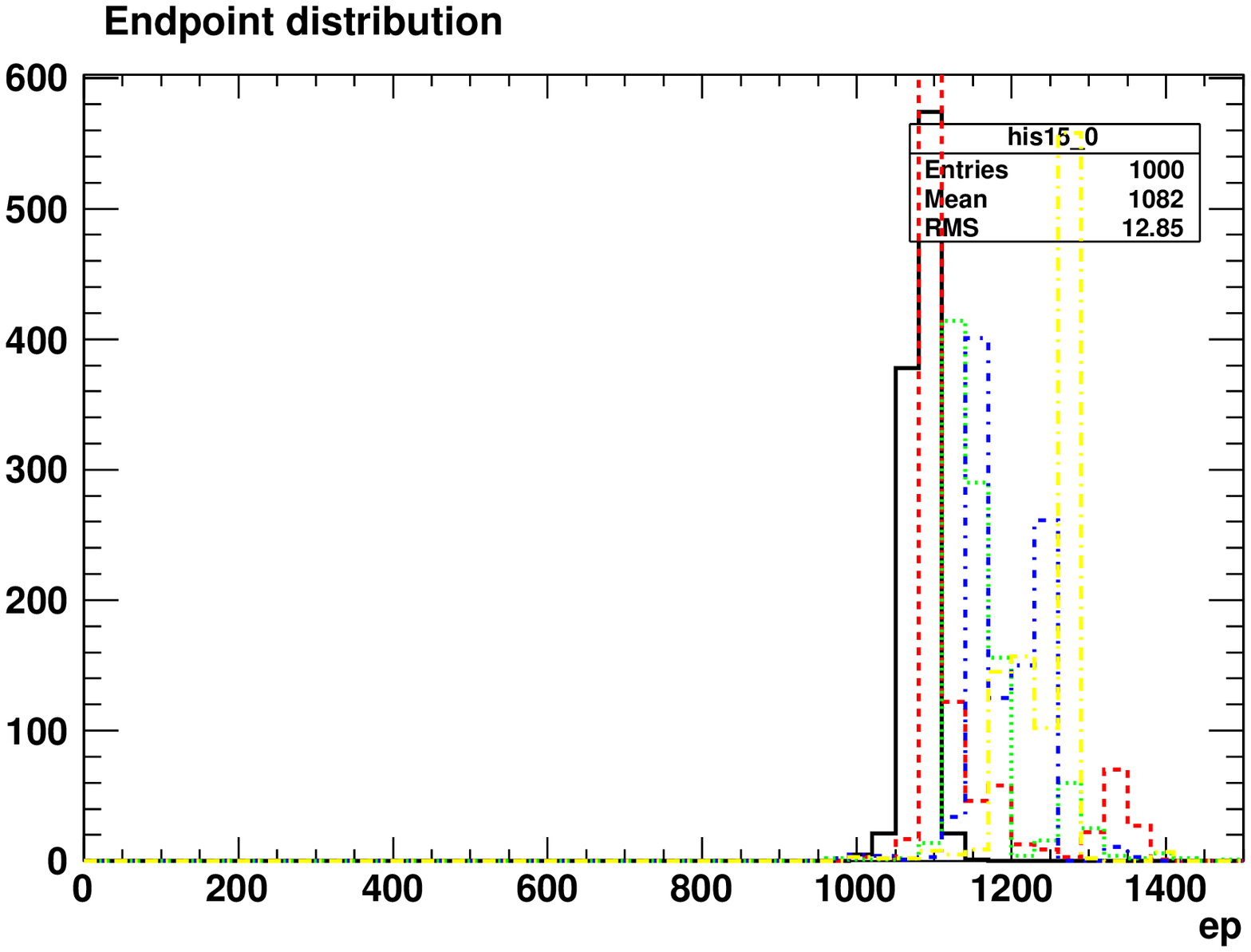}\\
  \caption{Left: Invariant mass of bottom quark(s) and leptons:
    $m_{b_n\ell\ell}$ (top) and $m_{bb\ell\ell}$ (bottom) for
    different values of $\gamma$. The black (solid),
    red (short-dashed), green (dotted), blue (short-dashed-dotted),
    and yellow (long-dashed-dotted) correspond to $\gamma =$ 0.5 \%,
    2.5 \%, 5.0 \%, 10.0 \%, and 15.0 \%, respectively. Right:
    Distribution of the corresponding endpoints obtained as fit
    parameters of the edge-to-bump method for $m_{b_n\ell\ell}$ (top)
    and $m_{bb\ell\ell}$ (bottom).}  
  \label{fig:m_bll}
\end{figure}
\begin{figure}[!ht]
  \includegraphics[width=0.47\textwidth]{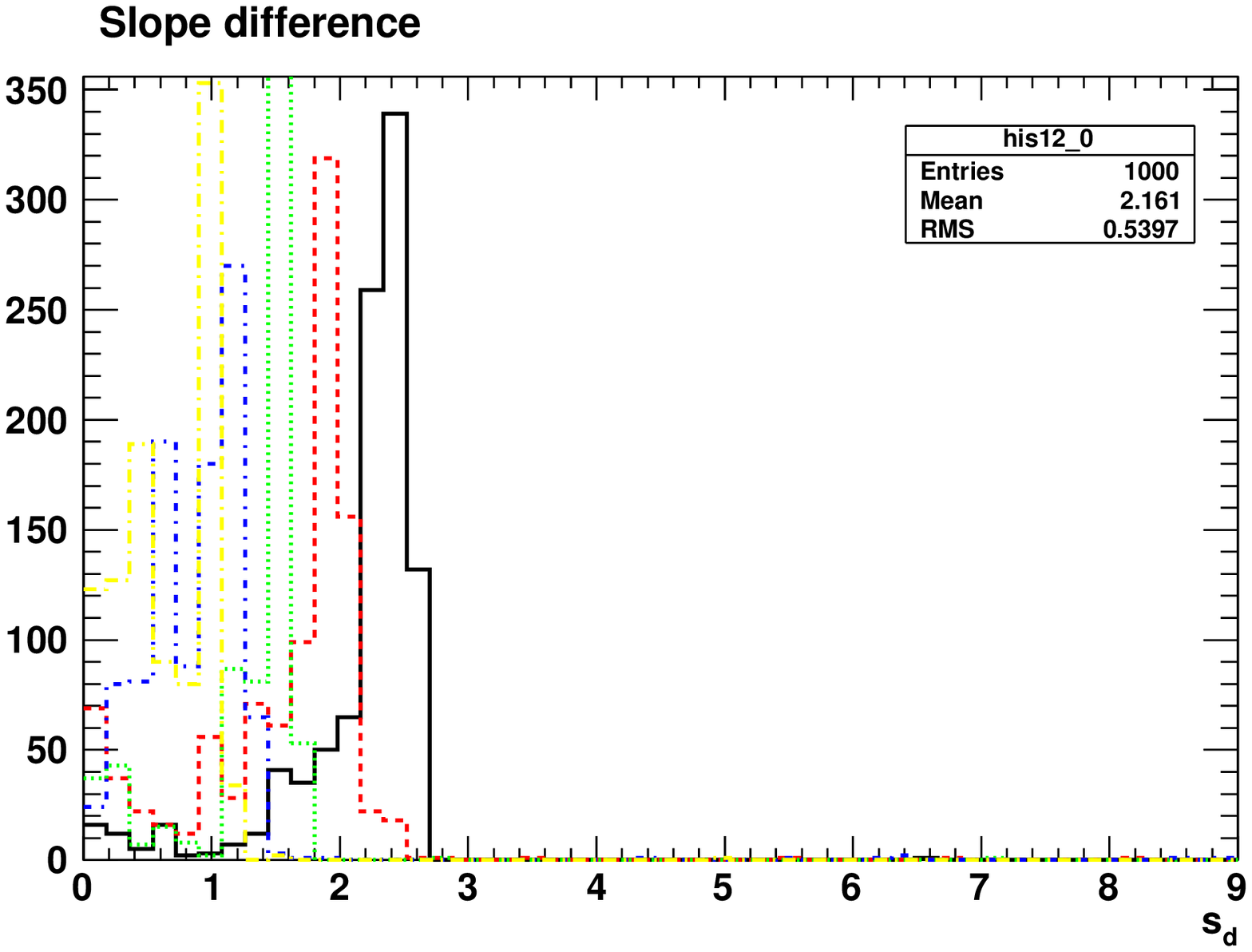}
  \includegraphics[width=0.47\textwidth]{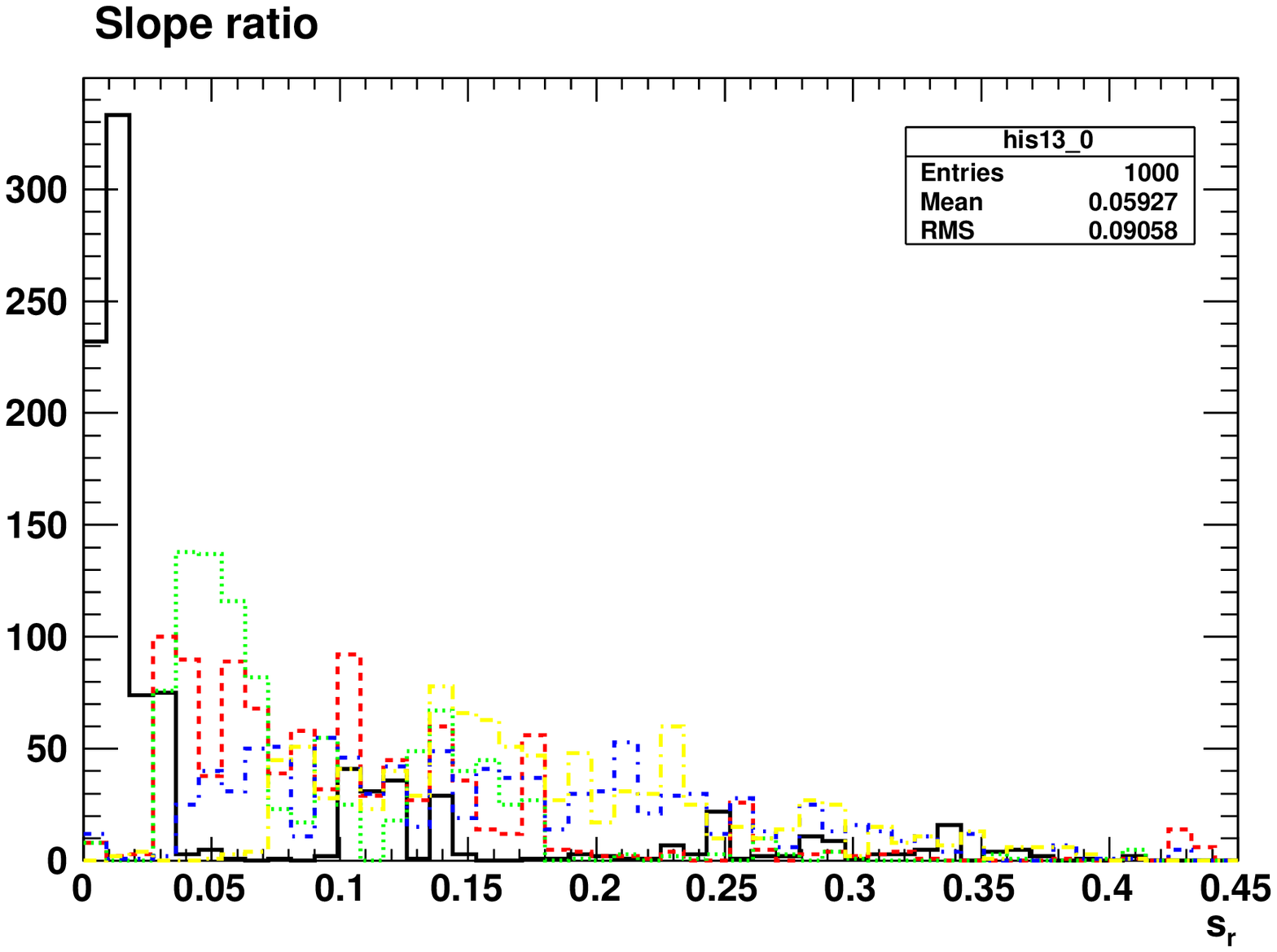}\\
  \includegraphics[width=0.47\textwidth]{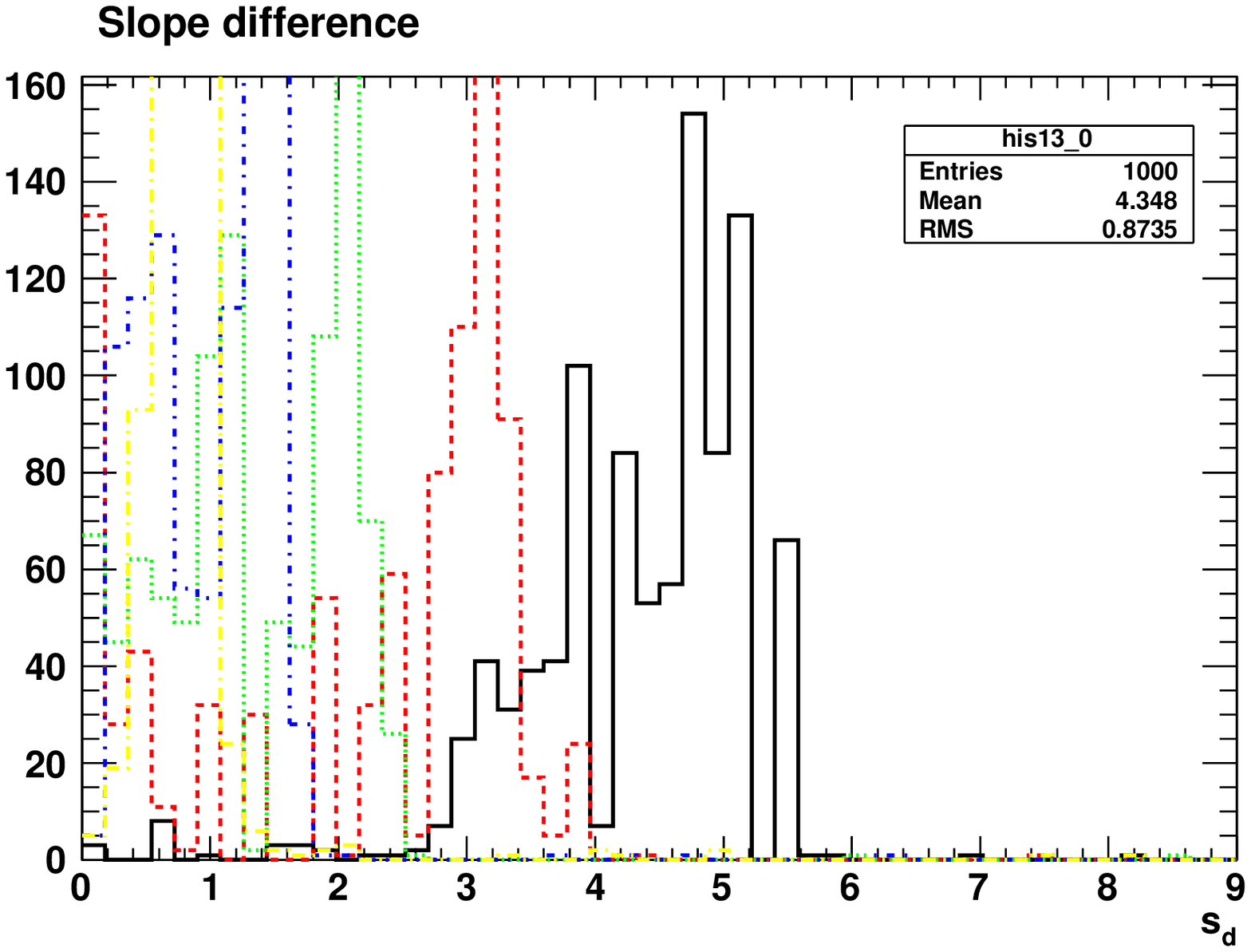}
  \includegraphics[width=0.47\textwidth]{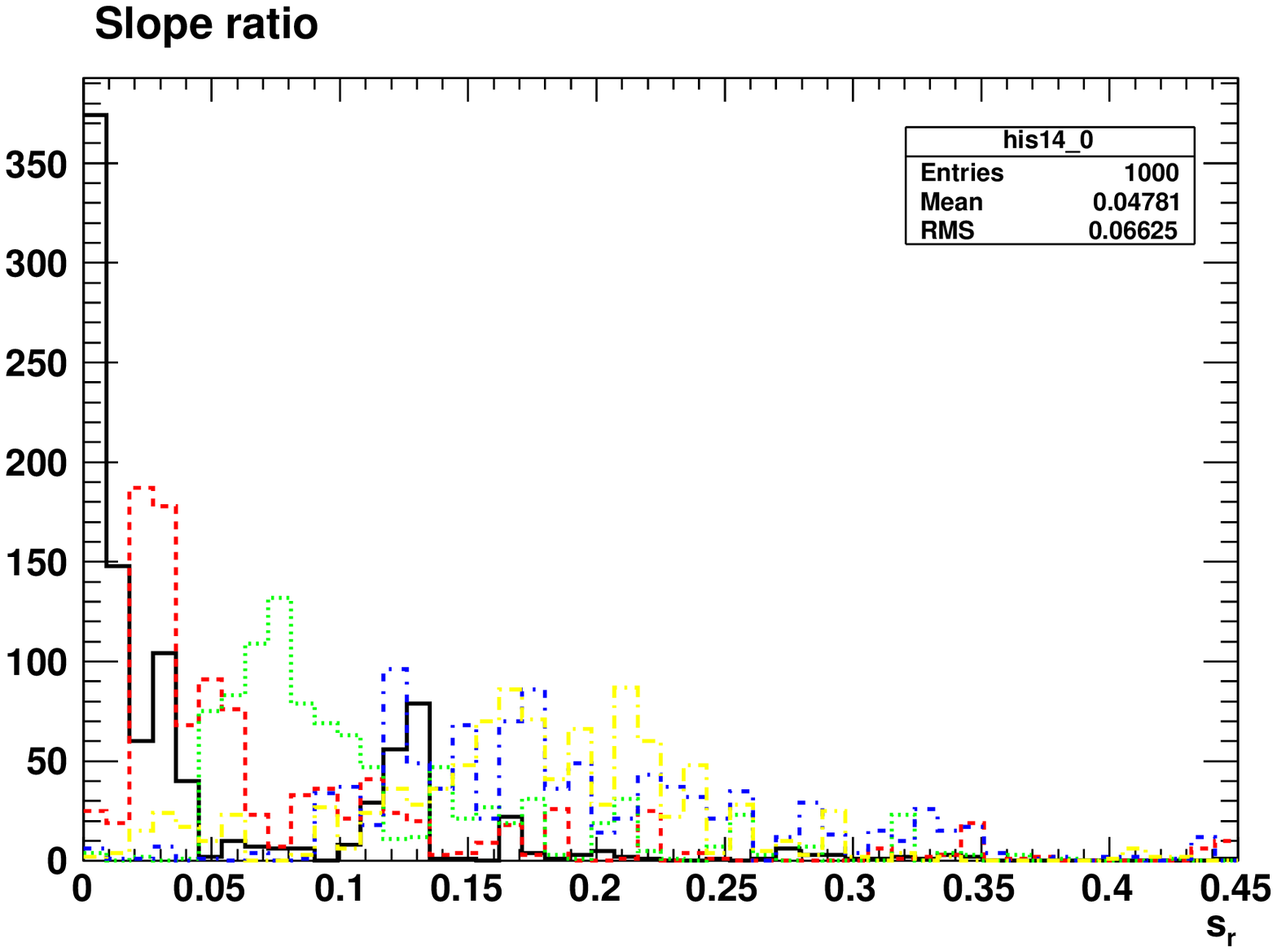}\\
  \caption{Statistically distributed slope differences (left) and
    ratios (right) for $m_{b\ell\ell}$ (top) and $m_{bb\ell\ell}$
    (bottom) obtained as fit parameters of the edge-to-bump method.}   
  \label{fig:m_bll_slopes}
\end{figure}

\paragraph{$\boxed{m_{b_n\ell\ell}}$}
The last three-particle invariant mass we investigate is the
combination of the near bottom quark and the lepton pair. Without the
particular need for distinction between the two leptons, we expect to
observe only a moderate distortion for large gluino widths, which
might be weakened by the fact that the first linear slope of the
undistorted distribution is not maximally steep since
$m_{b_n\ell\ell}$ is not built out of direct next neighbors in the
decay chain. An example of a maximally steep endpoint behavior is the
undistorted invariant di-bottom mass, whose first slope parameter
would ideally be infinite due to the triangular (phase-space)
shape. Fig.~\ref{fig:m_bll} depict in the upper line $m_{b_n\ell\ell}$
and the corresponding endpoint distributions, numerical values of
which are given on the left-hand side of
Table~\ref{tab:m_bll_fitvals}.    
Despite a fairly accurate endpoint estimate for up to $\gamma = 2.5\%$ 
(in comparison with the theoretical value of 578.8 GeV), there is
still a deviation of substantial size for large widths of up to 100
GeV. The two slope parameters both behave similarly to what we have
already seen with other invariant mass variables
(cf. Fig.~\ref{fig:m_bll_slopes}): an average slope difference $\bar
s_d$ of well above two is reduced to a value below one, that is
compatible with zero within two standard deviations. For small widths,
the slope ratio $\bar s_r$ is also close to zero, but increasing
$\gamma$ results in a steady growth of $\bar s_r$ of more than one
order of magnitude in size to values of up to 16 \%. 
\begin{table}
  \begin{center}
\begin{tabular}{rccc|ccc}
  $\gamma$ [\%] & $\bar{m}_{bll}^{\text{max}}$  & $\bar{s}_d$ & $\bar{s}_r$  & $\bar{m}_{bbll}^{\text{max}}$  & $\bar{s}_d$ & $\bar{s}_r$  \\
\hline
  0.5    & 565.4 $\pm$ 1.7 & 2.37 $\pm$ 0.15 & 0.010 $\pm$ 0.002   &
  1079.3 $\pm$ 2.4 & 4.64 $\pm$ 0.51 & 0.011 $\pm$ 0.009 \\  
  2.5    & 568.5 $\pm$ 3.4 & 1.92 $\pm$ 0.08 & 0.080 $\pm$ 0.037   &
  1105.1 $\pm$ 1.3 & 2.22 $\pm$ 1.22 & 0.035 $\pm$ 0.013 \\  
  5.0    & 594.5 $\pm$ 2.0 & 1.56 $\pm$ 0.04 & 0.050 $\pm$ 0.012   &
  1150.2 $\pm$ 19.9 & 1.34 $\pm$ 0.70 & 0.078 $\pm$ 0.020 \\  
  10.0   & 640.0 $\pm$ 12.0 & 0.94 $\pm$ 0.25 & 0.142 $\pm$ 0.064  &
  1193.7 $\pm$ 37.4 & 1.00 $\pm$ 0.44 & 0.163 $\pm$ 0.041 \\  
  15.0   & 668.6 $\pm$ 40.5 & 0.65 $\pm$ 0.33 & 0.157 $\pm$ 0.050  &
  1251.2 $\pm$ 30.9 & 0.78 $\pm$ 0.17 & 0.182 $\pm$ 0.036 \\  
\end{tabular}
\end{center}
\caption{Adapted mean values of endpoint positions (in GeV), slope
  differences (in 1/GeV) and slope ratios for $m_{b\ell\ell}$ and
  $m_{bb\ell\ell}$.} 
\label{tab:m_bll_fitvals}
\end{table}


\paragraph{$\boxed{m_{bb\ell\ell}}$}
Combining all objects from one cascade, we obtain the variable
$m_{bb\ell\ell}$, which has a very clear-cut endpoint and a steep edge 
structure, since it is constructed out of next neighbors only. The
sharper the drop of the distribution at the vicinity of the edge, the
larger we expect the impact of off-shell contributions to be, even so
if just one out of four propagators, in our case the signal gluino, is
affected. From the lower line of Fig.~\ref{fig:m_bll}, the amount of
distortion is already visible by eye. A more quantitative
statement is given in terms of numerical values on the right-hand side
of Table~\ref{tab:m_bll_fitvals}. While the estimated endpoints for
small values of $\gamma$ are in gross agreement with the theoretically
expected value of 1092.7 GeV, the largest deviation of
$m_{bb\ell\ell}^{\text{max}}$ for $\gamma = 15 \%$ is up to 172 GeV.  
A similar picture as for $m_{b_n\ell\ell}$ is found for the slope
parameters (cf. also Fig.~\ref{fig:m_bll_slopes}): ranging from just
below five down to well below one, the slope difference exhibits an
even larger spread of values. The slope ratios on one hand are
compatible with zero for the smallest width $\gamma = 0.5 \%$, but, on 
the other hand, increase by a factor of 15 for the largest off-shell
contribution.   


\subsection{Inclusive Approaches}

Up to now, we only studied exclusive invariant mass variables, 
i.e. combinations of objects arising from one cascade side from a
decay of just one single mother. In our case, this was taken to be the
gluino, which has a large variety of possible decay patterns and
consequently a plethora of interesting invariant mass
combinations. Restricting the analysis to just one particular decay
cascade entails several problems, the largest is presumably
combinatorics. Consider for example the symmetric case of two
identical decay chains of the type we have analyzed so far, where not
one, but two gluinos decay into two bottom quarks, two leptons and the
lightest neutralino. All variables we have just discussed assume that
a differentiation between the two cascades is somehow given
(e.g. the partitioning of four leptons into 2 $\times$ 2 leptons).
However, a priori, there is no general recipe that always allows for
such a correct assignment. This severely affects the usability of
these exclusive variables and is known as the combinatorial problem. 

\paragraph{$\boxed{M_{T2}}$}
An alternative, less exclusive approach is $M_{T2}$ as introduced
in~\cite{mt2}, which is a grouping of the measured objects into two
subsets and a following minimazation procedure over the invariant
transverse masses. Although $M_{T2}$ is similarly affected 
by the combinatorial problem (since here the visible momentum has to
be split into two separate sides as well) there are methods which
address this issue\footnote{e.g. $M_{T\text{Gen}}$
  \cite{Lester:2007fq}, which is the minimum of $M_{T2}$ for all
  possible momentum assignments into two partitions}. In the remainder
of this chapter we circumvent these kind of combinatorics through two
non-identical decay chains, which allow us to directly concentrate on
the off-shell effects. Hence, the partition of visible momenta into
two sides for the application of $M_{T2}
(p_{vis}^{(1)},p_{vis}^{(2)},\slashed{p}_T,m_\chi)$ is trivially given 
by 
\begin{align}
  p_{vis}^{(1)}& = \{b,\bar b,\ell^\pm,\ell^\mp\} \\
  p_{vis}^{(2)}& = \{q,\bar q\} \qquad ,
\end{align}
where the light-flavor partonic jets are assigned to one side and
the remaining objects to the other. The input test mass of the
invisible sparticle was set to $m_\chi = 150$ GeV close to the true
value of the benchmark scenario.   
\begin{figure}[!t]
  \includegraphics[width=0.47\textwidth]{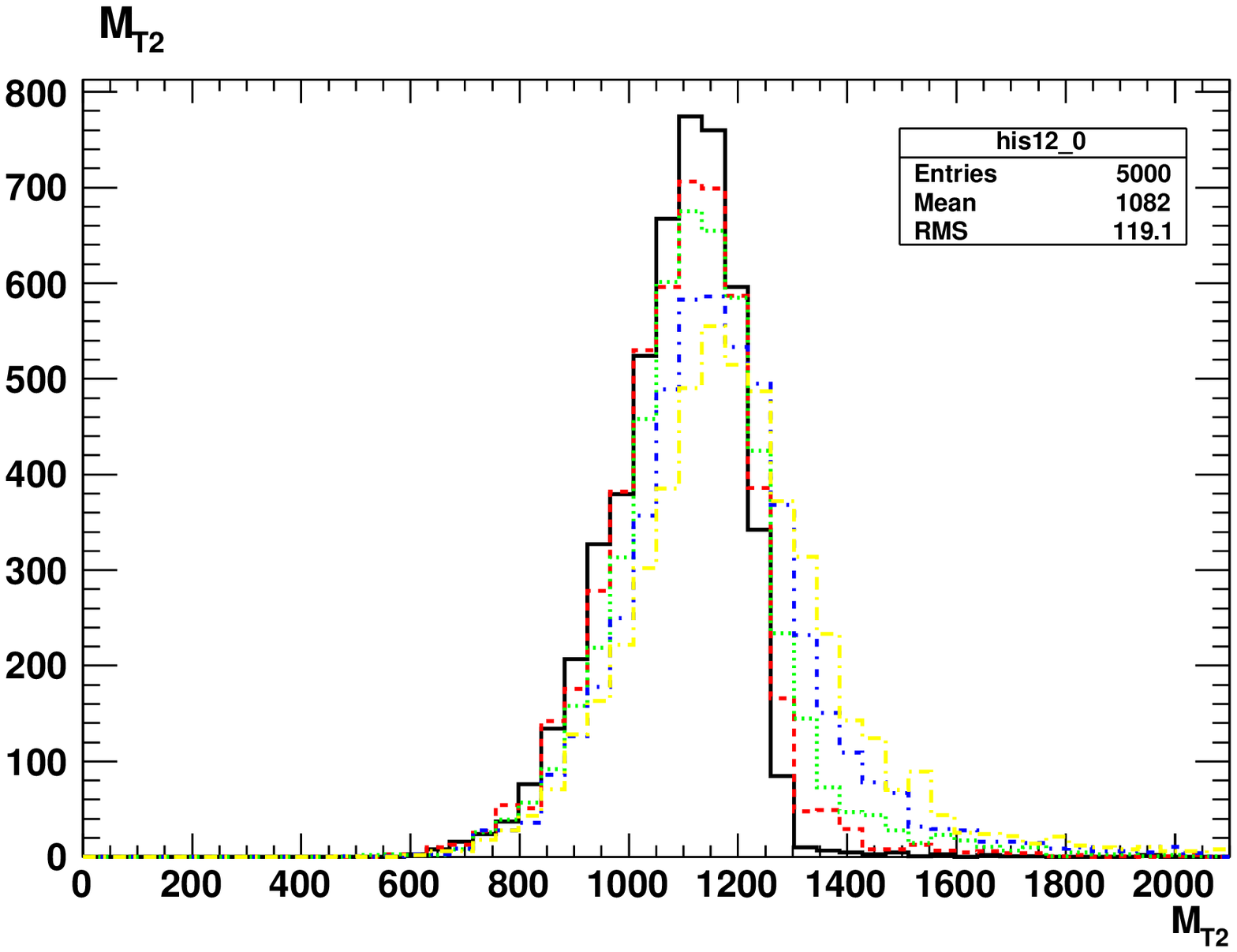}
  \includegraphics[width=0.47\textwidth]{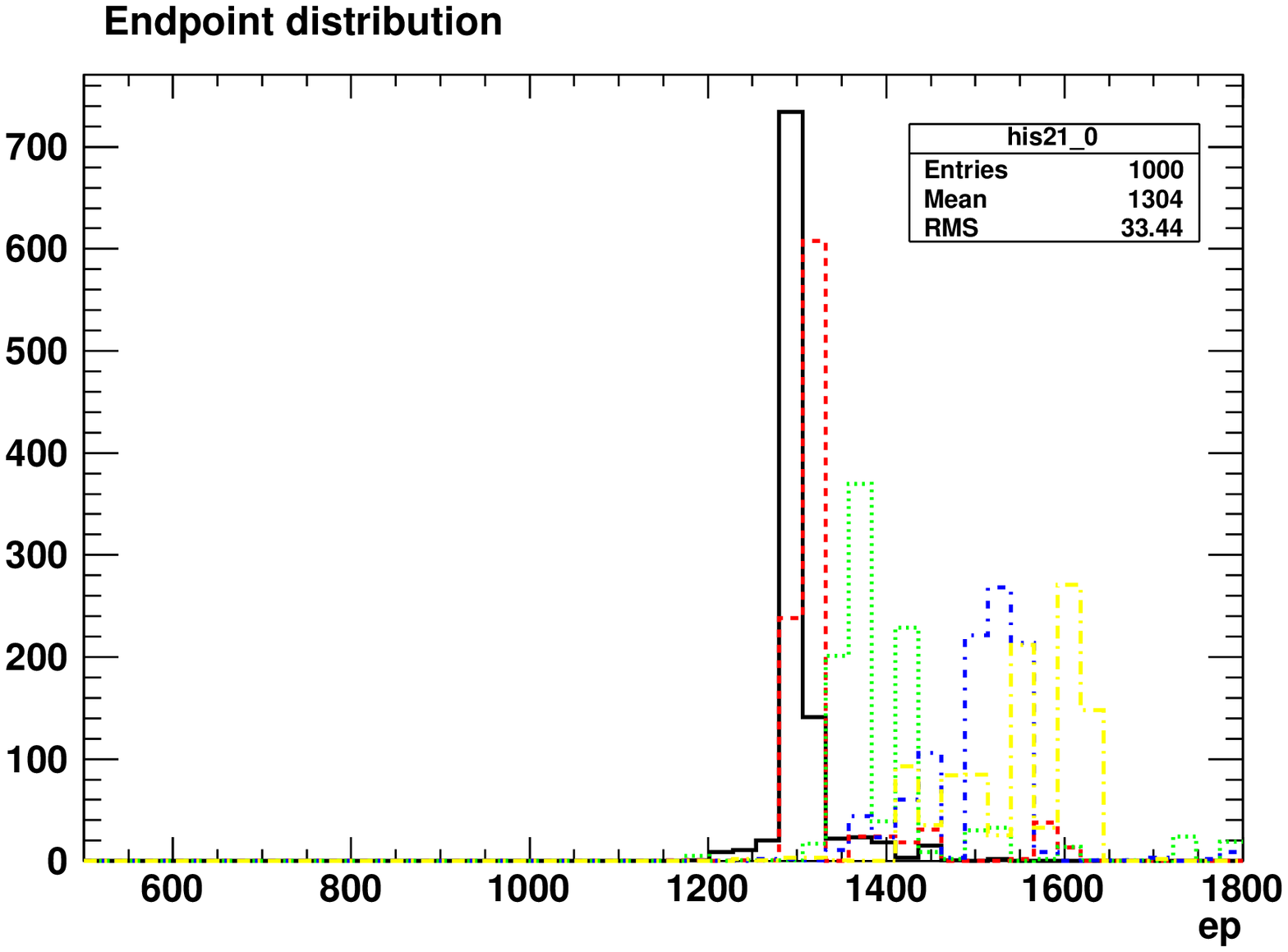}\\
  \includegraphics[width=0.47\textwidth]{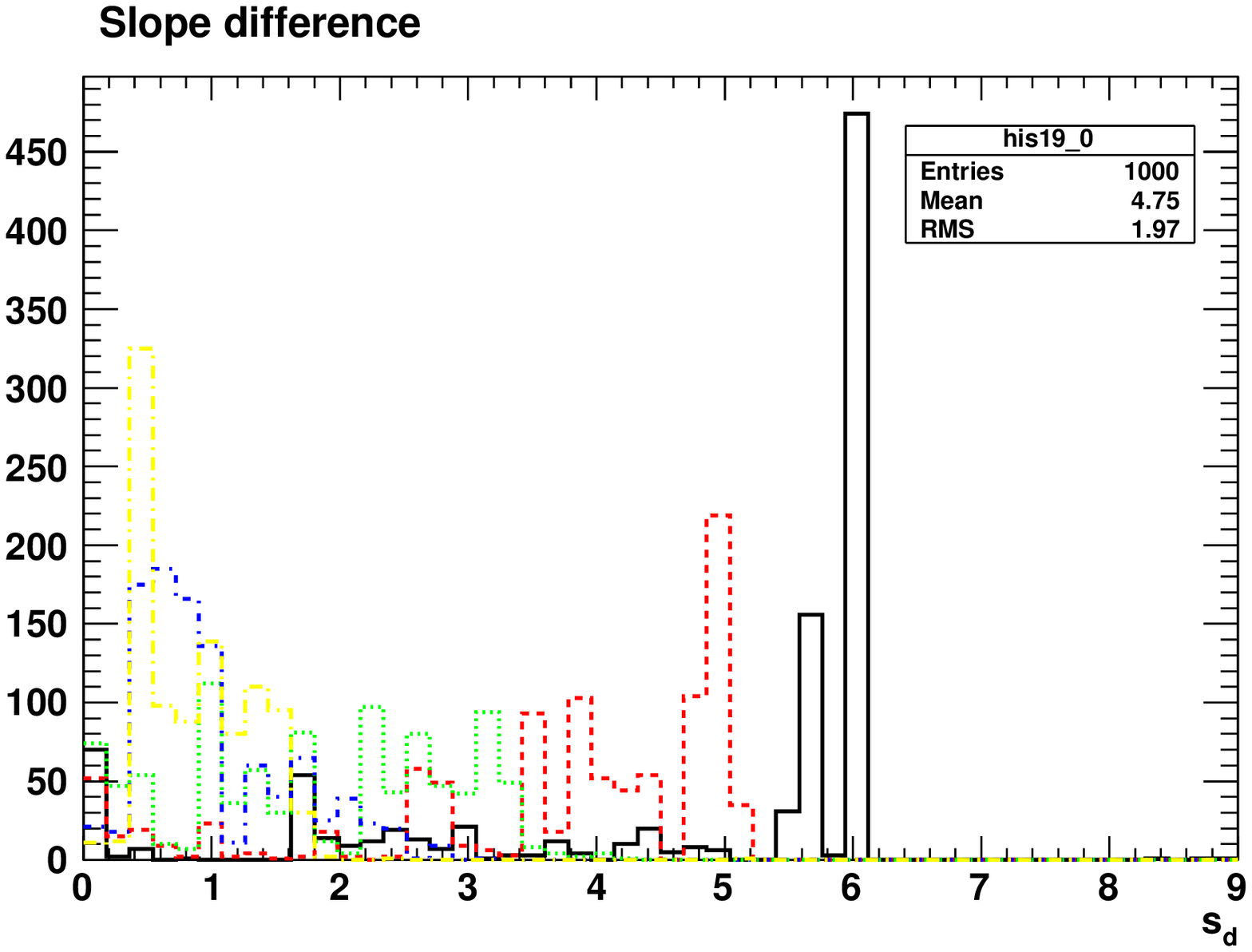}
  \includegraphics[width=0.47\textwidth]{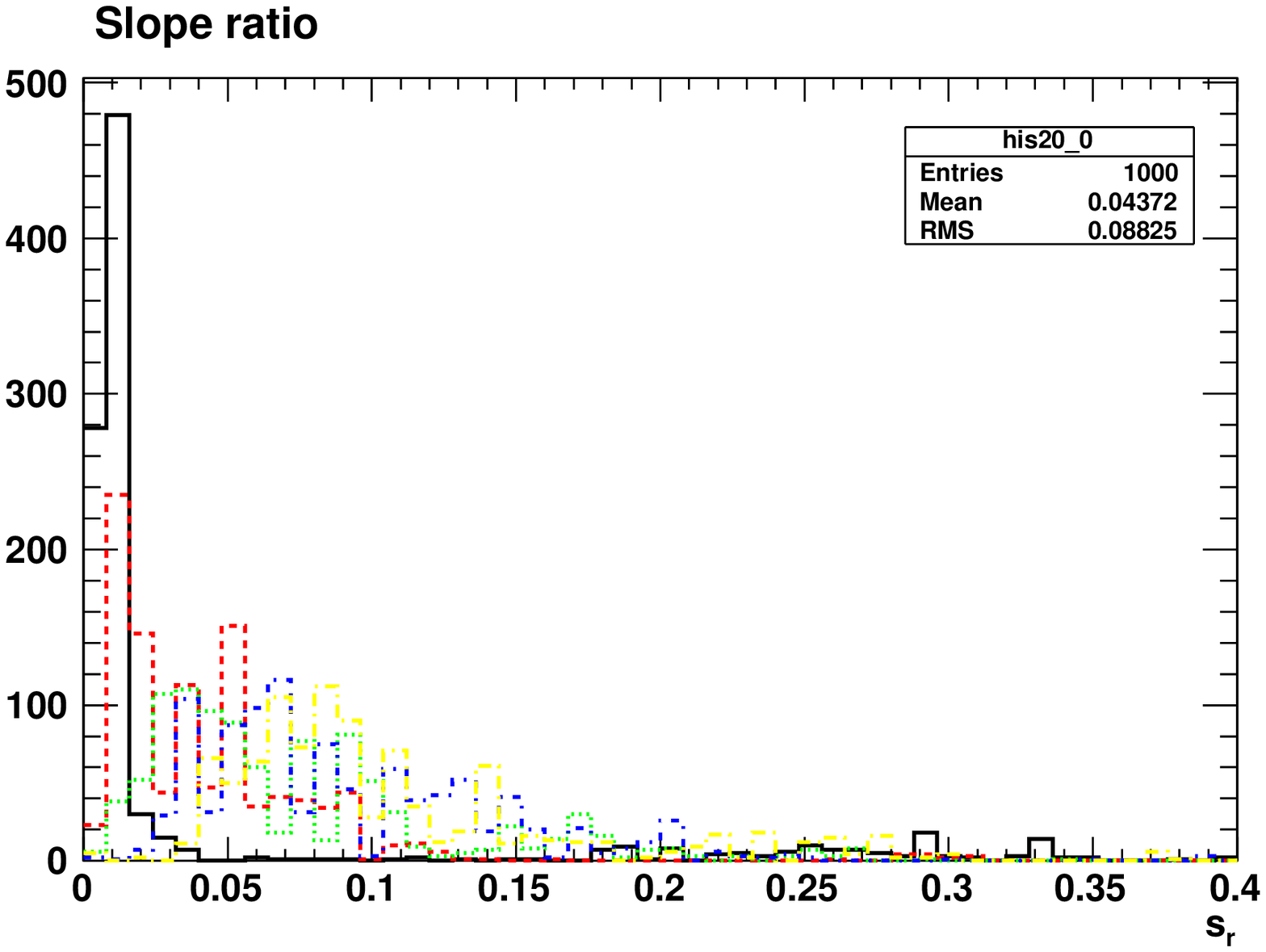}\\
  \caption{$M_{T2}$ distribution (upper left) and its statistically
    distributed endpoints (upper right), slope differences (lower
    left) and slope ratios (lower right) obtained as  
    fit parameters of the edge-to-bump method for five different 
    values of $\gamma$. The color scheme is the same as in
    Fig.~\ref{fig:m_bll}.}  
  \label{fig:m_mt2}
\end{figure}
Figure \ref{fig:m_mt2} illustrates in the upper left plot the impact
of the effective width parameter $\gamma$ on the distribution of
$M_{T2}$: similar to the exclusive invariant mass variables discussed
above, relative widths larger than 2.5 \% result in substantial
smearing and a long tail in the distributions, which is absent at
lower values. Owing to the general minimization procedure of $M_{T2}$,
the effect is somewhat smaller than for exclusive variables such as
$m_{bb}$ or $m_{bb\ell\ell}$, which is underlined by the numerical
results given in Table~\ref{tab:m_mt2_fitvals}.   
\begin{table}[!ht]
\begin{center}
\begin{tabular}{r|ccc}
  $\gamma$ [\%] & $\bar{M}_{T2}^{\text{max}}$  & $\bar{s}_d$ &
  $\bar{s}_r$  \\ 
  \hline
   0.5 & 1293.7 $\pm$ 0.8 & 5.92 $\pm$ 0.21 & 0.009 $\pm$ 0.003 \\  
   2.5 & 1315.0 $\pm$ 10.8 & 4.42 $\pm$ 0.54 & 0.029 $\pm$ 0.017 \\ 
   5.0 & 1381.0 $\pm$ 27.4 & 1.82 $\pm$ 1.03 & 0.054 $\pm$ 0.027 \\ 
   10.0 & 1526.0 $\pm$ 18.6 & 0.70 $\pm$ 0.20 & 0.072 $\pm$ 0.029 \\ 
   15.0 & 1589.5 $\pm$ 31.4 & 0.87 $\pm$ 0.40 & 0.079 $\pm$ 0.022 \\ 
\end{tabular}
\end{center}
\caption{Adapted mean values of endpoint positions (in GeV), slope
  differences (in 1/GeV) and slope ratios for $M_{T2}$.} 
\label{tab:m_mt2_fitvals}
\end{table}
The endpoint $M_{T2}^{\text{max}}$ is given by the parent sparticle
mass, thus we expect to observe a clear edge structure at the gluino
mass of 1277 GeV. Up to a slight overshoot, this is in gross agreement
with the values given for $\gamma < 5 \%$. However, effective widths
of up to 15 \% lead to an edge shift of nearly 300 GeV. This is the
largest translation of an endpoint we have obtained so far and can be
understood in terms of the underlying topology: in this inclusive
scenario both partitions are affected by off-shell contributions
through non-resonant parts of the gluino propagators, and hence they
have their share in the distribution by means of the definition of
$M_{T2}$. To this extent, it was necessary to create a special event
sub-sample, which slightly differs from the one introduced in the
beginning. With respect to Eq.~\eqref{eq:factorize_prod}, the
factorization of the spectator gluino was exchanged with by full
matrix-element calculation according to 
\begin{align}
    p p \rightarrow (b \bar b \tilde{\chi}^0_2) + (q \bar q
    \tilde{\chi}^0_1) \qquad ,
\end{align}
since the light quarks now also play a role in the construction of
$M_{T2}$. The six-particle final state (with 
subsequent factorized decay of the second-to-lightest neutralino
$\tilde \chi_2^0$) is necessary to be able to fully analyze all
off-shell effects in this inclusive scenario. The slope parameters in
Table~\ref{tab:m_mt2_fitvals} confirm the strength of the deviation:
while the differences $s_d$ exhibit a significant drop for rising
values of $\gamma$, the ratios $s_r$ show a minor but steady increase.  

In a first summary, mass determination methods generally all suffer
from off-shell effects. The amount of smearing in the distribution
depends on the exclusiveness or inclusiveness of the variable. In
general, the more particles originating from one or even two fat
gluinos are involved in the construction of the mass variable, the
more distortion or smearing appears in the distributions. The errors
in the mass determination already at parton-level can be as large as
10-15 per cent. 


\section{Effects on Spin Determinations}
\label{sec:spin}

After a possible discovery of any new physics beyond the SM, the next
steps in determining the underlying model characteristics are the
measurements of masses and spins of novel particles. While many mass
determinations rely on endpoint positions of invariant mass
distributions, the nature of the underlying spin is encoded in the
shape of those (and especially angular) distributions and as such more
delicate to differentiate. Hence, after discussing the effects of off-shell
contributions on mass determination variables, in this section we turn
to spin measurements and carefully analyze the effects of a fat gluino
onto several methods designed to distinguish a hypothetical SUSY
signal from an equivalent one of Universal Extra Dimensions (UED). In
general, spin studies compare shapes of distributions by choosing one
particular type of mass spectrum which is either of UED or
supersymmetric nature. Though typical UED spectra are far more
compressed resulting in softer decay 
products and thus require for more comprehensive analyses
(cf. e.g.~\cite{Smillie:2005ar,Alves:2006df}), we choose to 
stick to hierarchical SUSY like spectra to get a comprehensive
comparison. Moreover, as we want to
emphasize the difference of spin in intermediate propagators and their
impact on invariant mass distributions before comparing it to
contributions arising from off-shell effects, we construct a
particular UED model, which inherits all masses and width parameters
from our SUSY benchmark model and hence allows us to use the
particularly interesting decay chain already known from the mass
measurement section with the following replacements: 
$(\tilde{g},\tilde{b}_i,\tilde{\chi}_2,\tilde{l}_R,\tilde{\chi}_2)
\rightarrow (g^{(1)},b^{(1)},Z^{(1)},l^{(1)},\gamma^{(1)})$. Thus,
in this cascade, edges of invariant masses stay the same but shapes
thereof are expected to drastically change. In contrast to the gluino,
the Kaluza-Klein (KK) gluon will retain a small effective width of
$\gamma =$ 0.5 \% throughout this analysis. As for the technical side,
we use an adapted version of the minimal UED model
\cite{Cheng:2002ab}, implemented into \texttt{WHIZARD} using the
\texttt{FEYNRULES}~\cite{Christensen:2008py} interface for
\texttt{WHIZARD}~\cite{Christensen:2010wz}. In the
sequel, we start to analyze shape asymmetries based on
exclusive invariant quark-lepton masses before turning to hadronic
correlations inside a single cascade and finally investigating the
impact of non-resonant contributions on inclusive angular
distributions. 


\subsection{Shape Asymmetries}

Many studies of spin measurements rely on the specific decay topology
of the \textit{golden chain} and make inherent use of invariant mass
shapes therein as discussed above. Since we are
interested in effects emerging from a gluino, we concentrate on an 
extended version:
\begin{equation}
  \tilde{g} \to q_n \tilde{q}_L \to q_n q_f \tilde{\chi}_2^0 \to q_n
  q_f \ell^\pm \tilde{\ell}^\mp_R \to q_n q_f \ell^\pm \ell^\mp
  \tilde{\chi}_1^0 \quad .
\end{equation}
On the basis of this decay chain, we analyze to what
extent the spin determination methods proposed in~\cite{Alves:2006df}
are affected by the off-shell contributions from a fat gluino. The
approaches studied within that paper were designed to discriminate
signatures of a supersymmetric gluino from the ones emerging from a
Kaluza-Klein gluon excitation in models of UED. 

\paragraph{$\boxed{A^\pm (m_{b\ell})}$} At first, we investigate the
special bottom-lepton asymmetry with the definition:
\begin{align} 
  A^\pm (m_{b\ell}) = \frac{d\sigma / dm_{b\ell^+} - d\sigma /
    dm_{b\ell^-}}{d\sigma / dm_{b\ell^+} + d\sigma / dm_{b\ell^-}}  
  \quad .
\end{align}
We follow the assumptions in~\cite{Alves:2006df}, namely that the
bottom (instead of anti-bottom) quarks are uniquely identified through
a lepton charge tag of the $b$-tagging 
algorithm. Due to the Majorana nature of the gluino, all bottom quarks
are near bottom quarks in 50 \% of all decays, i.e. they are produced
in the first two-body decay step. Hence, visible effects from
off-shell contributions of a fat gluino propagator are expected to
influence half of the invariant mass shapes. 
\begin{figure}[!t]
  \includegraphics[width=0.47\textwidth]{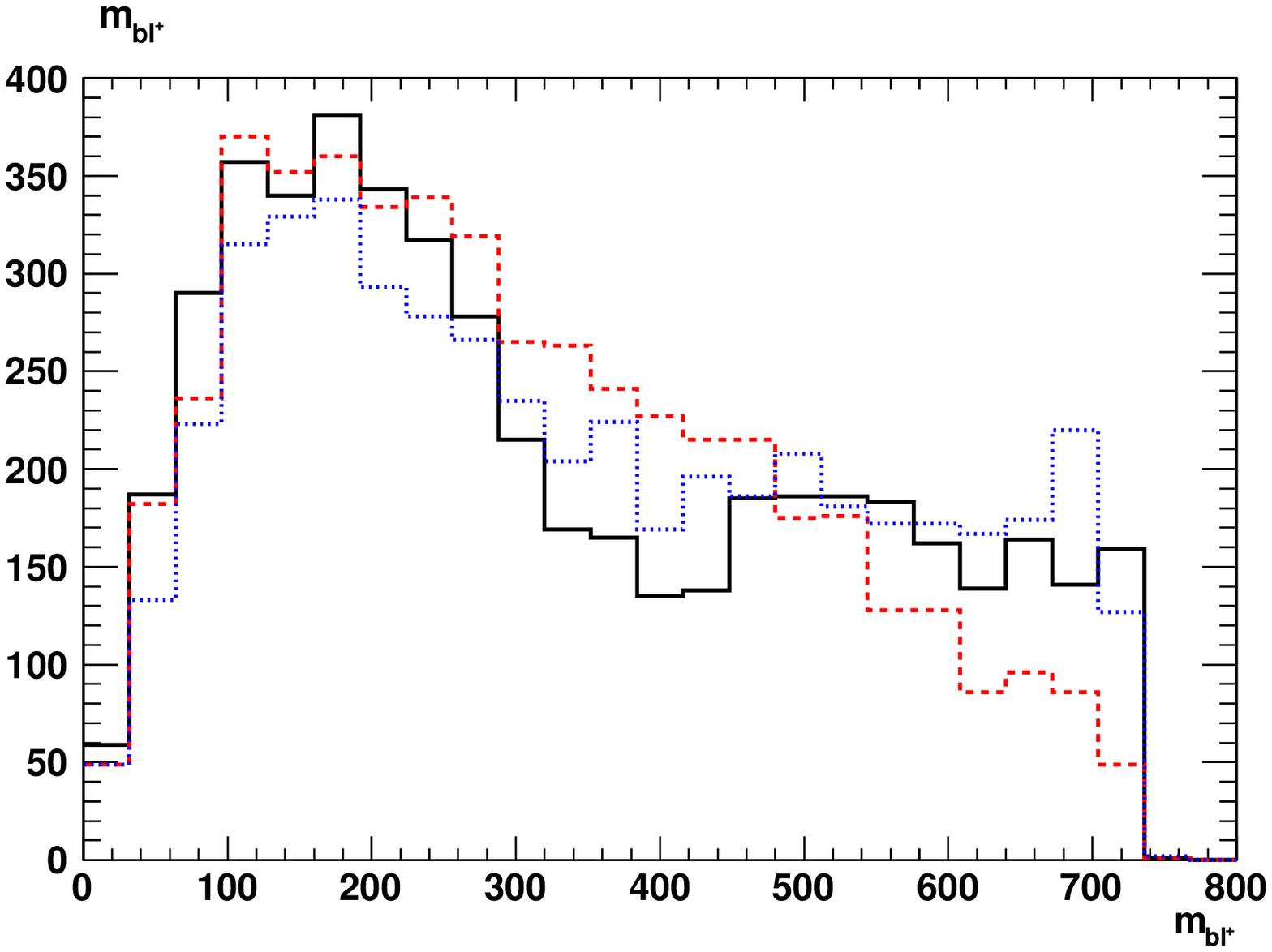}
  \includegraphics[width=0.47\textwidth]{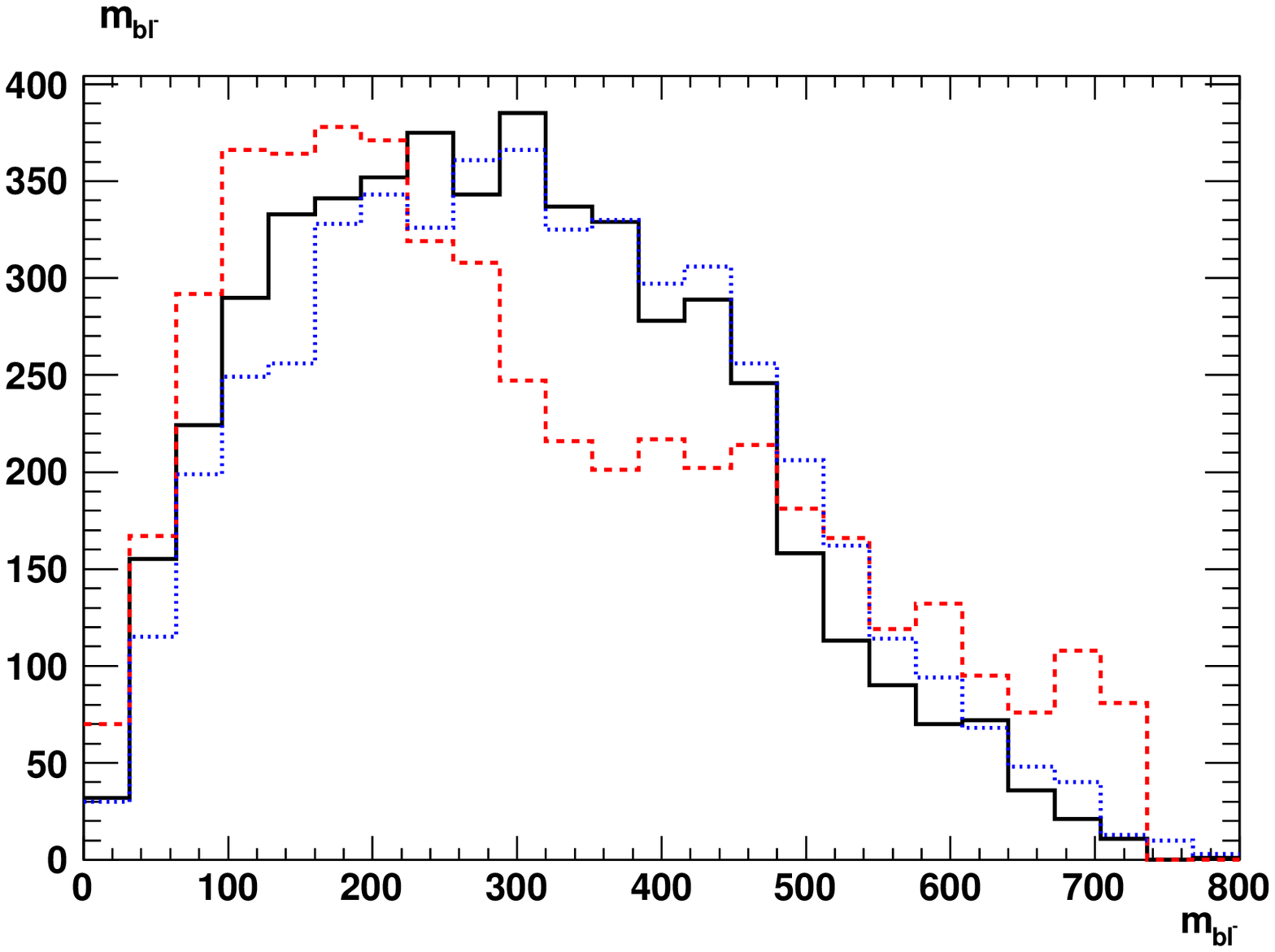}\\
  \includegraphics[width=0.47\textwidth]{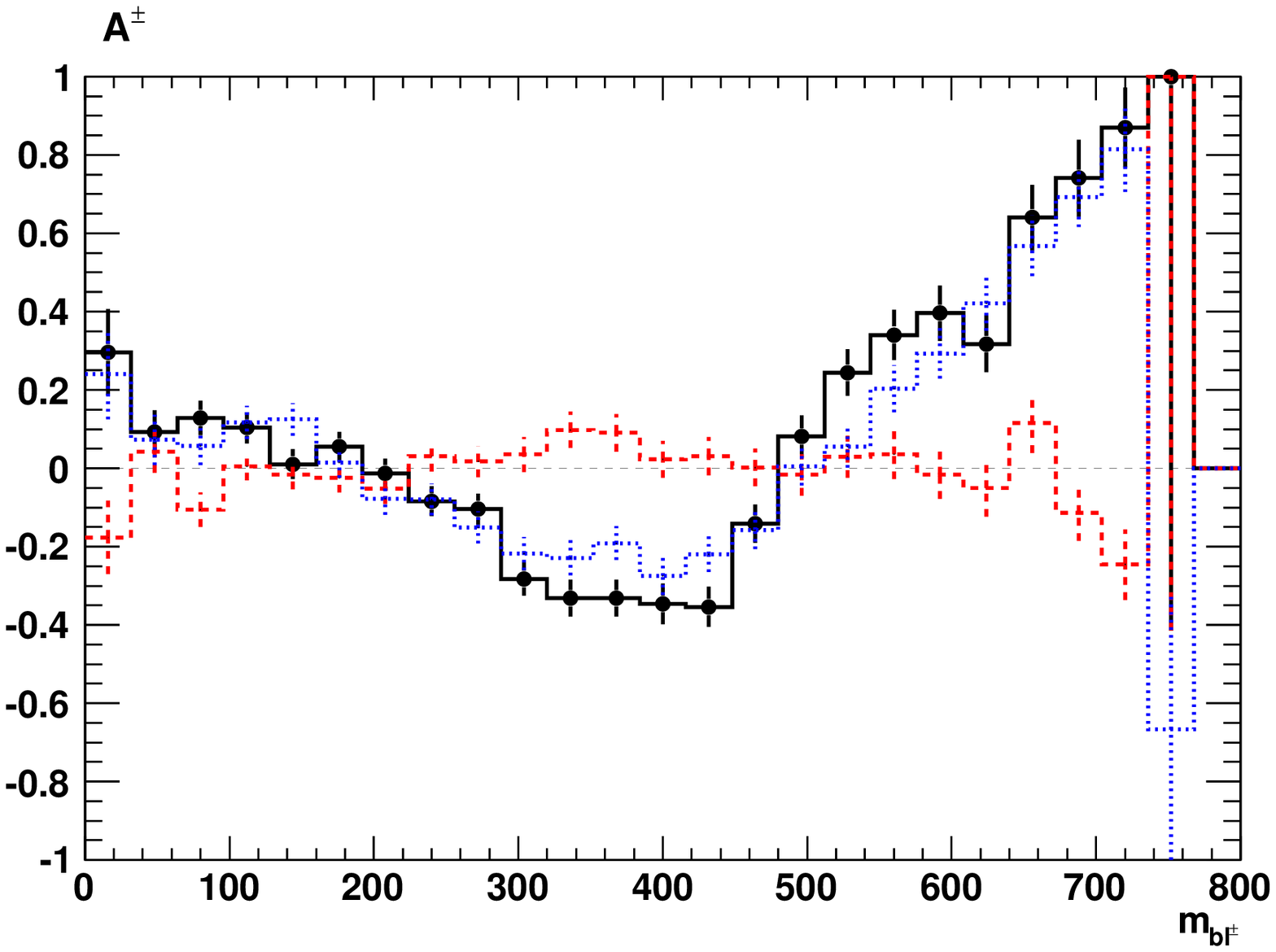}
  \includegraphics[width=0.47\textwidth]{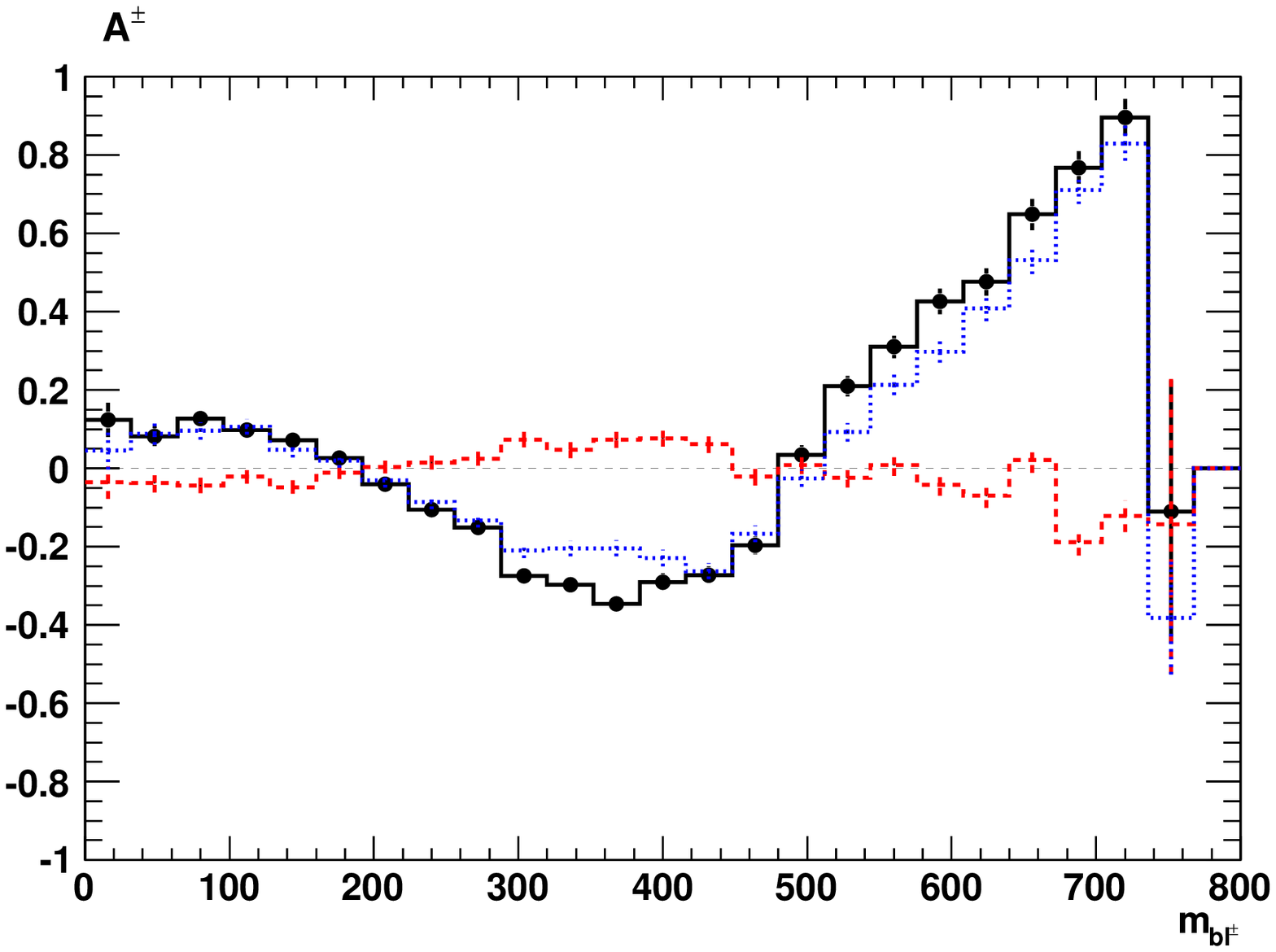}\\
  \caption{Invariant masses $m_{b\ell^+}$ (upper left) and
    $m_{b\ell^-}$ (upper right) with a bottom quark and their
    asymmetry $A^\pm$ for 5k (lower left) and 25k (lower right)  
    events. The black (solid) and blue (dotted) line
    correspond to $\gamma =$ 0.5 \%
    and 15.0 \%, respectively. The UED sample is given by the red
    (dashed) line.} 
  \label{fig:m_bl_asymm}
\end{figure}
Fig.~\ref{fig:m_bl_asymm} depicts both the invariant mass
distributions of $m_{b\ell^+}$ and $m_{b\ell^-}$ as well as the
bottom-lepton asymmetries, for 5k and 25k events. To keep the plots
clear and at the same time display all relevant information, we
refrain from using all widths, but rather restrict ourselves to the
most extreme values of $\gamma = 0.5$ and $ 15\%$, given by the black
(solid) and blue (dotted) lines as well as the UED sample, depicted by
the red (dashed) line. 

To clarify whether the deviation of the large-width sample is merely a 
statistical effect, we artificially increased the event number by a
factor of 5 up to 25k. As it turns out in Fig.~\ref{fig:m_bl_asymm}
(on the lower right) there is indeed a subtle effect that is 
observable in the large width SUSY sample, which is well beyond the
size of fluctuations, although both of the two invariant mass
distributions show no strong discrepancies with respect to the two
different values of $\gamma$. Nonetheless, the minimum plateau from
300 to 400 GeV as well as the subsequent rise from 500 to 800 GeV are
both reduced by up to one third in magnitude. As is evident through
direct comparison with the superimposed UED sample however, it is
obvious that large off-shell contributions are not endangering a
possible discrimination of the fundamentally different spin
scenarios, using this specific variable. 

\paragraph{$\boxed{A^\pm_s (m_{b\ell})}$} Next, we investigate the
impact another bottom-lepton asymmetry. The difference to the
first asymmetry discussed above is given by the spectrum dependent
property that the softer b-quark may coincide with the nearer
b-quark. The definition of this asymmetry is given by:
\begin{align}
  A^\pm_s (m_{b_s\ell}) = \frac{d\sigma / dm_{b_s\ell^+} - d\sigma /
    dm_{b_s\ell^-}}{d\sigma / dm_{b_s\ell^+} + d\sigma / dm_{b_s\ell^-}}. 
\end{align}
In fact, in our scenario, this assignment is true most of the time:
for small width $\gamma = 0.5 \%$ in $4557/5000 \sim 90 \%$ of all
events, the near bottom quark is also the softer one. Apparently, this
changes when the width is increased, as is illustrated in
Fig.~\ref{fig:pt_bottom_near_far}. However, even for the largest width
of $\gamma = 15 \%$, in $4109/5000 \sim 80\%$ of all events this
assumption is still correct. Obviously, this characteristic will
drastically change when the mass gap between the gluino and the sbottom is
increased. The nearer quark will become harder due to a larger phase
space in the gluino decay, and the value of $A^\pm_s$ will consequently
be reduced.  
\begin{figure}[!t]
  \includegraphics[width=0.47\textwidth]{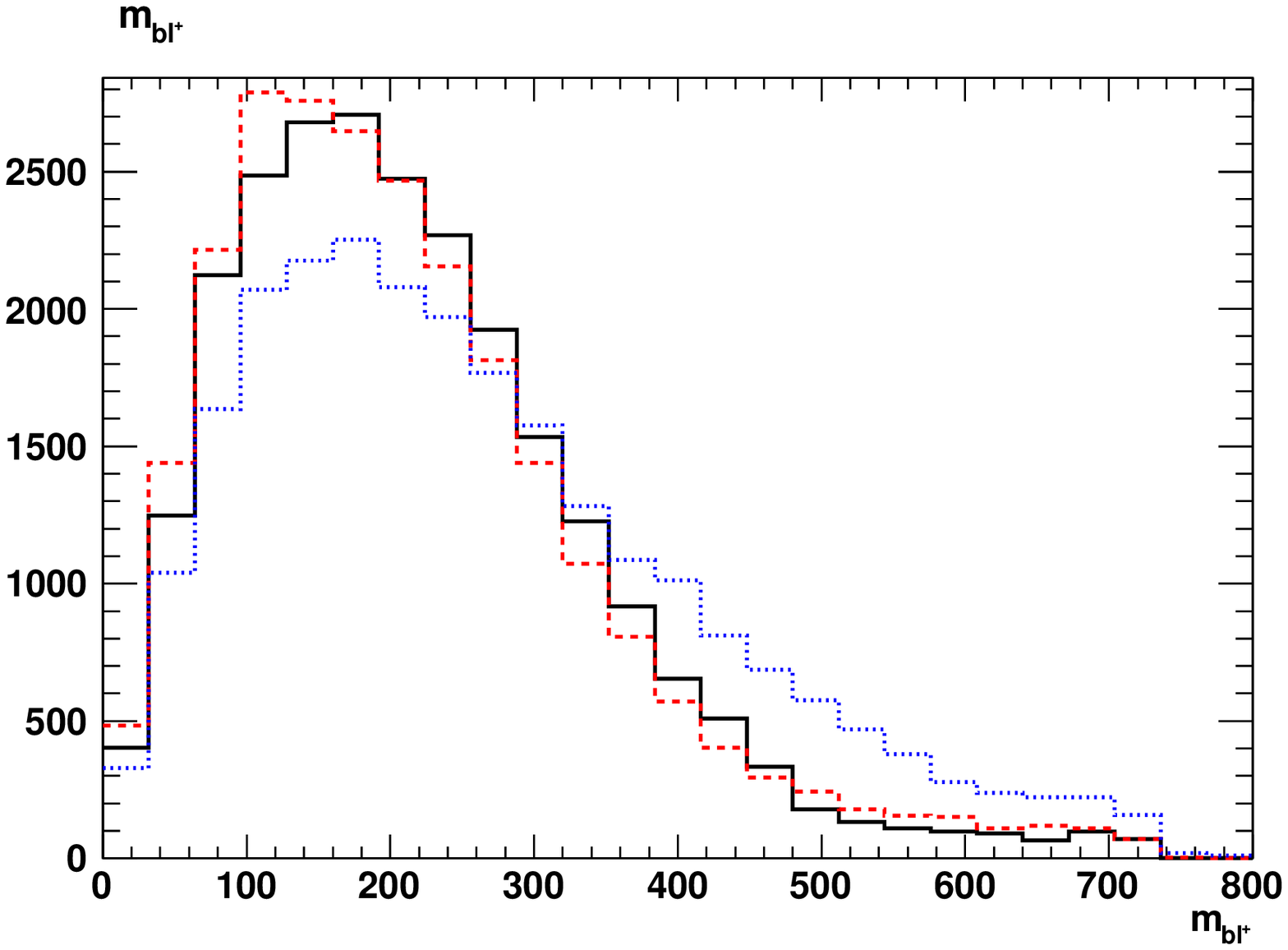}
  \includegraphics[width=0.47\textwidth]{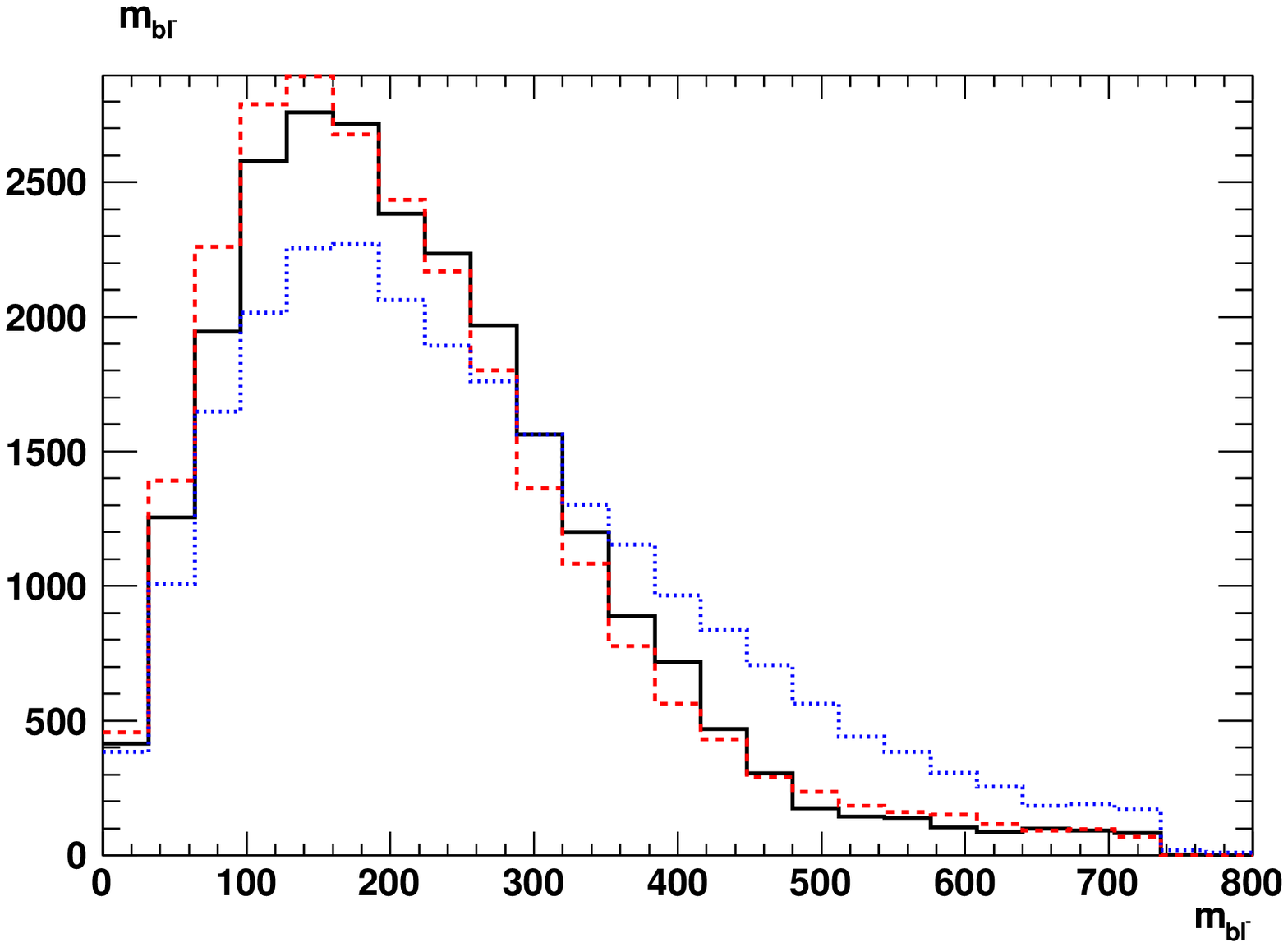}\\
  \includegraphics[width=0.47\textwidth]{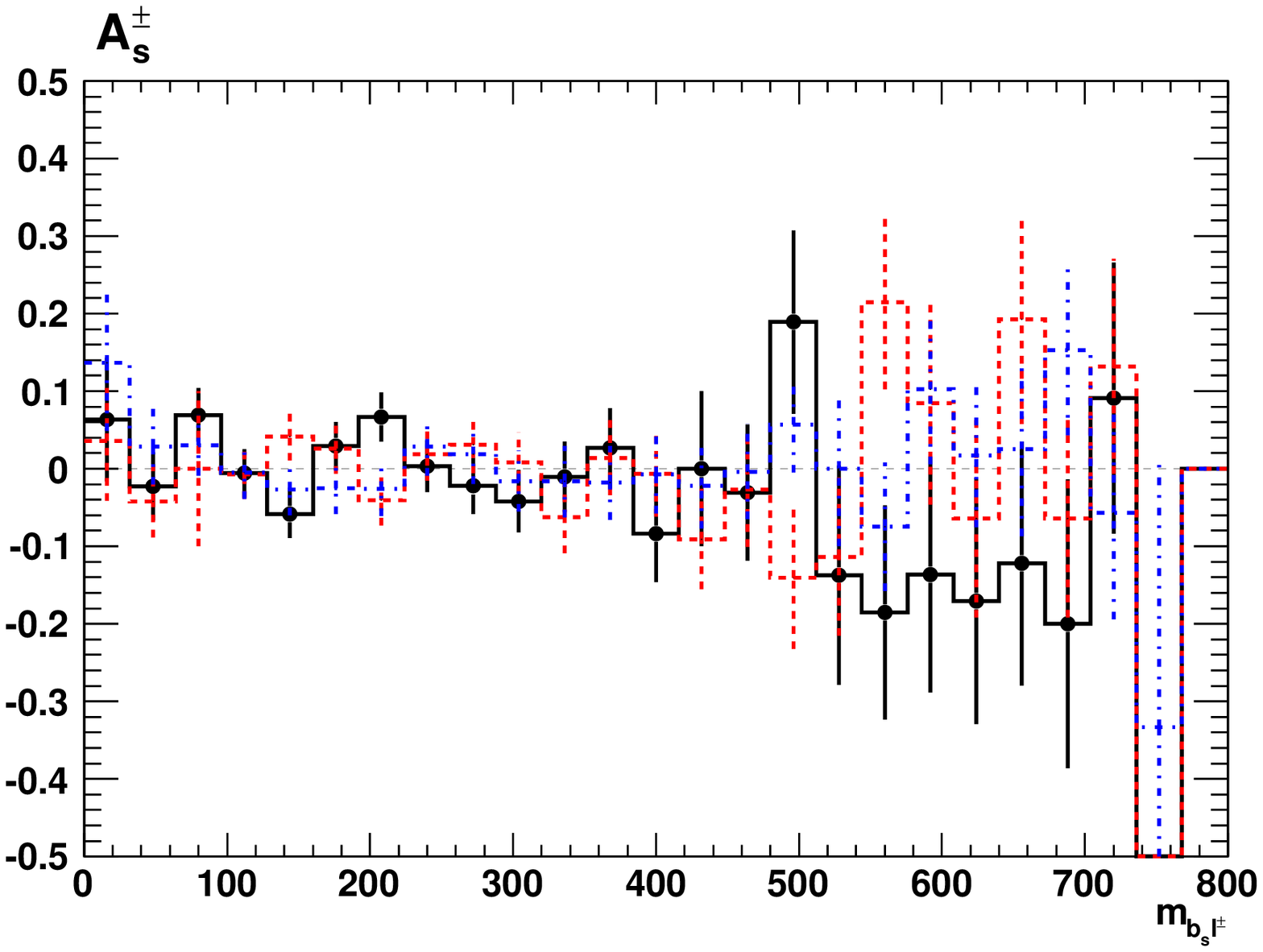}
  \includegraphics[width=0.47\textwidth]{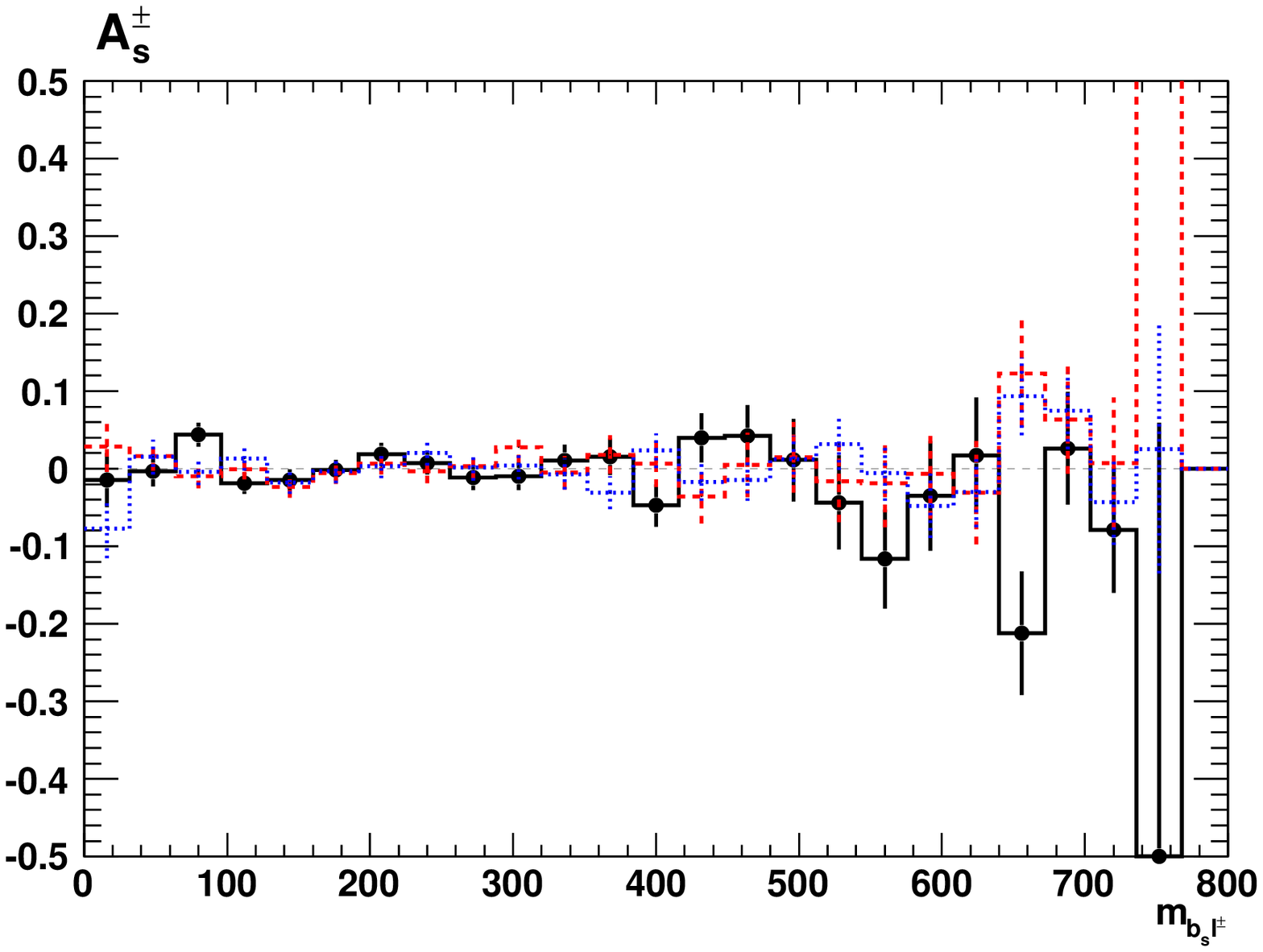}\\
  \caption{Invariant masses $m_{b_s\ell^+}$ (upper left) and
    $m_{b_s\ell^-}$ (upper right) with the softer of the two bottom
    quarks and their asymmetry $A^\pm_s$ for 5k (lower left) and 25k
    (lower right) events. The color scheme corresponds to the 
    one in Fig.~\ref{fig:m_bl_asymm}}. 
  \label{fig:m_bl_asymm_soft}
\end{figure}
In the upper line of Fig.~\ref{fig:m_bl_asymm_soft} we show both
invariant masses, $m_{b_s\ell^+}$ and $m_{b_s\ell^-}$, which exhibit the
typical smearing behavior for large widths encountered more often in
the mass measurement section above. Already by eye, the asymmetries in
the lower line of Fig.~\ref{fig:m_bl_asymm_soft} have a comparable
pattern: both invariant mass distributions have similar shapes and the
corresponding asymmetries are thus rather small. Regardless of the
size of the effective width $\gamma$, the distortion of $m_{b_s\ell^+}$
mimics the one of $m_{b_s\ell^-}$, and the same holds for the
asymmetries. Consequently, the small event sample of 5k is fully
compatible with a vanishing asymmetry throughout the complete range of
the histogram, not only for the two SUSY samples, but also for
UED. Moreover, the larger samples of 25k events also have only minor
deviations from $A^\pm_s \equiv 0$, which might be attributed to
statistical fluctuations. Altogether, we find the size of this
second bottom-lepton asymmetry to be of negligible size
compared to the already small deviations found
in~\cite{Alves:2006df}. Hence, it is not surprising that a steady 
distortion for both $m_{b_s\ell^+}$ and $m_{b_s\ell^-}$ results in a
negligible change of an already very small asymmetry. We conclude that
this last asymmetry is neither preferable in terms of discriminative
power between SUSY and UED nor for suffering from smearing due to
off-shell effects. After all, this should be attributed to the specific
kind of underlying mass spectrum. 


\subsection{Hadronic Angular Correlations} 

It was further proposed to analyse purely hadronic
correlations such as the average pseudo-rapidity $\bar \eta_{bb}$ or
the difference of azimuthal angles $\Delta \phi_{bb}$ of the two
bottom quarks~\cite{Alves:2006df}.  
\begin{figure}[h!]
  \includegraphics[width=0.47\textwidth]{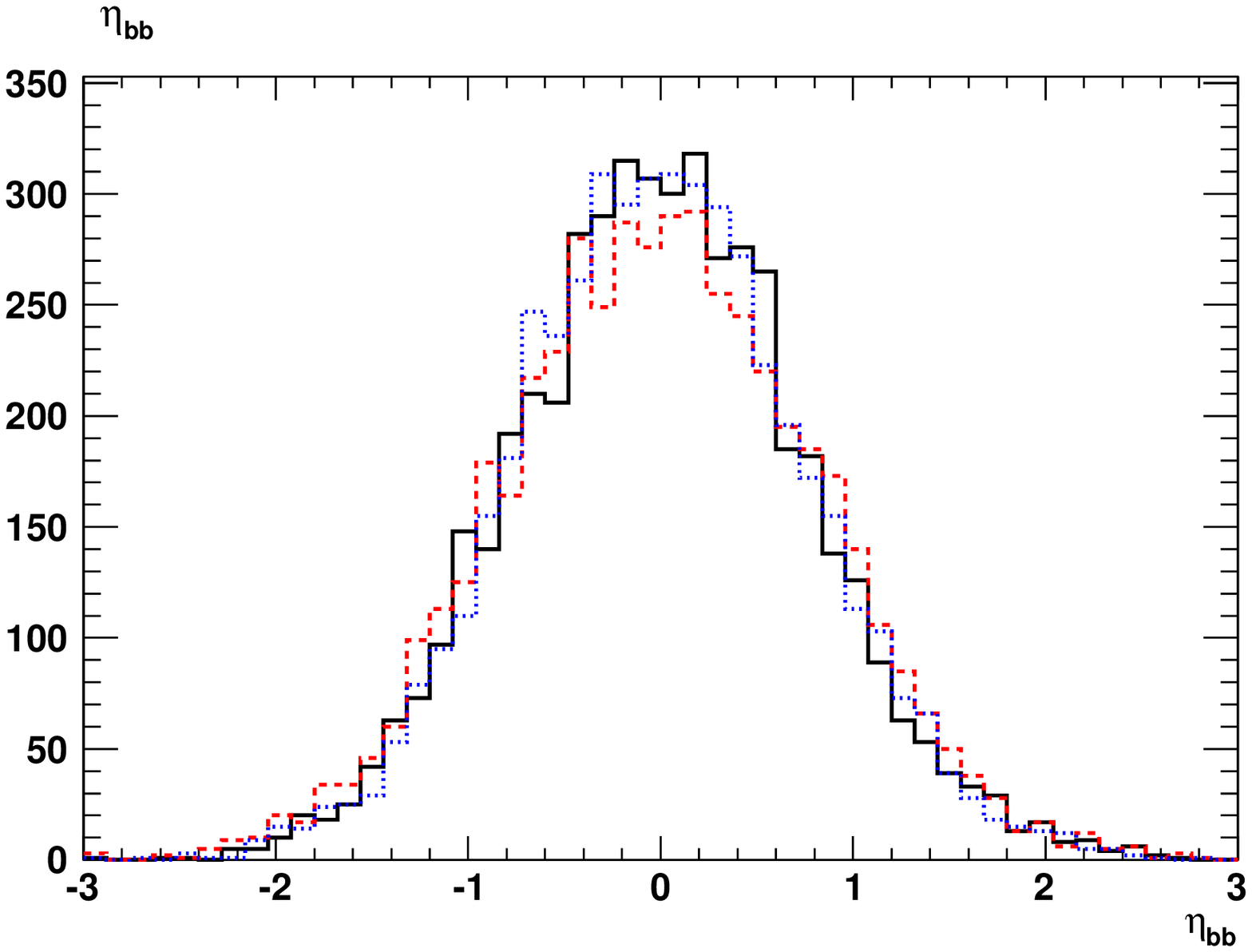}
  \includegraphics[width=0.47\textwidth]{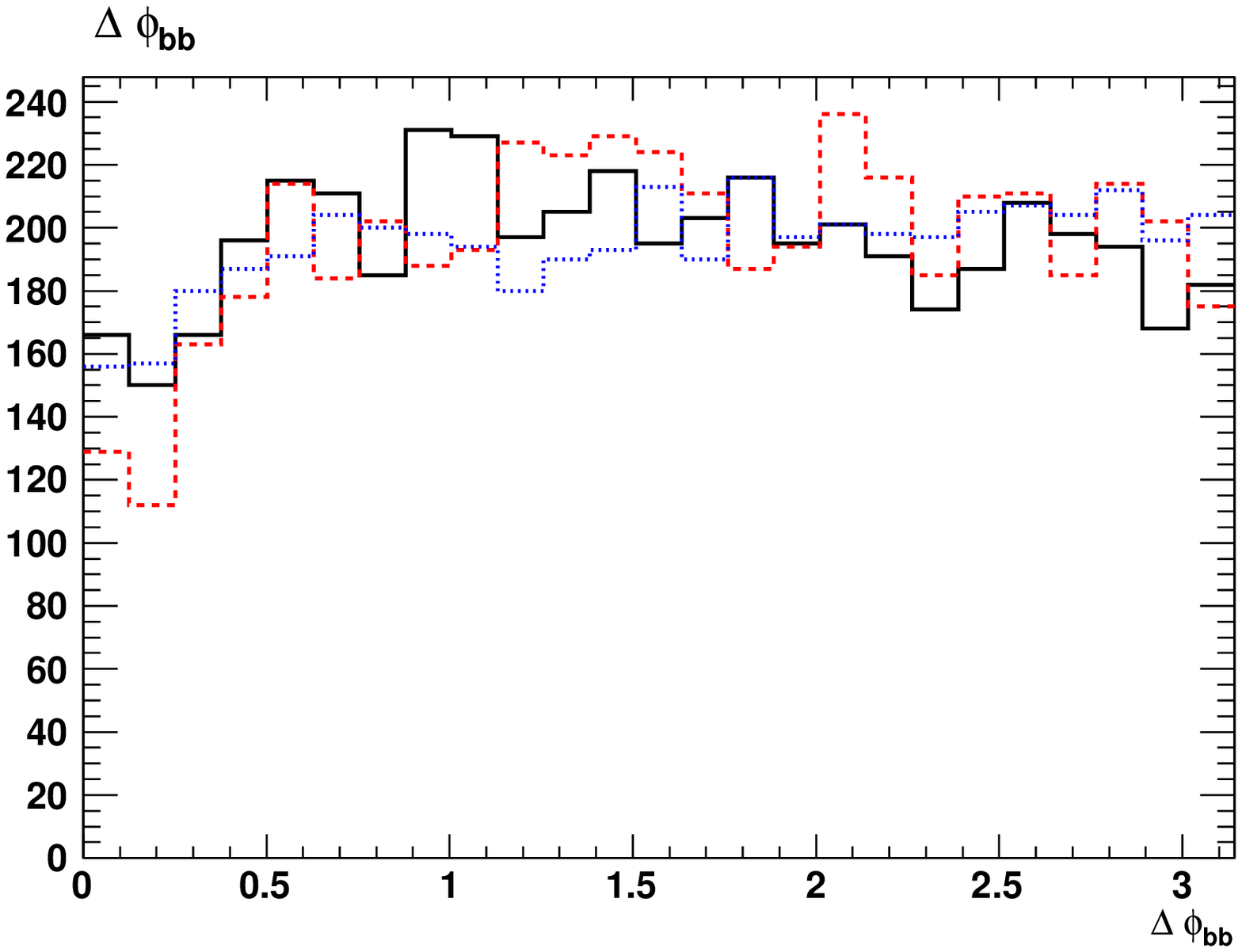}\\
  \includegraphics[width=0.47\textwidth]{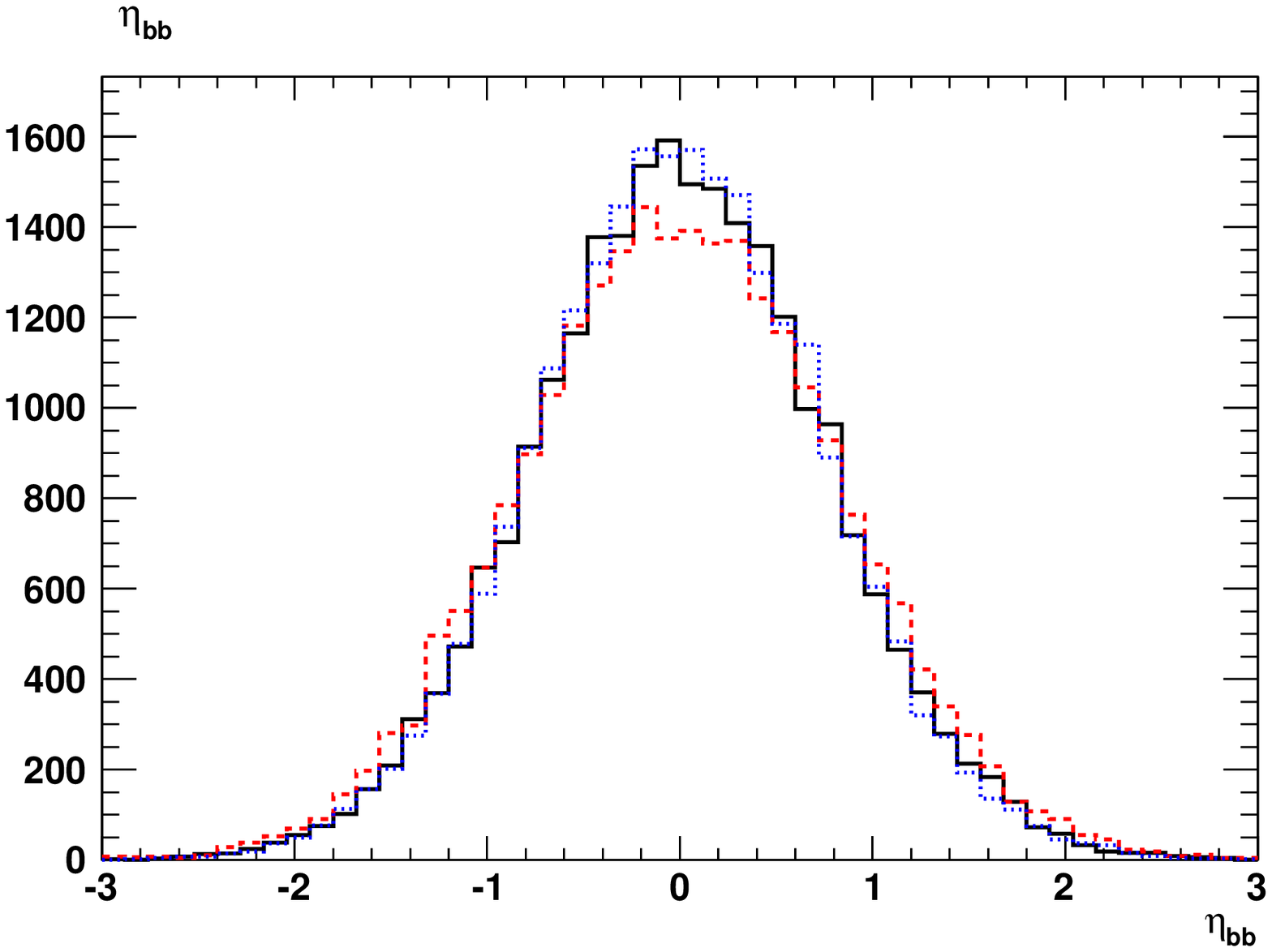}
  \includegraphics[width=0.47\textwidth]{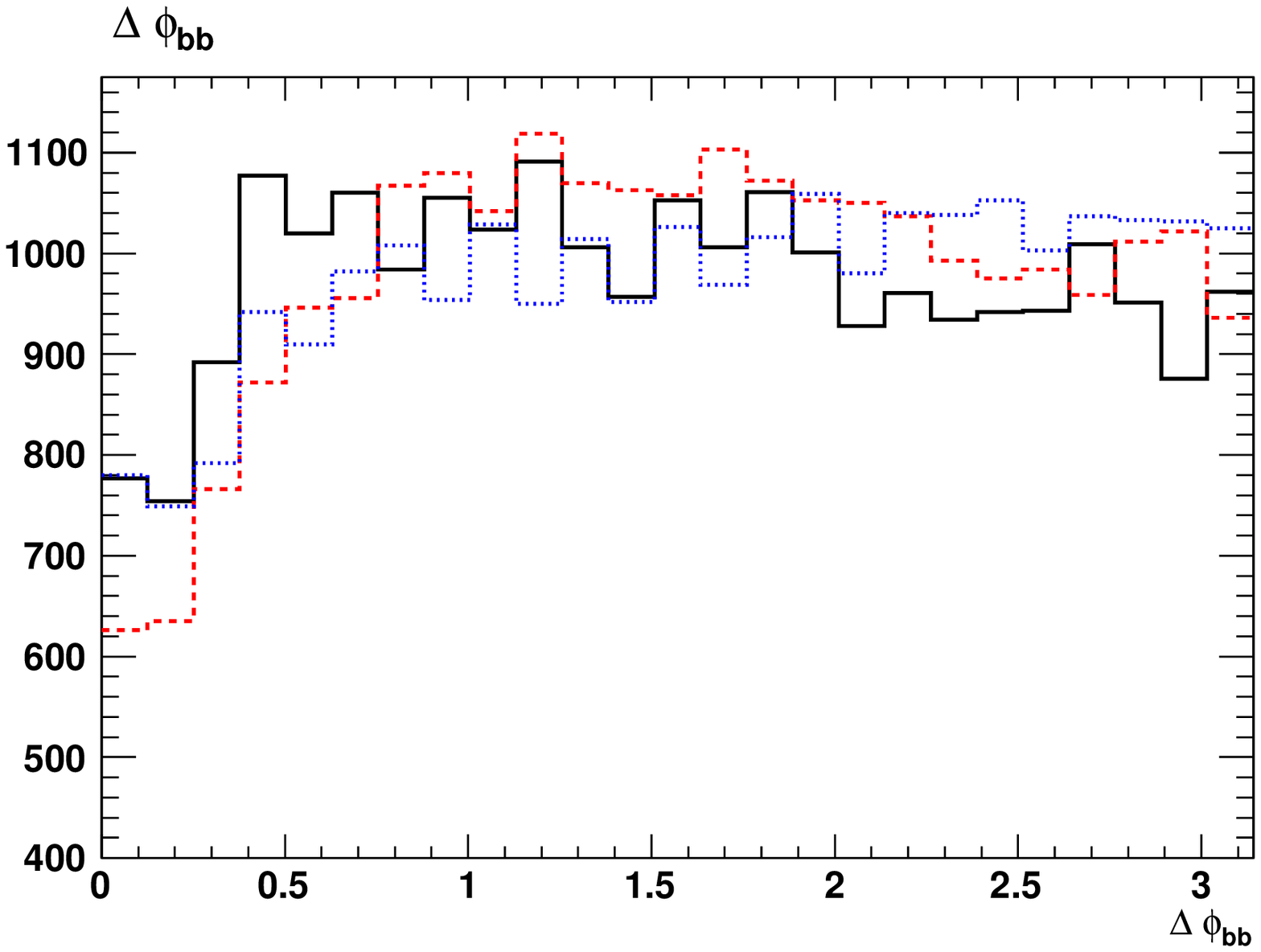}\\
  \caption{Average pseudo-rapidity (left) and azimuthal distance
    (right) of both bottom quarks for 5k (top) and 25k (bottom)
    events. The coloring corresponds to the one in
    Fig.~\ref{fig:m_bl_asymm}.}   
  \label{fig:dphi_eta_bb}
\end{figure}
The first of these observables exhibits no visible distortion with
respect to the maximal width $\gamma$ in the supersymmetric sample,
although a small difference to the UED case is perceptible. We confirm
the results of~\cite{Alves:2006df}, where a slightly more central
behavior of the two $b$-quarks was found for SUSY signals. For the
azimuthal distance, we find a non-negligible deviation
(cf. Fig.~\ref{fig:dphi_eta_bb}) of the large width SUSY sample from
the standard SUSY sample. In addition, the 
former turns out to resemble the shape of UED events. As a next step,
for each of the two variables the following asymmetries are
defined~\cite{Alves:2006df}:  
\begin{align}
  A^\pm_\eta &= \frac{N (|\bar \eta_{bb}| < 1) - N (|\bar \eta_{bb}| >
    1) }{N (|\bar \eta_{bb}| < 1) + N (|\bar \eta_{bb}| >
    1)} \label{eq:asym_eta}\\ 
  A^\pm_\phi &= \frac{N (\Delta \phi_{bb} < \pi/2) - N (\Delta
    \phi_{bb} > \pi/2) }{N (\Delta \phi_{bb} < \pi/2) +  N (\Delta
    \phi_{bb} > \pi/2)}, \label{eq:asym_phi} 
\end{align}
which were proposed to obtain an additional measure that allows for a
discrimination between a standard SUSY and a UED signal. We apply these
variables to our three scenarios, namely the standard and off-shell
SUSY as well as the UED samples. Numerical values for all of these are
given in Table~\ref{tab:phi_eta_bb_vals} for 5k and 25k events,
respectively.  
\begin{table}[!ht]
\centering
\begin{tabular}{c|cc}
sample & 5k & 25k\\
\hline
 $A^\pm_\eta$ (std) &0.627 $\pm$ 0.017  &0.628 $\pm$ 0.008  \\
 $A^\pm_\eta$ (ofs) &0.645 $\pm$ 0.017  &0.645 $\pm$ 0.008  \\
 $A^\pm_\eta$ (ued) &0.567 $\pm$ 0.016  &0.557 $\pm$ 0.007  \\
\hline
 $A^\pm_\phi$ (std) &0.014 $\pm$ 0.014  & 0.005 $\pm$ 0.006 \\
 $A^\pm_\phi$ (ofs) & -0.047 $\pm$ 0.014& -0.052 $\pm$ 0.006\\
 $A^\pm_\phi$ (ued) & -0.042 $\pm$ 0.014& -0.039 $\pm$ 0.006\\
\hline
 $A^\pm_{ct}$ (std) &0.194 $\pm$ 0.015  &0.180 $\pm$ 0.007\\
 $A^\pm_{ct}$ (ofs) &0.125 $\pm$ 0.014  &0.129 $\pm$ 0.006\\
 $A^\pm_{ct}$ (ued) &0.003 $\pm$ 0.014 &0.008 $\pm$ 0.006\\
\end{tabular}
\caption{Numerical figures for the three asymmetries $A^\pm_\phi$,
  $A^\pm_\eta$ and $A^\pm_{ct}$ defined in equations
  \eqref{eq:asym_phi}, \eqref{eq:asym_eta} and \eqref{eq:asym_ctqq}
  for different scenarios: std, ofs and ued correspond to the standard
  ($\gamma = 0.5\%$) and off-shell ($\gamma = 15\%$) SUSY as well as
  UED event samples. Errors are purely statistical.} 
\label{tab:phi_eta_bb_vals}
\end{table}
The value of the asymmetry $A^\pm_\eta$ of the average pseudo-rapidity
$\bar \eta_{bb}$ only exhibits a marginal increase for the off-shell
SUSY sample compared to the undistorted standard one. Both coincide
within at most two sigma of the purely statistical error. The UED case
on the other hand has less central values than the off-shell SUSY
parts. This difference to both the standard SUSY sample as well as to
the UED sample should be attributed to the fact that the off-shell
contributions tend to harden (at least the first) decay
product(s) (recall the $p_T$ distributions of the near and far 
bottom quarks in Figure \ref{fig:pt_bottom_near_far}), and hence
allow for even more central values of $\eta$.  

The situation looks a lot less promising for the azimuthal distance
$\Delta \phi_{bb}$ of the two bottom quarks. Here, we find that
off-shell contributions in the large-width SUSY sample drive the
asymmetry $A^\pm_\phi$ such that a discrimination between SUSY and UED
is no longer possible. Although the shapes of the distributions differ
only by a moderate amount visible by eye in all three scenarios
(cf.~Fig.~\ref{fig:dphi_eta_bb}, upper right), the off-shell SUSY
sample is washed out in such a way so as to drive the quantitative
numerical estimate of the asymmetry negative by the same amount as in
the case of UED (cf.~Table~\ref{tab:phi_eta_bb_vals}).  


\subsection{Inclusive Angular Distributions}

Finally, we investigate angular correlations of the initially produced
particles, which in our case corresponds to gluinos or
KK gluons. Although we are not able to reconstruct the complete mother
particle momenta due to missing energy, the first emitted partons of
each decay cascade should still possess an observable angular
correlation among each other. This was first used in the variable 
\begin{align}
 \cos \theta_{ll}^* = \tanh \left( \frac{\Delta \eta_{ll}}{2}\right)
\end{align}
in a study of slepton pair production \cite{Barr:2005dz}, and later
adapted to general colored SUSY
production~\cite{MoortgatPick:2011ix}. The adapted method was 
applied to fully hadronized inclusive signal event samples with gluino
and squark contributions or the corresponding equivalent for UED,
where the largest discriminative power is attributed to the squark and
KK quark signatures. Our study on the other hand is based on a
parton-level analysis, and we restrict ourselves to subsamples with
gluinos, since we aim to assess the impact of their off-shell
contributions on the method. More precisely, we apply the variable to the
\textit{exclusive} gluino benchmark process in our scenario introduced
in the beginning in the following way: 
\begin{align}
 \cos \theta_{qq}^* = \tanh \left( \frac{\Delta \eta_{q_lq_r}}{2}\right)
\end{align}
where $q_l = \min (b_1,b_2)$ and $q_r = \min (q_1,q_2)$ are the softer
of the two quarks from each cascade side. While
in~\cite{MoortgatPick:2011ix} the largest contributions arise from
prompt squark decays to quarks and the lightest neutralinos, and thus
$\cos\theta_{qq}^*$ is chosen to be applied to the two hardest objects, we
make particular use of the (unfortunately spectrum-dependent) approach
of selecting the softer quarks to be attributed to the first (near)
gluino or KK-gluon decay products. These are furthermore assumed to
inherit features of the initially produced mother particles. 
\begin{figure}[!t]
  \includegraphics[width=0.47\textwidth]{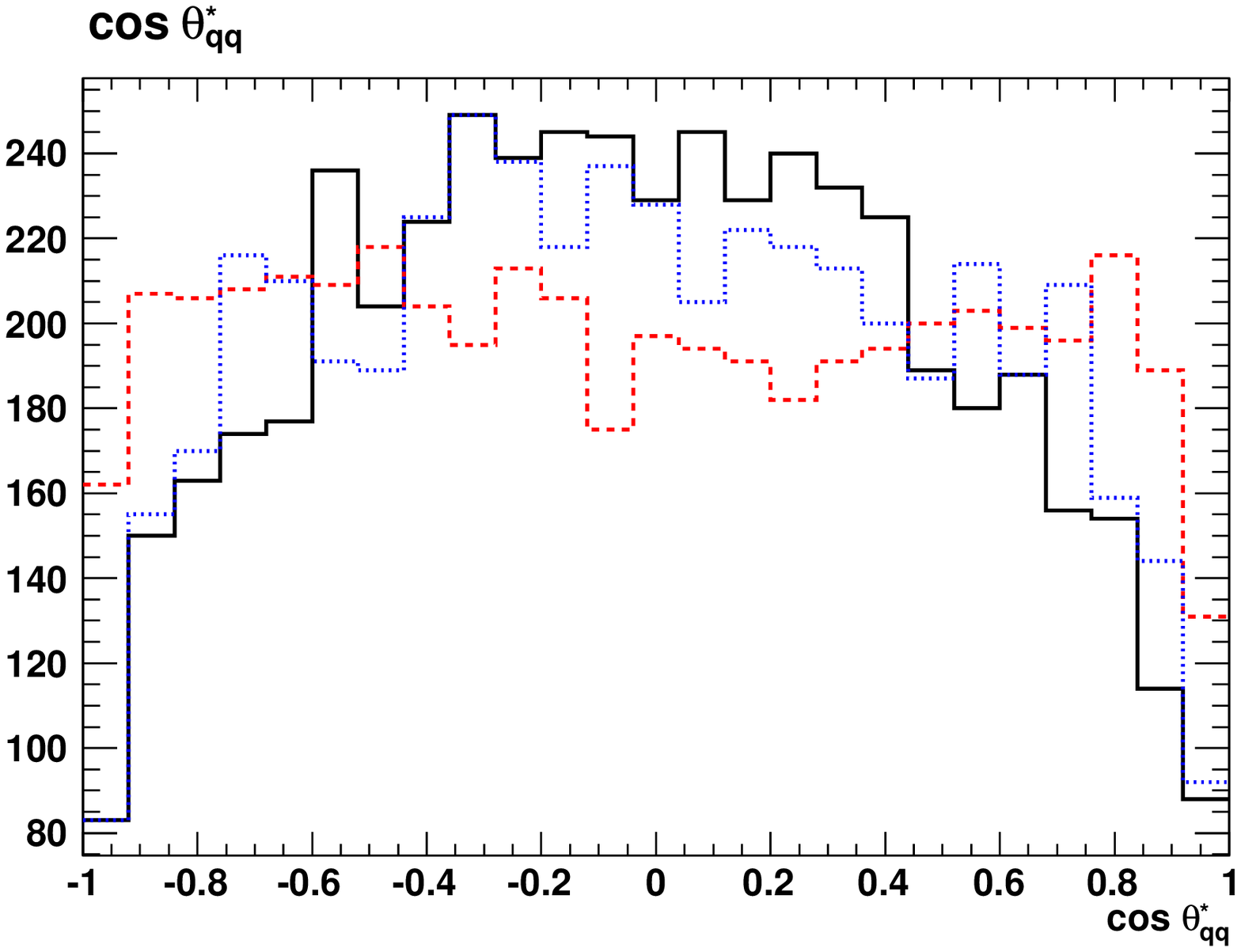}
  \includegraphics[width=0.47\textwidth]{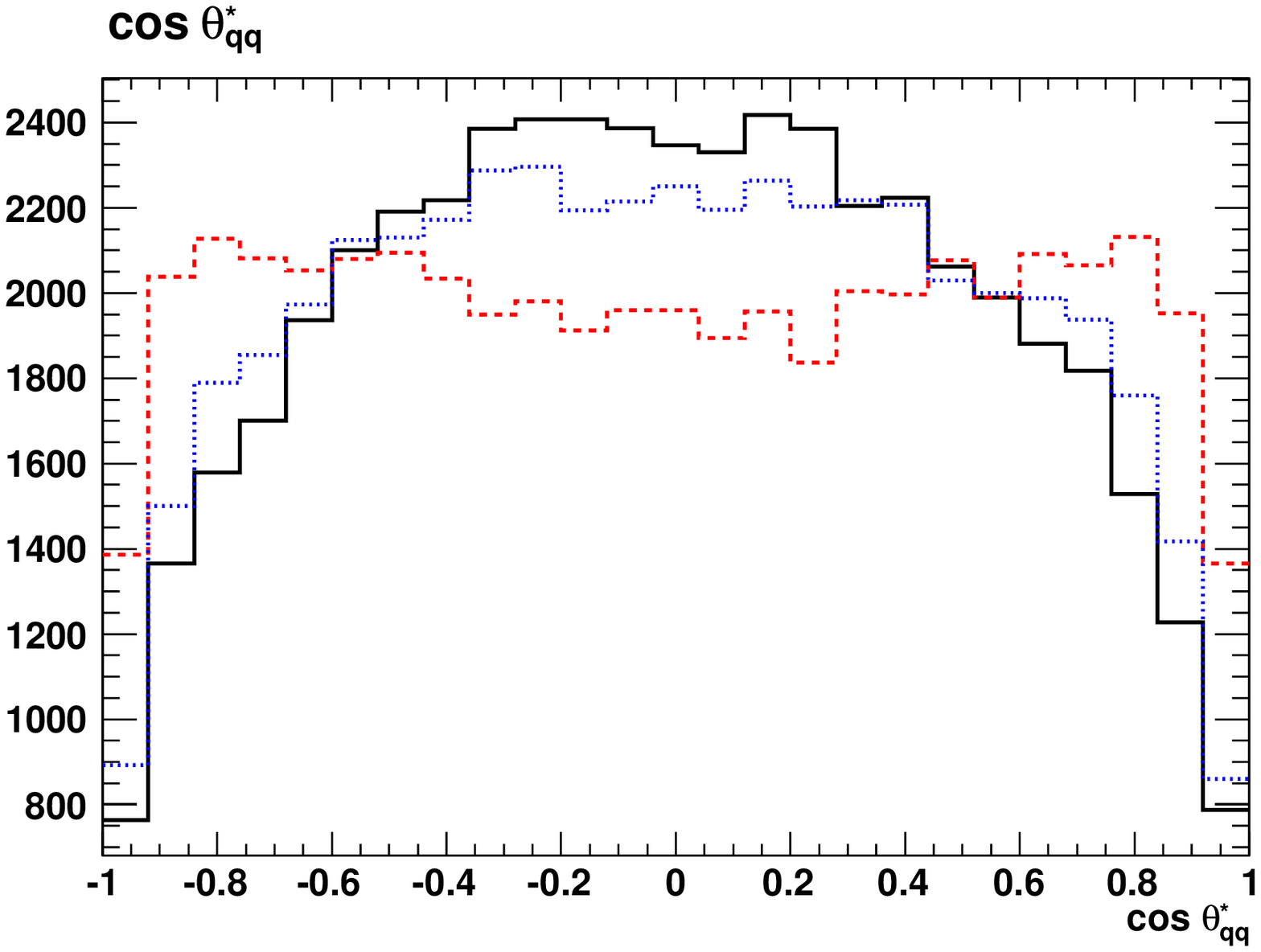}\\
  \caption{$\cos \theta_{qq}^*$ for 5k (left) and 50k (right)
    events. The color scheme is the same as in
    Fig.~\ref{fig:m_bl_asymm}.}  
  \label{fig:cos_theta}
\end{figure}
Fig.~\ref{fig:cos_theta} illustrates the behavior of $\cos
\theta_{qq}^*$ for the two SUSY scenarios with different widths (black
and blue) and the UED scenario (red) for 5k and 50k events,
respectively. We included the high statistics sample to show the
asymptotic behavior of the fundamentally different models. The
distortion due to off-shell contributions is apparent, and although
they tend to wash out the distribution to less central values, the
differentiation with respect to UED is not endangered in the exclusive
gluino subsample. A quantification of this statement can be obtained
through the definition of the following
asymmetry~\cite{MoortgatPick:2011ix}:  
\begin{align}
  A^\pm_{ct} = \frac{N (|\cos \theta_{qq}^*| < 0.5) - N (|\cos
    \theta_{qq}^*| > 0.5)}{N (|\cos \theta_{qq}^*| < 0.5) + N (|\cos
    \theta_{qq}^*| > 0.5)} \qquad ,
  \label{eq:asym_ctqq} 
\end{align}
whose values for the three cases are given in
Table~\ref{tab:phi_eta_bb_vals}. The initial observations from
Fig.~\ref{fig:cos_theta} are confirmed: while the value of
$A_{ct}^\pm$ is indeed reduced by one third in the \textit{standard}
SUSY sample with respect to the off-shell SUSY sample, the UED case is
compatible with a value of zero. Hence we conclude, that although an
there is an apparent modification of $\cos\theta_{qq}^*$ due to
finite-width effects, it is not threatening the discrimination of SUSY
and UED models.  


\section{Conclusions}
\label{sec:conclusions}

In this paper, the impact of off-shellor finite-width effects from a
broad (``fat'') gluino on a representative selection of observables
for mass and spin determination is investigated with a discrete set of
effective width-to-mass ratios $\gamma = \Gamma / M \in \{0.5 \%, 2.5 \%, 5.0
\%, 10.0 \%, 15.0 \%\}$. Such larger widths of gluinos are not
uncommon in certain regions of parameter space, and are e.g. quite
natural in GMSB scenarios. The basis for this analysis are SUSY
signals from gluinos where production and the first decay step are 
simulated with a full matrix-element calculation to account for
non-resonant contributions in the gluino propagator and interferences,
distorting kinematic observables. All subsequent decays are then
factorized using the NWA, including full spin correlations. An adapted 
edge-to-bump method is employed to quantify deformations in kinematic
distributions, which arise in scenarios with large effective
width-to-mass ratios $\gamma$. In this approach, several parameters
from a naive linear kink function fitted to the corresponding
variables are extracted and utilized to quantify errors in the mass
and spin determination arising from finite-width effects. In general,
mass measurement observables are much more 
obviously affected in that the endpoint smearing dominates the
distributions of otherwise sharp edge structures already at moderate
width-to-mass ratios of $\gamma =
5\%$, and steadily increases throughout higher values of $\gamma$. As a
result, the endpoint positions obtained with the edge-to-bump method
are drastically shifted already at the parton level. The difference
and the ratio of the two slope parameters in the fit are additional
measures of distortion which similarly indicate a washed-out endpoint 
behavior. 

Methods of spin determination are on the other hand less
affected by propagator contributions far off the mass-shell. Shape
asymmetries as well as angular correlations exhibit only small
deviations of event samples with broad gluinos from the ones with a
narrow resonance. In the special case of azimuthal
distance, the difference of these two event samples however is of
considerable size, and although the shapes of the broad SUSY as well
as the UED sample still differ, the numerical values of the associated
asymmetry $A_{\phi}^\pm$ coincide within their purely statistical
errors. 

In summary, our studies show that broader BSM resonances as
in our case a fat gluino lead to severe distortions of
kinematic distributions which are the basis for many mass and spin
determination methods. These deviations from non-resonant
contributions arise at the fundamental parton level in simulations
where production and decay are simulated with the full matrix
element. Many existing studies on the other hand utilise the NWA,
whose predictions vigorously differ from the full calculation. Hence a
correct treatment of such effects is of crucial importance for both
mass and spin measurements so as to not misinterpret the potential
signals of much sought-after new physics. While the spin determination
might not suffer too much in such cases, broader particles like a fat
gluino might make life hard for mass determinations at the LHC.


\subsubsection*{Acknowledgements}

We would like to thank N.~Pietsch, K.~Rolbiecki, and K.~Sakurai for
fruitful and enlightening discussions. JRR wants to thank
D.~Rainwater, who partially initiated the idea for the studies
performed here. 


\end{document}